\documentclass[twocolumn]{aastex631}
\usepackage{booktabs} 
\usepackage{multirow}
\usepackage{makecell}
\usepackage{array} 
\usepackage{xcolor}
\usepackage{gensymb}
\usepackage{multirow}
\usepackage{amsmath}
\usepackage{enumitem}
\usepackage{placeins}
\usepackage{soul}
\usepackage{float} 

\hypersetup{
	colorlinks=true, 
	linktoc=all,     
	linkcolor=blue,  
}

\shorttitle{AGN jet neutrinos}
\begin{document}

\title{Multi-messenger emission derived from relativistic magnetized jet dynamics using a multi-zone framework}

\author[0009-0000-6497-3336]{Harshita Bhuyan}
\affiliation{Department of Astronomy, Astrophysics and Space Engineering, Indian Institute of Technology Indore, India}

\author[0000-0001-5424-0059]{Bhargav Vaidya}
\affiliation{Department of Astronomy, Astrophysics and Space Engineering, Indian Institute of Technology Indore, India}

\author[0000-0002-3528-7625]{Christian Fendt}
\affiliation{Max Planck Institute for Astronomy, K\"onigstuhl 17, 69117 Heidelberg, Germany}

\author[0009-0006-9732-4248]{Aditya Sharma}
\affiliation{Department of Astronomy, Astrophysics and Space Engineering, Indian Institute of Technology Indore, India}

\date{\today}
 
\begin{abstract}
Relativistic jets from Active Galactic Nuclei (AGN) are highly energetic and emit radiation across a wide range of frequencies. 
Despite several observational studies, their particle composition still remains a key open question. 
The detection of high-energy neutrinos from blazar sources such as TXS 0506+056 has highlighted the plausibility of 
hadronic/lepto-hadronic models for AGN jets. 
To understand the origin of high-energy neutrinos from such sources, it is imperative to capture the complex interplay between
the jet dynamics, their composition, and the mechanism of particle acceleration and cooling in relativistic jets. 
In this pilot study, we have coupled a numerical multi-zone framework for lepto-hadronic modeling, with 3D relativistic 
magneto-hydrodynamic simulations of AGN jets, including external photon fields. 
Our framework provides synthetic multi-wavelength and neutrino flux by spatially sampling the simulated jet into multiple zones. 
We investigate the implications of such a framework in exploring the different intrinsic and extrinsic pathways for proton-enrichment in jets. 
Essentially, we find that for low proton-to-electron number density ratios, 
producing a substantial jet neutrino flux, requires the underlying proton energy distribution to have a relatively flat spectrum with 
a power-law index of $\simeq 2.0$. 
We further find that while intrinsic shocks triggered by kink-instabilities in the jet can accelerate electrons to high energies, they may not be sufficient to produce such flat particle energy distributions for the chosen set of parsec-scale jet parameters. 
Finally, to produce a significant jet neutrino emission, 
our simulations suggest the need to consider particle acceleration mechanisms through alternative pathways, either internal or external.
\end{abstract}


\section{Introduction}
Active Galactic Nuclei (AGN) launch highly collimated relativistic jets, which can be observed over a wide range of length scales (from pc to Mpc). Radiation emitted by these outflows, are predominantly non-thermal, encompassing almost the entire electromagnetic spectrum. AGN jet plasma, propagating with bulk Lorentz factors ranging from a few up to $\sim 50$ \citep{Lister_2009} and threaded by magnetic fields, gives rise to a range of small and large-scale structures, which are causally associated with variability timescales spanning a range from minutes to years \citep{HOVATTA2019101541}. Recent high-resolution radio imaging and multi-wavelength/messenger campaigns are aiding in unraveling the conditions under which these jets form, evolve and produce the observed emissions \citep[e.g.,][and references therein]{Walker_2018, Bottcher_2019, Blandford_2019, Lister_2021}.

Blazars are a class of AGN with jets oriented close to the line of sight of the observer. The Spectral Energy Distribution (SED) of blazars consists of two distinctly separated non-thermal peaks (the double hump structure). 
The first peak lies between the infrared and X-ray frequencies, the second peak lies between the MeV to TeV range in the gamma ray band \citep{URRY1999159, Ghisellini2008, Abdo_2010}. 
Although the low-frequency peak has a well-established origin, the mechanisms driving the emissions contributing to the second peak are still a mystery. 
The source of this high energy peak of the blazar SED, at its core is a problem fundamentally associated with the composition of the AGN jet which 
remains elusive till date, and is an open question in astrophysics. 

Different models have been proposed in order to probe into the matter composition of the jet, starting from leptonic models \citep{Maraschi1992, Dermer1993, Bloom1996, Blazejowski_2000, DILTZ201463}, to hadronic/lepto-hadronic models \citep{Mastichiadis_Kirk1995, Mucke2003, AHARONIAN2000, Bottcher_2013, Murase2014, 10.1093/mnras/stu2691, Petropoulou2015, Zacharias_2022}.
These models incorporate various acceleration and radiative cooling mechanisms relevant to the particle species they propose as constituents of the jet. 
While all of these models can replicate the observed electromagnetic emissions from blazars, neutrino production is exclusive to models involving hadrons.

In September 2017, the IceCube Neutrino Observatory detected a $\sim290$\, TeV neutrino, from a direction consistent with the blazar TXS 0506+056. This triggered a follow-up analysis of $9.5$ years of archival IceCube data, which revealed an excess of $13\pm5$ muon neutrinos from the same direction during the 2014 - 2015 period, with a significance of $3.5 \sigma$ \citep{IceCube2018}.
Further, Baikal GVD also detected a neutrino event from this blazar source in 2021 \citep{Allakhverdyan2023}.
Additionally, PKS B1424-418 \citep{2016Kadler_nu}, GB6 J1040+06170617 \citep{Garrappa_2019}, and PKS 0735+178 \citep{Sahakyan_2022, Omeliukh:2024kgk} 
are other blazar sources identified as possible candidates for neutrino events detected in IceCube. 
Evidence demonstrating that neutrinos are statistically associated with radio-bright blazars have been shown by utilizing a total of 71 track like IceCube events from 2009 - 2022 \citep{Plavin2020, Plavin2021, Plavin2023}
In subsequent analysis of IceCube archival data for different AGN sources, a substantial neutrino excess was found from a Seyfert galaxy NGC 1068 as well,  with a  $4.2 \ \sigma$ significance \citep{IceCube:2022der}.
These observations have naturally reignited the interest in hadronic and lepto-hadronic models for AGN jets as unlike their purely leptonic counterparts, they can account for neutrino production.

High-energy neutrino production in AGN jets requires proton acceleration to very high energies, enabling photo-meson and proton-proton ($pp$) interactions.
These interactions initiate hadronic cascades through the decay of charged pions and muons, ultimately producing neutrinos with a characteristic flavor ratio of $\nu_e:\nu_\mu:\nu_\tau = 1:2:0$ at the source \citep{Kai_Zuber}. Neutrino oscillations over huge cosmological distances effectively average out, gradually transforming the initial flavor ratio during propagation to an arrival ratio of $1:1:1$ \citep{IceCube_flavorOsc_2015}. 
  
Particle acceleration in AGN jets may arise either through internal shocks driven by jet dynamics and fluid instabilities
\citep[for e.g][and references therein]{Marcowith:2020}, or via external shocks created as the jet interacts with its surroundings
\citep{Barkov2012, Palacio, Britzen:2019}. 
Several state-of-the-art particle acceleration models have also been developed in the last few years to study non-thermal emission from relativistic jets \citep{Liu_2017, MATTHEWS2020, Kundu_2021, UPRETI2024146, Dubey_2023, Jerrim2025}.
Specifically, in the case of accelerating hadrons, additional mechanisms like magnetic reconnection \citep{Giannios, Ball_2018, Fiorillo_2024} and the espresso mechanism \citep{Mbarek_2021} can also contribute significantly, potentially leading to ultra-high-energy cosmic-ray (UHECR) and neutrino production. This raises a key question: is the hadronic component (if any) in neutrino-emitting blazars an intrinsic part of the jet, or is it entrained from the surrounding medium through jet–environment interactions?

To disentangle the question above, it is important to capture the intricate relationship between jet composition, its interactions and dynamical behaviour, and the microphysics governing particle transport. This is a task where computational simulations can serve as indispensable tools. 
 
Throughout the years, there have been a few attempts to computationally implement the Blazar lepto-hadronic models. 
Some of the noteworthy implementations include the AM3 code \citep{Gao_2017}, ATHE$\nu$A \citep{Mastichiadis1995, Petropoulou:2014lja}, B\"{o}ttcher13(B13) \citep{Bottcher_2013}, LeHa-Paris \citep{10.1093/mnras/stu2691}, and \texttt{Katu} \citep{Katu2020}. A common feature among most of these lepto-hadronic codes is that they are typically utilized for single emission zone modeling, which assumes that the majority of emissions from an AGN jet are arising from a single homogeneous emission region. On the other hand, there are multi-zone models \citep{Xue_2021, Aguilar_Ruiz_2023}, which advocate for the possibility of multiple emission regions inside the jet contributing to the observed emissions. 

Most lepto-hadronic codes however do not incorporate the dynamical evolution of the jet, ignoring how the interplay of various phenomena like shocks, 
instabilities and magnetic reconnection across the jet can produce several regions where different particle acceleration and cooling processes may dominate resulting in multiple emission zones that may significantly impact the observed emissions. These multi-zone models, despite being a more realistic representation of the actual physics happening inside AGN jets, are challenging to implement due to their complexity as well as an extensive free-parameter space. Hence, only a limited number of studies have utilized these models to successfully explain the observed multi-wavelength and neutrino emissions coming from different blazar jets. \cite{Xue_2021} is one such study which uses a two zone approach, whereas \cite{Zacharias_2022} explores a lepto-hadronic model for an extended jet. 

To address these notable challenges in modeling AGN jets and its emissions, in this study we introduce a computational framework for lepto-hadronic multi-zone modeling that builds upon the foundation of existing codes that are suitable for one-zone modeling. In this unique approach, we couple the multi-zone framework with 3D relativistic magneto-hydrodynamic (RMHD) simulations with particle transport, to draw inference about the free parameters for each emission zone in the jet. This reduces the huge free parameter space typically required for multi-zone modeling. 

Hence, our multi-zone framework essentially acts as a bridge between dynamical simulations and microphysics of particle evolution within the jet. This can serve as a virtual testbed to model and study different extrinsic and intrinsic scenarios for proton enrichment in jets and its subsequent multi-messenger emissions. For the purposes of this study, we restrict our focus to exploring the emissions from a particle population intrinsic to the jet, the extrinsic scenarios will be explored in our subsequent works.

This paper is organized as follows.
In Section 2, we describe the numerical setup and outline the methodology behind our multi-zone framework. 
Section 3 presents the key results for two jet simulations differing in initial bulk Lorentz factor, together with the outcomes from their multi-zone analyses. 
In Section 4 we discuss the implications of our findings, and present the conclusions in Section 5. 

%----------------------------------------------------------------------
\section{Numerical approach }
\label{Numerical_Approach}
In this section we delve into the numerical setup for our lepto-hadronic multi-zone framework. We will explain in details the methodology we use to integrate the dynamical evolution of the relativistic jet with this framework in order to model
the photon SED and neutrino flux. 
We will also elaborate on the two computational codes we utilize for our work, along with the initial parameters and the boundary conditions used for the simulation setup.

%----------------------------------------------------------------------------------------------------
\subsection{Dynamical modeling} 
\label{PLUTO_setup}
For this work, we utilize the PLUTO code \citep{Mignone_2007}, a multi-algorithm fluid dynamics code, to perform 3D RMHD simulations for a section of an AGN jet. Performing such simulations involve numerically solving the following system of equations in its conservative form,
\begin{equation}
\frac{\partial}{\partial t}
\begin{pmatrix}
\Gamma \rho \\ \mathbf{m} \\ \mathcal{E}_t \\ \mathbf{B}
\end{pmatrix}
+ \nabla \cdot 
\begin{pmatrix}
\Gamma \rho \mathbf{v} \\ w_t \Gamma^2\mathbf{vv} - \mathbf{bb} + I p_t \\
\mathbf{m} \\ \mathbf{vB} - \mathbf{Bv}
\end{pmatrix}^{T}
= 0,
\label{PLUTO_RMHD}
\end{equation}
where
$\rho$ is the rest-mass density, 
$\mathbf{v}$ is the bulk velocity, 
$\Gamma$ is the bulk Lorentz factor, 
and $I$ is the identity matrix. 
Here, $\mathbf{B}$ is the observer frame magnetic field while the co-moving frame magnetic field is given by
$b^{k} = \Gamma \ \{ \mathbf{v} \cdot \mathbf{B} \ , \ B^i / \Gamma^2 + v^i(\mathbf{v}\cdot \mathbf{B}) \}$, 
the relativistic total enthalpy is $w_t \equiv (\rho h + b^2)$, 
$\mathbf{m} = w_t \Gamma^2 \mathbf{v} - b^0\mathbf{b}$ is the momentum density, 
$\mathcal{E}_t = w_t\Gamma^2 - b^0b^0 - p_t$ represents the total energy density, and $p_t = P + b^2/2$ is the total pressure where $P$ is the gas pressure \citep{Del_zanna2003,Mckinney}.

%----------------------------------------------------------------------------
\subsubsection{Numerical setup}
\label{Num_setup}
We model a representative section of an AGN jet, taking a cylindrical plasma column with an initial radius of $R_{\rm j} \equiv l_0 = 1$\,pc, in which the material propagates along the z-axis. 
We apply a uniform 3D Cartesian grid. The size of the entire simulation domain is
$[-L_x <x < L_x]$, $[-L_y < y < L_y] $, and $[0 < z < L_z]$, where $L_x = L_y = L_z = 24\,l_0 = 24$\,pc. 
The domain numerical grid has $(768 \times 768 \times 384)$ grid cells resulting in a resolution of 16 cells per pc. 
Therefore, the simulation domain represents a parsec-scale segment of the AGN jet, capturing the jet’s early propagation phase near the broad-line region 
and the central engine.

All the quantities in our dynamical simulations are defined in dimension-less code units, with density being scaled by $\rho_0 = 1.661 \times 10^{-26}$ g cm$^{-3}$, magnetic field being scaled by $B_0 = v_0 \sqrt{4 \pi \rho_0}= 0.014$ G, while $P_0 = \rho_0 v_0^2 = 1.492$ dyne cm$^{-2}$ and $t_0 = l_0/v_0 = 3.26$ years are used to normalize pressure and time respectively. 
Velocities are normalized to the speed of light,  $v_0 = c$.

 We apply a divergence cleaning algorithm prescribed to preserve the $\nabla \cdot B = 0$ condition \citep{Dedner2002}.
A total variation diminishing (TVD) piece-wise linear reconstruction scheme \citep{MIGNONE2014784} which is 2nd order accurate in space, 
has been used for the primitive variables along with a second order accurate Runge-Kutta algorithm for time stepping. Additionally, we have adopted the Taub-Matthews equations of state as defined in \citet{Mckinney} to close the respective conservation laws in Equation \ref{PLUTO_RMHD}.

%-----------------------------------------------------------------------------------
\subsubsection{Initial conditions and boundary conditions}
\label{PLUTO_ini_bc}
We assume a rotating, inviscid, infinitely conducting flow constituting our jet. The jet section is initialized as an axisymmetric cylindrical plasma column by prescribing radial profiles of various flow variables following \cite{Bodo_2019}. Choosing the Lorentz factor $\Gamma_z (r)$ profile of the form
\begin{equation}
    \Gamma_z(r) \equiv \frac{1}{\sqrt{1 - v_z^2}} = 1 + \frac{\Gamma_{\rm c} - 1}{\cosh(r / r_{\rm j})^6},
\end{equation}
where, $\Gamma_{\rm c}$ is the Lorentz factor along the central axis, $r_{\rm j}$ is the jet radius, and $v_z$ is the axial velocity. The azimuthal velocity $v_{\phi}$, toroidal $(B_{\phi})$ and poloidal magnetic field $(B_z)$ profiles taken for our simulations are defined in Appendix~\ref{A_Profiles}. 
To trigger the onset of current-driven instabilities, a perturbation following the formulation specified in \citet{Rossi2008} is applied to the radial velocity. The corresponding expression can be defined as
\begin{equation}
\label{eq_perturb}
    v_r = \frac{K_0}{24} \sum_{m=0}^{2} \sum_{l=1}^{8} cos(m\phi + \omega_l t + o_l),
\end{equation}
which sums up waves of different frequencies with $\omega_l = c_s(0.5,1,2,3,0.03,0.06,0.12,0.25)$.
 This considers different jet instabilities with $m=0,1,2$ representing sausage, kink and fluting modes respectively. 
The $o_l$ represents random phase shifts and the amplitude $K_0$ denotes the fractional change of the bulk Lorentz factor given 
by $K_0=\frac{\sqrt{(1+\psi)^2 - 1}}{\Gamma (1 + \psi)}$,
where $\psi$ is a small dimensionless parameter characterizing the relative strength of the perturbation, typically set to a value of $0.05$.

We perform jet column simulations for two different initial bulk Lorentz factors along the central axis: $\Gamma_{\rm c, ini} = 5,\, 10$. 
For both simulations, the initial density inside the jet column is defined as $\rho_{\rm j} = \rho_0$, while ambient medium density is $\rho_a = 1000\,\rho_0$. 
Initially, the pressure is taken to be uniform across the domain $P_{\rm j} = P_a = 0.1\,P_0$.
To completely specify the magnetic field profiles, 
we define the pitch parameter along the central axis
\begin{equation}
    \mathcal{P}_{\rm c} = \left|\frac{r B_z}{B_\phi}\right|_{r=0}.
\end{equation}
We apply $\mathcal{P}_{\rm c} = 0.9$ for both jet simulations,
while the central axis poloidal magnetic field $B_{\rm zc}$ and the jet angular velocity along the central axis $\Omega_{\rm c}$ are different with values specified as,
\begin{itemize}
\item
    $B_{\rm zc} = 1.710~B_0$, $\Omega_{\rm c} = 0.434$ \quad (for $\Gamma_{\rm c, ini} = 5$),
\item
    $B_{\rm zc} = 2.148~B_0$, $\Omega_{\rm c} = 0.192$ \quad (for $\Gamma_{\rm c, ini} = 10$). 
\end{itemize}

The initial profiles defined in cylindrical coordinates are subsequently transformed into Cartesian coordinates to initialize the 3D Cartesian simulation grid.

To track the evolution of the jet material, we also define a tracer variable $\mathcal{T}$ such that initially $\mathcal{T} = 1$ for the material in
grid cells inside the jet, and $\mathcal{T} = 0$ for the ambient medium.
To avoid sharp transitions or discontinuities at the jet boundary, each fluid variable is multiplied by a smoothening function. 

Standard outflow boundary conditions are applied along the $x$ and $y$ direction vertical boundaries.
The simulation domain is sufficiently large, placing these boundaries far enough from the jet core to prevent spurious reflections from influencing the solution. 
Periodic boundary conditions are imposed on the transverse planes at $z = 0$ and $z = 24\, l_0$, which define the domain boundaries along the z-axis.

%-------------------------------------------------------------------------------------------------------
\subsection{Hybrid particle module}
\label{Hybrid_module}
The Lagrangian particle module for PLUTO code as described in \cite{Vaidya_2018}, utilizes a hybrid approach for modeling leptonic particles within fluid simulations, and its respective non-thermal emissions. It utilizes the concept of a Lagrangian macro-particle which is an ensemble of leptons in spatial proximity but with a finite distribution over an energy range. The macro-particles propagate through an Eulerian grid and get advected with the flow. 

For each macro-particle, a time-dependent distribution function $N_{\rm LP}^{\rm e} (\gamma, \tau)$ can be defined, 
representing the number density of electrons over an energy range\footnote{Subscript LP denotes
the variables specified for Lagrangian particles, while superscripts $e$ or $p$ denote the particle species consisting of electrons/leptons or protons, respectively.}
For numerical convenience, we define a normalized number density ratio function $\chi_{\rm LP}^{\rm e}$ for the Lagrangian macro-particles such that:
\begin{equation}
    \label{eq_chi}
    \chi_{\rm LP}^{\rm e} (\gamma, \tau) = \frac{N_{\rm LP}^{\rm e} (\gamma, \tau)}{n_{\rm F}},
\end{equation}
where $n_{\rm F}$ represents the fluid number density at the location of the macro-particle, $\gamma$ represents the energy of the particle normalized to its rest mass energy, and $\tau$ is the proper time which is related to the lab-frame time $t$ by $d\tau = dt/\gamma$.

It is important to note that, all the variables are defined in the co-moving frame unless explicitly stated otherwise, a convention which will be followed throughout the text.

A simplified form of the cosmic ray transport equation, derived under the assumptions specified in \cite{Vaidya_2018}, is used to describe the spatial and temporal evolution of Lagrangian macro-particles and the associated number density distribution function of the non-thermal particles within it. This particular form of the transport equation can be expressed as
\begin{equation}
\label{eq_DE_chi}
 \frac{d\chi_{\rm LP}^{\rm e}}{d\tau} + \frac{\partial}{\partial \gamma}\left[\left( - \frac{\gamma}{3} \Delta_{\mu} u^{\mu}+ \dot{\gamma_{l}} \right)\chi_{\rm LP}^{\rm e}\right] = 0\, , 
\end{equation}
where $\dot{\gamma_{l}} < 0$ represents the radiative losses of the particles.
The first term within the brackets accounts for the energy losses caused by adiabatic expansion.
The particle module solves Equation \ref{eq_DE_chi} separately for the spatial and energy dependent parts. 

The transport step handles the update of the spatial coordinates $\mathbf{x}_{\rm LP}$ for each macro-particle using the equation
\begin{equation}
    \frac{d\mathbf{x}_{\rm LP}}{dt} = \mathbf{v}(\mathbf{x}_{\rm LP})\, , 
\end{equation}
where $\mathbf{v}(\mathbf{x}_{\rm LP})$ is the interpolated fluid velocity at the position of the Lagrangian particle.
This equation is solved consequently with the respective fluid equations. 

This is followed by the spectral evolution step which governs the evolution of the particle distribution inside each macro-particle. 
Radiative losses of the particles via processes like synchrotron and Inverse Compton \citep[Sharma et. al. in prep]{Khangulyan_2014, Acharya_2023}, 
as well as energy gains due to acceleration processes such as diffusive shock acceleration (see Section \ref{DSA}), are incorporated within the particle module. 

The target photon field required for the inverse Compton losses may originate from various regions around the central supermassive black hole (SMBH), such as the broad line region (BLR), the 
accretion disk or the dusty torus (DT). 
In this study, for two of the simulations, we adopt simplified external photon fields for our jet, considering that in the AGN frame, the dissipation region is at a distance of $\gtrsim 10$\, pc from the SMBH (assuming an opening angle of $\approx 2-4^\circ$). In particular, we consider a BLR field in one EC simulation and a DT field in the other. The BLR and DT radiation fields are approximated as blackbody radiation peaking respectively at frequencies $\approx 2 \times 10^{15}$\, Hz \citep{Ghisellini2008}, and $\approx 3 \times 10^{13}$\,Hz \citep{Cleary_2007} in the AGN frame. The corresponding AGN frame photon-field luminosities are parameterized as $L^{\rm AGN}_{\rm BLR/DT}=\mathcal{G}_{\rm BLR/DT}\,L^{\rm AGN}_{\rm disc}$, where we adopt $L^{\rm AGN}_{\rm disc}=6\times10^{45}$\,erg\,s$^{-1}$ and covering factors $\mathcal{G}_{\rm BLR}\approx 0.1$ and $\mathcal{G}_{\rm DT}\approx 0.5$.

We use the approximation provided in \citet{Schlickeiser_2010} for the inverse Compton electron energy loss rate for an isotropic graybody photon field in the Klein-Nishina limit which can be expressed as follows,
\begin{equation}
    |\dot{\gamma}_{\rm EC}| \approx \frac{4 \sigma_{\rm T} c U_{\rm rad} \gamma^2}{3 m_{\rm e} c^2} f_{\rm KN}(\gamma)
\end{equation}
where the subscript EC represents external Compton losses in our system, $f_{\rm KN} (\gamma) = \frac{\gamma_{\rm K}^2}{\gamma_{\rm K}^2 + \gamma^2}$ and $\gamma_{\rm K}$ is the critical Klein-Nishina Lorentz factor given by 
\begin{equation}
    \gamma_{\rm K} \equiv \frac{3\sqrt{5}\, m_e c^2}{8\pi \, k_{\rm B}T}
\end{equation}

The co-moving energy density of the external photon field is determined in terms of its luminosity and its characteristic radial scale from the SMBH. The dilution of the photon field with the distance of the emission zone is captured by rescaling the energy density with the respective geometric dilution factor
$\kappa^{\rm AGN}_{\rm BLR/DT}$ \citep{Khangulyan_2014, Acharya_2023} (Described in Appendix \ref{Appendix_EC}).

At initial time, we inject a total of $10^6$ macro-particles that optimally samples the plasma column.
Each of these leptonic macro-particles is initialized with a power-law particle distribution $\propto \gamma^{-6}$ with $\gamma_{min} = 10^2$ and $\gamma_{max} = 10^8$ 
for both the simulated jets. 
The initial number density of electrons within each macro-particle is set to be $2.14 \times 10^{-6}$\, cm$^{-3}$ for the $\Gamma_{c,ini} = 5$ jet 
and $8.52 \times 10^{-6}$\, cm$^{-3}$ for the $\Gamma_{c,ini} = 10$ jet.
The initial power-law spectra for these macro-particles are evolved with time, as they propagate within the jet column owing to radiation losses and diffusive acceleration due to shocks. The synchrotron emissivity in the observer's frame is then quantified taking into account the evolving spectra.
The emissivity can be further utilized to generate the synthetic synchrotron SEDs directly from the hybrid PLUTO jet simulations using the prescription for $\nu F_{\nu}$ defined in \citet{DermerMenon} (see Appendix~\ref{Appendix}). 

%------------------------------------------------------------------------------------------------------
\subsubsection{Diffusive Shock acceleration}
\label{DSA}
PLUTO's hybrid particle module employs analytical estimates for diffusive shock acceleration (DSA) within the test-particle limit, which is valid for highly turbulent relativistic shocks. The module uses a shock detection algorithm to flag shocked regions based on a threshold compression ratio, which is followed by a corresponding change in the post-shock particle distribution associated with each macro-particle. 

The changes in the particle distribution due to DSA depends on the compression ratio $\mathcal{C}$ of the shock encountered and the angle $\varphi$ between
the shock normal and the magnetic field vector. 
The detailed algorithm employed for DSA in the particle module is provided in \cite{Vaidya_2018}. 
The compression ratio for the grid cells flagged by the shock detection algorithm is determined as
\begin{equation}
    \mathcal{C} = \frac{\boldsymbol{\beta}_{1}^{'} \cdot \mathbf{\hat{n}_{\rm s}} }{\boldsymbol{\beta}_{2}^{'} \cdot \mathbf{\hat{n}_{\rm s}} },
\end{equation}
where $\beta_{1}^{'}$ and $\beta_{2}^{'}$ represent the upstream and downstream velocities in the shock rest frame, respectively,
while $\hat{n}_{\rm s}$ is the shock normal vector. 
The particle distribution $\chi_{\rm LP}^{\rm e}(\gamma, \tau)$ after the shock, is updated by injecting a power-law 
spectrum of the electrons in the form,
\begin{equation}
    \label{DSA1}
    \chi_{\rm LP}^{\rm e}(\gamma, \tau) = \mathcal{A} \left(\frac{\gamma}{\gamma_{\rm min}}\right)^{-q+2},
\end{equation}
where $\mathcal{A}$ is the normalization factor, $\gamma_{\rm min}$

is the lower limit of the normalized particle energy for the electron distribution, 
and $q$ is the power-law index which is a function of the compression ratio $\mathcal{C}$. 

The updated spectra depend on two free parameters, namely, the non-thermal to thermal number density ratio $\zeta$, and the ratio of total energy of 
injected particles to fluid internal energy density $\zeta_{\rm E}$, and also takes into account the old pre-shock particle population. 
 We calculate $\gamma_{\rm min}$ and $\mathcal{A}$ by solving
\begin{align}
   \label{DSA2}
    \mathcal{A} \int_{\gamma_{\rm min}}^{\gamma_{\rm max}} \left( \frac{\gamma}{\gamma_{\rm min}} \right)^{-q+2} d\gamma & = \zeta \frac{\rho}{m} + n^{\mathrm{old}}, \\
    \label{DSA3}
    \mathcal{A} \int_{\gamma_{\rm min}}^{\gamma_{\rm max}} \left( \frac{\gamma}{\gamma_{\rm min}} \right)^{-q+2} \gamma d\gamma &= \zeta_{\rm E}  \frac{\mathcal{E}}{m_e c^2} + \frac{E^{\mathrm{old}}}{m_e c^2}.
\end{align}
where we interpolate the mass density $\rho$ and the internal energy $\mathcal{E}$ of the fluid at the particle position, 
while normalized energy upper limit $\gamma_{\rm max}$ for the electron distribution is estimated by inserting the  following constraints related to particle escape. 

In particular, for electrons the first constraint is 
$\gamma^{\rm e} \le \gamma^{\rm e}_{\rm max} = eBr_{L}^{max}/(\beta_{\perp}m_e c^2)$, 
where the maximum allowed Larmor radius is calculated as 
$r_{L}^{max} = r_{LP} \approx 0.5 {\rm min}(\Delta x, \Delta y, \Delta z)$ (cell dimensions), 
where $e$ is the electron charge, $B$ is the magnetic field at the location of the macro-particle and $\beta_{\perp}$ is the ratio of electron
velocity to light speed. For the second constraint, 
$\gamma_{max}$ is determined by equating the acceleration and cooling timescales. 
The minimum of $\gamma_{max}$ determined through both these conditions is taken as the high energy cutoff for Equations \ref{DSA1}, \ref{DSA2} and \ref{DSA3}. 

%------------------------------------------------------------------------------------------------------
\subsection{Lepto-hadronic modeling setup}
The methodology of our work requires a one-zone lepto-hadronic (OZLH) code that is suitable for modeling radiative processes and neutrino emissions for an individual blazar emission zone. 
For this purpose, we utilize the code used in \citet{Katu2020}, which is publicly available at \url{https://github.com/hveerten/katu}, where it is referred to as Katu \footnote{For this study, a few small modifications were made locally to fit the code to our model requirements. Additionally, secondary SED components are generated and stored on a dedicated secondary grid using a local wrapper code}.

This code solves the kinetic equation for a range of different species of particles which involves leptons 
(electrons, positrons, muons and anti-muons, electron neutrinos, muon neutrinos and their respective anti-neutrinos)
as well as hadrons (protons, neutrons, charged and neutral pions) and photons.
The governing kinetic equation is written as,
\begin{equation}
\label{FP_Katu}
    \frac{\partial N(\gamma)}{\partial \tau} = Q_{\rm ex} + Q_{\rm in} + \mathcal{L} - \frac{N}{T_{\rm esc}} - \frac{N}{T_{\rm dec}} 
+ \frac{\partial}{\partial \gamma} \left( \dot{\gamma} N \right),
\end{equation}
where all the quantities are measured in the co-moving frame unless stated otherwise. Here, $N(\gamma)$ represents the number density distribution for a particular species of particles, 
the term $Q_{\rm ex}$ accounts for the external injection of particles into the emission region,
$Q_{\rm in}$ accounts for the internal injection of the particular particle species created as secondaries inside the region,
$\mathcal{L}$ represents particle loss due to internal processes, while $N/T_{\rm esc}$ and $N/T_{\rm dec}$ accounts for the particle loss due to particle escape and decay respectively. 
Here $T_{\rm esc}$ represents the escape timescale which depends on zone geometry and $T_{\rm dec}$ depicts the decay timescale. 

The last term in Equation \ref{FP_Katu} accommodates the changes due to different radiative emission mechanisms. 
This term contains $\dot{\gamma}$, accounting for the particle energy losses due to processes such as synchrotron and inverse Compton emission. 
Here, the radiation energy density adopted for the BLR photon field is done in a similar manner to that of the hybrid particle module (See Section~\ref{Hybrid_module}) 
and the IC emission is calculated as described in \citet{Katu2020} using \citet{Jones1968} . 

The utilized OZLH code can also model electromagnetic processes like Bethe-Heitler (BH) pair production{ \citep{Kelner_Aha_2008}, photon-photon pair production (implemented following \citealt{Boettcher1997}), 
and hadronic processes like photo-meson interaction \citep{Hummer_2010}} along with pion and muon decays which are important processes for astrophysical neutrino production \citep{Lipari2007, Hummer_2010}.

We want to emphasize that the wide range of processes described above that can be modeled via this code, makes it a useful tool for the modeling of individual 
emission zones in our framework (see Section~\ref{MZ_sec}).
However, we point out that our framework is compatible with any comparable code capable of one-zone lepto-hadronic modeling, hence to generalize, we will use the acronym OZLH to refer to the usage of such a code for our analysis throughout the text. 

%----------------------------------------------------------------------------------------------------
\subsection{The multi-zone framework}
\label{MZ_sec}
\begin{figure*}
    \centering
    \includegraphics[width=0.80\textwidth]{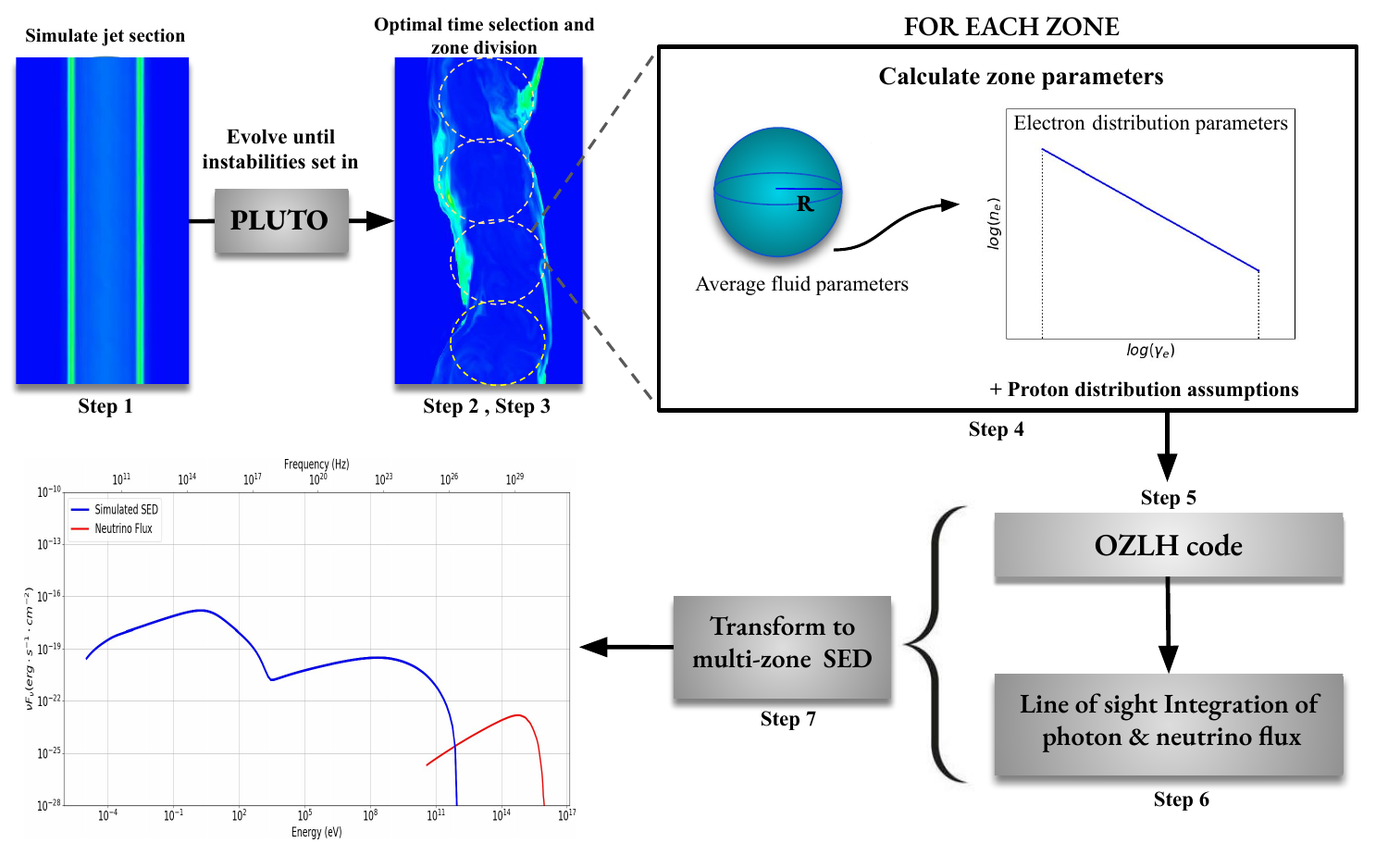}
    \caption{\small Flowchart depicting the methodology used in our lepto-hadronic multi-zone framework. 
    The step numbers directly correspond to the order of steps for the algorithm as described for the framework in the paper text.}
    \label{Method_flow}
\end{figure*}

Our multi-zone framework serves as a bridge between the jet dynamics and the micro-physics of the jet.
It harnesses the robust dynamical modeling capabilities of the PLUTO code along with its hybrid particle module while building upon the foundation of the single emission zone modeling capabilities of the OZLH code. 
The flowchart shown in Figure~\ref{Method_flow} describes the methodology of our work. 

\noindent We apply the following algorithm to accomplish our objectives:
\begin{enumerate}
    \item \textbf{Dynamical simulation of AGN jet:} We start by simulating the dynamics of a relativistically moving magnetized section of an AGN jet using PLUTO code, and incorporate electron macro-particles which advects with the flow. We provide a perturbation to the radial velocity of the jet and evolve it until current-driven instabilities set in.
    
    \item \textbf{Optimal time selection:} Based on the evolution of various energy components within the jet, we select an optimal time for conducting our multi-zone analysis. This time corresponds to an intermediate point during the linear growth phase of the kink instability formation process in the jet. This phase is particularly suitable, as it generates complex magnetic field configurations facilitating a dynamic interplay between particle acceleration and cooling mechanisms, while also ensuring that cooling is not yet dominant on the system.
    
    \item \textbf{Multi-zone division:} At the selected dynamical time, we segment the jet into multiple spherical emission zones, with the zone radius decided by the local radius of the jet at the position of the emission zone.  
    
    \item \textbf{Calculating zone parameters:} For each zone we compute the fluid parameter averages and net electron distribution \footnote{Throughout the text, the terms \textit{electron distribution}, \textit{proton distribution}, or \textit{particle distribution} refer to the distribution of the respective particle species' number density over an energy range.} parameters using the jet simulation data. Since the particle module used in the jet simulations does not provide us with a proton distribution, we rely on approximations to estimate it. (Detailed discussion in Section~\ref{Approx_MZ}). 
    
    \item \textbf{Bridging jet dynamics and microphysics:} The average fluid and particle distribution parameters, calculated for each zone from the jet simulation in the previous step, are provided as input parameters to our multi-zone lepto-hadronic framework, which is modeled and executed using an OZLH code. Leveraging fluid simulations to inform emission zone modeling parameters significantly narrows the dimensions of the multi-zone parameter space.
    
    \item \textbf{Zonal simulations:} We use the OZLH code used for \citet{Katu2020} to perform steady-state lepto-hadronic simulations for individual zones by incorporating the relevant leptonic and hadronic processes. This provides us with the photon and neutrino emissions from each zone.
    
    \item \textbf{Multi-zone SED generation:} We stack the emissions that are contributed by each zone to generate the multi-zone photon SED and neutrino flux.
\end{enumerate}

%------------------------------------------------------------------------------------------------------
\subsubsection{Initial parameters and approximations for the multi-zone framework}
\label{Approx_MZ}
Most OZLH codes including the one we use do not incorporate the fluid dynamical evolution of the jet.
However, connecting such a code with 3D RMHD simulations via our multi-zone framework, provides us with an ideal platform for simulating blazar emissions. 
We may therefore model emissions from individual zones of the jet stream by using the OZLH code, following the algorithm we have described above. 

We perform a lepto-hadronic analysis, for the RMHD jet column simulations having 
$\Gamma_{\rm c,ini} = 5$ and $10$ at the dynamical times $t=t_{\rm s1}=782$ years  and $t=t_{\rm s2}=743$ years, respectively.
These times align with the initial phase of the instability formation process (see detailed discussion in Section~\ref{sec_dynamical_evol}). 

At times $t_{\rm s1}$ and $t_{\rm s2}$, we divide the respective jet columns into six distinct, spherical emission zones. 
The zones are distributed along the jet axis, denoted by the numbers 1 to 6 from the bottom to the top of the column, respectively. 
We hereby assume the individual zones in our multi-zone framework to be independent emission regions. 

For each zone, we model the synchrotron emission from both leptons and hadrons, synchrotron self-Compton (SSC) emissions, B-H pair production, photon-photon ($\gamma\gamma$) pair production and photo-meson interaction. Each zone hosts $\approx 10^4 -10^5$ macro-particles. The normalized spectra for leptons, $\chi_{\rm LP}$, for each macro-particle is defined via Equation \ref{eq_chi} and evolved following Equation \ref{eq_DE_chi}. 
This evolution does not explicitly account for energy losses due to SSC to avoid high computational cost of implementing SSC losses in our dynamical runs. 

Further, in the majority of the zones, synchrotron cooling timescales are much shorter than SSC cooling timescales.
Therefore, in these zones the final electron distributions will be primarily governed by synchrotron losses. 
However, we observe that in one or two zones SSC cooling may become important.
In such cases, not accounting for SSC cooling, overestimates the number density of high energy electrons, and consequently the photon flux from these zones in certain frequency bands.  As a consequence, this can impact the photon field available for the photo-meson process, and hence the neutrino fluxes that are estimated for these zones may constitute an upper limit only.

In summary, there are a total of 11 initial parameters required for the steady-state lepto-hadronic modeling for each zone.
By these parameters we are able to connect the emission zone modeling with the OZLH code to the dynamics of the RMHD simulations.
In our approach, these zonal input parameters\footnote{All input parameters labeled with the subscript “zone” are zone-dependent, i.e., they may take on different values for each emission zone.} 
for a selected dynamical time are determined as follows:
\begin{itemize}
   \item[-] \textbf{Non-thermal density}, $\rho^{\rm NT}_{\rm zone}$:  
    Derived from the spatially averaged fluid density in each zone of the jet simulations, 
    \begin{equation}
        \rho^{\rm NT}_{\rm zone} = \left(\frac{\zeta}{1+\zeta}\right) <\rho>_{\rm zone}
    \end{equation}
    where $\zeta$ represents the ratio of non-thermal density to the thermal density in the zone. This is a free-parameter which we choose to be $0.15$ for all zones across the jet.

    \item[-] \textbf{Proton to electron number density ratio} ($\eta_{\rm zone}$): The ratio given by $\eta_{\rm zone} = n^{\rm p}_{\rm zone}/n^{\rm e}_{\rm zone}$, is constrained by $\rho^{\rm NT}$ and $\zeta$ for each zone in the jet as specified below:
    \begin{equation}
        \rho^{\rm NT}_{\rm zone} = \frac{\zeta}{1+\zeta} <\rho>_{\rm zone} = n^{\rm e}_{\rm zone}(m_{\rm e} + \eta_{\rm zone} m_{\rm p})
    \end{equation}
    where $m_{\rm e}$ and $m_{\rm p}$ are electron and proton mass, $n^{\rm e}_{\rm zone}$ and $n^{\rm p}_{\rm zone}$ are electron and proton number densities within the zone respectively.
    
    \item[-] \textbf{Magnetic field} ($B_{\rm zone}$): Determined from the average co-moving magnetic field in each zone of the jet simulations.

    \item[-] \textbf{Zone radius} ($R_{\rm zone}$): The radius of the spherical zones, $R_{\rm zone}$, is derived from the average radius of the jet cross-section, $R_{\rm jet}(z_{\rm c})$,
     at $z=z_{\rm c}$, the location of the center of the spherical zone along the $z$-axis.
     Hence, the number of zones we take for our multi-zone modeling is dependent on the radius of the jet at the chosen time.

    \item[-] \textbf{Doppler factor} ($\delta_{\rm zone}$): The Doppler factor for each grid cell of the 3D jet simulation is calculated as follows: 
    \begin{equation}
    \label{Eq_delta}
        \delta_{\rm cell} = \frac{1}{\Gamma_{\rm cell}(1-\boldsymbol{\beta_{\rm cell}} \cdot \hat{\mathbf{s}})}
    \end{equation}
    where $\Gamma_{\rm cell}$ is the bulk Lorentz factor for the grid cell while $\boldsymbol{\beta_{\rm cell}}$ represents the grid cell velocity normalized to light speed, $\hat{\mathbf{s}}$ is the unit vector for the line of sight. 
    We then average over all the grid cells within the zone to obtain $\delta_{\rm zone}$,
    \begin{equation}
        \delta_{\rm zone} = \frac{\int_{\rm zone} \mathcal{T}\delta_{\rm cell} dV}{\int_{\rm zone} \mathcal{T}dV}
    \end{equation}
    where $\mathcal{T}$ is the tracer variable, $dV$ is the co-moving volume of the cell.

    \item[-] \textbf{Electron distribution - } $(\alpha^{\rm e},\gamma^{\rm e}_{\rm min}, \gamma^{\rm e}_{\rm max})$: The electron distributions from all the macro-particles within a particular zone of the jet are used to estimate the net electron distribution for the zone as follows:
    \begin{equation}
    \label{Eq_net_dist}
        N^{\rm e}_{\rm net} (\gamma) = \sum_{i} n_{{\rm F},i} \ \chi^{\rm e}_{{\rm LP},i} (\gamma)
    \end{equation}
    where $ N^{\rm e}_{\rm net} (\gamma)$ represents the net number density function for the electron distribution within an emission zone defined over the energy range $[\gamma^{\rm e}_{\rm min}, \gamma^{\rm e}_{\rm max}]$, $n_{{\rm F},i}$ is the fluid number density at the location of the $i^{\rm th}$ Lagrangian particle in the zone, $\chi^{\rm e}_{{\rm LP},i}$ is the normalized number density distribution for the $i^{th}$ Lagrangian particle. Here, we calculate $\gamma^{\rm e}_{\min} = \min (\{ \gamma^{\rm e}_{\min,i} \}_{\rm zone} )$ and $\gamma^{\rm e}_{\max} = \max ( \{ \gamma^{\rm e}_{\max,i} \}_{\rm zone} )$, where $\{ \gamma^{\rm e}_{\min,i} \}_{\rm zone}$ and $\{ \gamma^{\rm e}_{\max,i} \}_{\rm zone}$  denotes the sets of minimum and maximum $\gamma$ for each Lagrangian particle in the zone.

    Depending on the zonal net distribution of electrons, we fit $N^{\rm e}_{\rm net}$ with either a \textit{power-law} (PL) or a \textit{power-law with an exponential cutoff} (PLEC) and get the required electron distribution parameters:
    \begin{align}
    \begin{split}
        N^{\rm e}_{\rm fit,PL} &= A^{\rm e}_{\rm fit}\, \, \gamma^{- \alpha_{\rm fit}^{\rm e}}\\
        N^{\rm e}_{\rm fit,PLEC} &= A^{\rm e}_{\rm fit}\, \, \gamma^{- \alpha_{\rm fit}^{\rm e}}\, \, e^{(- \gamma/\gamma^{\rm e}_{\rm break})}
    \end{split}
    \end{align}
    where $A^{\rm e}_{\rm fit}$ is the normalization factor for the fitted curve to the electron distribution which can either be PL or PLEC depending on the zone, $\alpha^{\rm e}_{\rm fit}$ is the power-law index for the fitted curve and $\gamma^{\rm e}_{\rm break}$ represents the energy for the exponential cutoff for the PLEC fitted curve. From this we can get the input parameters for electron distribution like $\alpha^{\rm e} = \alpha^{\rm e}_{\rm fit}$ while the $\gamma^{\rm e}_{\rm min}$ and $\gamma^{\rm e}_{\rm max}$ remains the same for the fitted curve and the net distribution within the zone.
    
    \item[-] \textbf{Proton distribution - } $(\alpha^{\rm p},\gamma^{\rm p}_{\rm min}, \gamma^{\rm p}_{\rm max})$: Since the jet simulations with the hybrid particle module does not provide us with a proton distribution, we have taken 4 cases of assumptions for the proton distribution parameters for each zone as depicted in table \ref{tab_proton}. 

\end{itemize}

\begin{deluxetable*}{  c  c  c  c  c  }
\renewcommand{\arraystretch}{1.4}
\tablecaption{Model assumptions for the proton distribution.\label{tab_proton}}
\tablehead{
\colhead{\makecell[c]{
}}& 
\colhead{Case 1} & 
\colhead{Case 2} & 
\colhead{Case 3} & 
\colhead{Case 4}
}
\startdata
 \makecell{Distribution\\ type $N^{\rm p} (\gamma)$ }& $\eta_{\rm zone} \, N^{\rm e}_{\rm fit}(\gamma)$ & $\eta_{\rm zone} \, N^{\rm e}_{\rm fit,PL}(\gamma)$ & $\eta_{\rm zone} \, N^{\rm e}_{\rm fit,PL}(\gamma)$ & $\eta_{\rm zone} \, N^{\rm e}_{\rm fit,PL}(\gamma)$ \\ \hline
 $\alpha^{\rm p}$ & $\alpha^{\rm e}_{\rm fit}$ & $\alpha^{\rm e}_{\rm fit}$ & $\alpha^{\rm e}_{\rm fit}$ & $2.0$ \\ \hline
 $\gamma_{\rm max}^{\rm p}$ & $\gamma_{\rm max}^{\rm e}$ & $\gamma_{\rm max}^{\rm e}$ & \makecell{Constrained by \\ Hillas criterion} & $\gamma_{\rm max}^{\rm e}$ \\
\enddata
\tablecomments{
These parameters are used as input for the OZLH code, based on the fitted electron distributions from PLUTO simulations.
In Case 1, $N^{\rm e}_{\rm fit}(\gamma)$ can have either a \textit{power-law} (PL) or \textit{power-law with exponential cutoff} (PLEC) form, 
depending on the zone. Case 3 applies the Hillas condition to constrain $\gamma_{\rm max}^{\rm p}$.
}
\end{deluxetable*}

%The following is the for point 11)
By using the initial parameters determined from our jet simulations at the chosen optimal times $t_{\rm s1}$ and $t_{\rm s2}$, respectively, 
the lepto-hadronic analysis parameters for each zone can been derived as detailed above. 
For each jet simulation, Table~\ref{tab_zonal_params} reports the range of these zonal parameters, from the minimum to the maximum of the parameter values across the six zones.

\begin{deluxetable*}{lcccccccc}
\tablecaption{Input parameters for the lepto-hadronic multi-zone modeling.\label{tab_zonal_params}}
\tablehead{
\colhead{Jet} & 
\colhead{$\rho^{\rm NT}_{\rm zone}$} & 
\colhead{$\eta_{\rm zone}$} & 
\colhead{$B_{\rm zone}$} & 
\colhead{$R_{\rm zone}$} & 
\colhead{$\delta_{\rm zone}$} & 
\colhead{$\alpha^{\rm e}_{\rm fit}$} & 
\colhead{$\gamma^{\rm e}_{\rm min}$} & 
\colhead{$\gamma^{\rm e}_{\rm max}$} \\
\colhead{} & 
\colhead{($10^{-26}$ g cm$^{-3}$)} & 
\colhead{($10^{-2}$)} & 
\colhead{($10^{-2}$ G)} & 
\colhead{($10^{18}$ cm)} & 
\colhead{} & 
\colhead{} & 
\colhead{} & 
\colhead{($10^{6}$)} 
}
\startdata
$\Gamma_{\rm c,ini}=5$  & $2.7 - 4.9$    & $0.42 - 1.01$ & $1.69 - 2.14$ & $6.172$        & $1.78 - 2.77$ & $2.65 - 3.10$ & $14.76 - 23.24$ & $49.66 - 205.15$ \\
$\Gamma_{\rm c,ini}=10$ & $1.34 - 5.41$  & $0.07 - 1.34$ & $1.83 - 3.15$ & $6.172$        & $1.95 - 3.54$ & $2.47 - 3.17$ & $8.21 - 14.65$  & $8.64 - 333.41$ \\
\enddata
\tablecomments{
The values show the range (minimum to maximum) across the six zones at the selected dynamical times ($t_{\rm s1}$ and $t_{\rm s2}$) when the multi-zone analysis is done for each jet simulation.
}
\end{deluxetable*}

In our model the temporal evolution of the primary electron population 
is computed with the dynamical evolution of the jet, 
accounting for the different acceleration and cooling processes using the particle module in PLUTO.
Note that this approach also helps us draw inference for the primary proton distribution (see below).

For the lepto-hadronic analysis however, we adopt a hybrid/quasi steady-state model where the initial primary electron and proton populations are assumed to have settled in a steady state. 
This can be interpreted as a scenario where the underlying dynamics and associated acceleration processes do continuously inject non-thermal particles in each zone,
which maintain the primary particle distribution (given to us by the hybrid RMHD simulation) 
for the duration of the lepto-hadronic analysis 
(i.e for the time needed for the solution of the time-dependent kinetic equations 
to reach a steady state, thus $T_{stop} = T_{ss}$).

Our hybrid-particle module that is coupled to the fluid dynamics, applied to derive the evolving electron spectra, should be understood as a pilot study that is able to bridge the complex jet dynamics with associated 
shocks and fluid instabilities that is required to shape realistic electron spectra.
Our approach is in contrast to existing emission zone modeling which assumes a fixed power-law particle injection (as e.g. adopted in \citet{Xue_2019}) throughout the evolution. 
Since, we are essentially estimating the neutrino and photon flux by post processing from an MHD simulation snapshot, we are inherently assuming that the primary electron distribution obtained from our hybrid model does not substantially evolve within one dynamical time step.

The spectra of the other particle species that are generated as secondaries are evolved with time by solving their respective kinetic equations until a 'final' steady state condition is achieved - marked by both the photon, and neutrino populations simultaneously satisfying the following steady state 
saturation criteria, 
\begin{equation}
    \sqrt{\int \left[ \frac{1}{N^{ i}(\gamma)} \frac{\partial N^{ i}(\gamma)}{\partial t} \right]^2 d\gamma} < 10^{-8}
    \label{SS_eq}
\end{equation}

 where, $N^{i}(\gamma)$ represents the number density at energy $\gamma$ of a particle species '\textit{i}'. This criterion is the $L^2$-norm of the fractional time derivative of $N^{ i}(\gamma)$, i.e. it quantifies the relative rate of change of the particle population with respect to time for each particle species.

The SED is evaluated only after both the photon and neutrino components have reached steady-state, because the SED depends directly on their steady-state populations and additionally neutrinos are among the last particle species to form.

There are three important timescales that are relevant for our analysis,
which are (i) the time needed for the system to achieve the 'final' steady state  in our lepto-hadronic framework, $T_{\rm ss}$,
(ii) the light crossing/escape time $\sim R_{\rm b}/c$, and 
(iii) the dynamical timescale of the jet $T_{\rm dyn} \sim R_{\rm b}/<v_{\rm fluid}>$.
We have seen that for our specific simulations for each zone $ T_{\rm ss} \sim R_{\rm b}/c$, which is much smaller than $T_{\rm dyn}$ for the chosen jet snapshot used for the lepto-hadronic analysis.  

Further, since the snapshots of the jet from the RMHD simulations are taken at times $t_{s} \gg R_{\rm b}/c$, 
we can safely assume that the particle distribution has already achieved steady-state by then. 
Thus, we may hold, under this approximation, the primary electron and proton populations fixed during the window of our lepto-hadronic analysis within 
$[t_{\rm s},\, \, t_{\rm s} + T_{\rm ss}]$ and evolve only the secondary species to steady state. 
This is justified because the fluid dynamics that is governing the Lagrangian particle distributions, evolves appreciably only over 
the larger dynamical timescale $T_{\rm dyn}$.

In summary, we initially prescribe a primary electron and proton population, 
with the number density of all secondary particle species set to zero. 
The secondary species populations are then progressively built up till the 'final' steady state condition is satisfied. 

As briefly mentioned earlier, PLUTO's hybrid particle module only has provisions to incorporate electrons so far, we therefore adopt heuristic approximations for the proton distribution parameters based on the electron distribution parameters of each zone. This is based on the idea that the evolved electron distribution at the selected dynamical times $t=t_{\rm s1}$\, and $t_{\rm s2}$ for the respective jets, carries information about the history of acceleration and cooling processes the region has undergone during the dynamical evolution of the jet. 

For Case 1, we assume that the proton distribution has a similar functional form to the electron distribution obtained from the fitted spectra which can either be PL or PLEC. In Case 2, we take into consideration the fact that when compared to electrons, synchrotron and inverse Compton cooling is significantly less efficient for protons because of higher proton mass, resulting in much longer cooling timescales in protons. Hence, in this case we assume a simple power-law for protons but with no exponential cut-off for cooling. Additionally, we assume the proton distribution to have the same slope (in log scale) as the power-law portion of the fitted spectra for the electron distribution in each zone. Case 3 is similar to case 2 but we determine the $\gamma_{\rm max}$ of the proton distribution using the Hillas criterion which constraints the maximum allowed Larmor radius of the protons using the size of the emission zone. For Case 4, all other parameters are similar to Case 2 but this time we choose a much flatter power-law index ($\alpha^{\rm p} = 2.0$).

In all four cases, we assume the normalization constant of the proton distribution, $A^{\rm p}$, is scaled by the zonal proton-to-electron number density ratio ($\eta_{\rm zone}$), as compared to the electron distribution. The proton distribution normalization constant is defined as follows:
\begin{equation}
        A^{\rm p} = \eta_{\rm zone} \, A^{\rm e}_{\rm fit}.
\end{equation}
This ensures consistency with the proton number density considered for the zone, based on the $\eta_{\rm zone}$ constraint described earlier.
%--------------------------------------------------------------------------
\subsubsection{Multi-zone SED modeling}
Following the steps specified in our multi-zone framework algorithm and utilizing the initial parameters as described in Section~\ref{Approx_MZ} for the framework, we get a resultant photon and neutrino distribution when the steady state convergence criteria (Equation \ref{SS_eq}) is satisfied for individual zones. These photon and neutrino distributions can be used to generate the observed photon SED and neutrino flux for each zone i.e. the "zonal SEDs" using the following expression \citep{DermerMenon, Bottcher:2019ole, Katu2020},
\begin{equation}
    \nu F_\nu = \epsilon'^2 N'_{\rm ph} (\epsilon') m_{\rm e} c^2 \frac{\delta_{\rm zone}^4}{4 \pi (1 + z) d_{\rm L}^2} \frac{V'}{T'_{\rm esc}}
    \label{SED_eq}
\end{equation}
For this equation, to illustrate the transformation between the co-moving and observer frames, we use primed quantities for the co-moving frame and un-primed quantities for the observer frame \footnote{Previously, we refrained from using primed notation for the sake of convenience, as most quantities were defined in the co-moving frame.}.
Here, $\epsilon'$ is the photon energy, normalized to the rest mass energy of an electron, $N'_{\rm ph} (\epsilon')$ is the photon population density as a function of energy $\epsilon'$, $m_{\rm e} c^2$ is the rest mass energy of an electron, $z$ is the redshift of the source, $d_{\rm L}$ is the luminosity distance of the source, $V'$ is the volume of the emission region, and $T'_{\rm esc}$ is the escape timescale for photons.  A similar approach is applied to neutrinos, adopting the standard consideration that they arrive at Earth in a $1:1:1$ flavor ratio \citep{IceCube_flavorOsc_2015}. This leads us to the final step of the algorithm where we combine the simulated emissions contributed by each zone by stacking the zonal SEDs. This process generates the desired lepto-hadronic multi-zone SED consisting of both multi-wavelength and neutrino fluxes for a section of the blazar jet. 

%---------------------------------------------------------------
\section{From dynamics to synthetic emission signatures}
In this study, we have performed a detailed analysis of morphologically intriguing regions that arise from the interdependence between instabilities and shock structures in the AGN jet simulations. 
We begin by presenting results from our non-resistive 3D RMHD simulations, which model cylindrical, relativistic, and magnetized section of an AGN jet. 
This is followed by a zonal analysis of the different regions of interest. 
We then present findings from our lepto-hadronic multi-zone analysis using the algorithm described in Section~\ref{MZ_sec}, that yields photon and neutrino SEDs which offer fundamental insights into the internal dynamics of the jet and its correlation with emission signatures.

%---------------------------------------------------------------------------------------
\subsection{Dynamical evolution}
\label{sec_dynamical_evol}
 We have simulated cylindrical sections of parsec-scale jets with two different bulk Lorentz factors, $\Gamma_{\rm c,ini}=5$ and $10$ along the jet axis. 
The jets are initialized with a rotational velocity profile along with both poloidal and toroidal magnetic field components. 
A detailed description of the initial parameters and boundary conditions for the jet simulations is provided in Section~\ref{PLUTO_setup} and Appendix~\ref{A_Profiles}.

\begin{figure*}[t]
    \centering
    \includegraphics[width=0.9\textwidth]{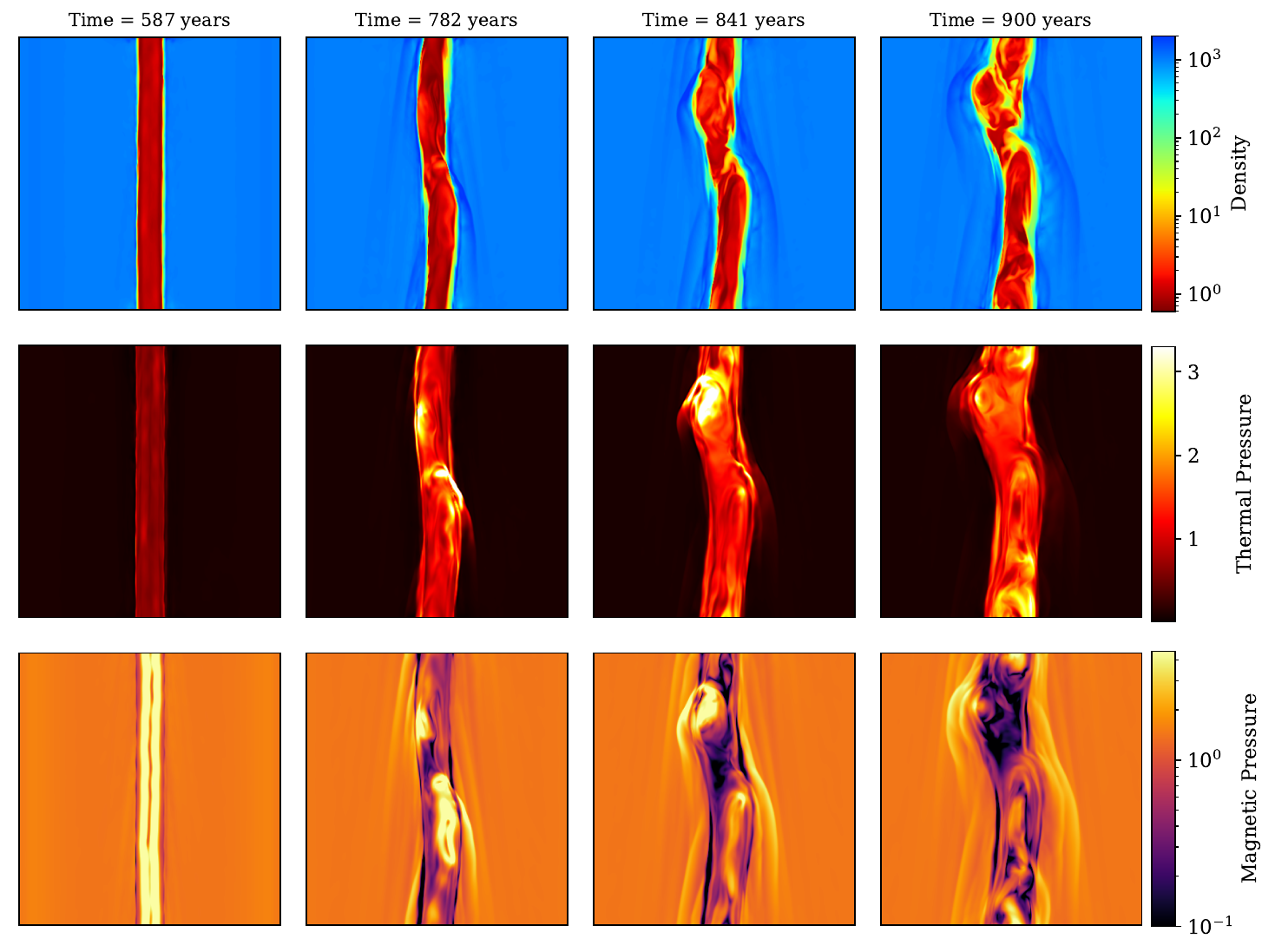}
    \caption{\small Magneto-hydrodynamic evolution of the jet. Shown are \textit{yz}-plane slices for the density (top), thermal pressure (middle) and magnetic pressure (bottom) at several 
    simulation times for the jet with $\Gamma_{\rm c, ini} = 5$. 
    All quantities are in code units.
    }
    \label{jetplots}
\end{figure*}

The relativistic and magnetized jet simulations presented here are particularly susceptible to current-driven instabilities due to the presence of a prominent toroidal magnetic field along with a radial velocity perturbation of the form described in Equation (\ref{eq_perturb}). Such a perturbation incorporates sausage, kink and fluting modes of instabilities, each competing for dominance. It also subsumes waves of different frequencies as well as randomized phase shifts. 
Therefore, proving to be well-suited for simulating realistic conditions where the jet maybe subjected to a variety of disturbances which may stem from sources like disk winds, jet precession, etc.

In our work, we primarily observe the emergence of a helical instability as the jet simulations evolve with time for both $\Gamma_{\rm c,ini} = 5$ and $10$ jets. 
A good diagnostic tool to differentiate whether the helical displacement is dominated by the current-driven kink instability, or the Kelvin-Helmholtz instability, 
is the Alfv\'en Mach number $\mathcal{M}_{\rm A}$, which can be defined as \citep{Bodo_2019},
\begin{equation}
    \mathcal{M}_{\rm A} = \frac{\Gamma_{\rm c}^2(1-v_{\rm A}^2)}{v_{\rm A}^2}.
\end{equation}
Here, $\mathcal{M}_{\rm A}$ essentially gives the ratio between matter energy density and magnetic energy density, $v_{\rm A} = B/\sqrt{\rho h +B^2}$ is the Alfv\'en speed (for $c=1$), and $h$ is the specific enthalpy.
Both jets start out with an average Alfv\'en Mach number of unity, but quickly transition to a sub-Alfv\'enic flow.
The onset of the helical instability, coincides with a rise in $\mathcal{M}_{\rm A}$, consistent with a decline in magnetic energy and a rise in transverse (\textit{xy}-plane) kinetic energy density. 

Despite the increase in $\mathcal{M}_{\rm A}$ during the growth of the kink instability, the $\Gamma_{\rm c,ini} = 5$ jet still remains sub-Alfv\'enic throughout the simulation time, 
while the $\Gamma_{\rm c,ini} = 10$ jet transitions from sub-Alfv\'enic during the instability formation process to marginally trans-Alfv\'enic in 
the later stages of the simulation. This is most probably due to faster hydrodynamic evolution (larger Lorentz factor) in the latter, as both jets exhibit similar magnetic field strengths during the later stages. 

These trends suggest that the helical deformations in both jets are primarily due to current-driven kink instabilities, owing to their predominantly sub-Alfv\'enic flow. 
However, the faster jet begins to exhibit some shear-driven features in the later stages, possibly 
indicating a minor contribution from Kelvin-Helmholtz instability dynamics as the jet velocity approaches the Alfv\'en speed.

\begin{figure*}[t]
    \centering
    \includegraphics[width=\linewidth]{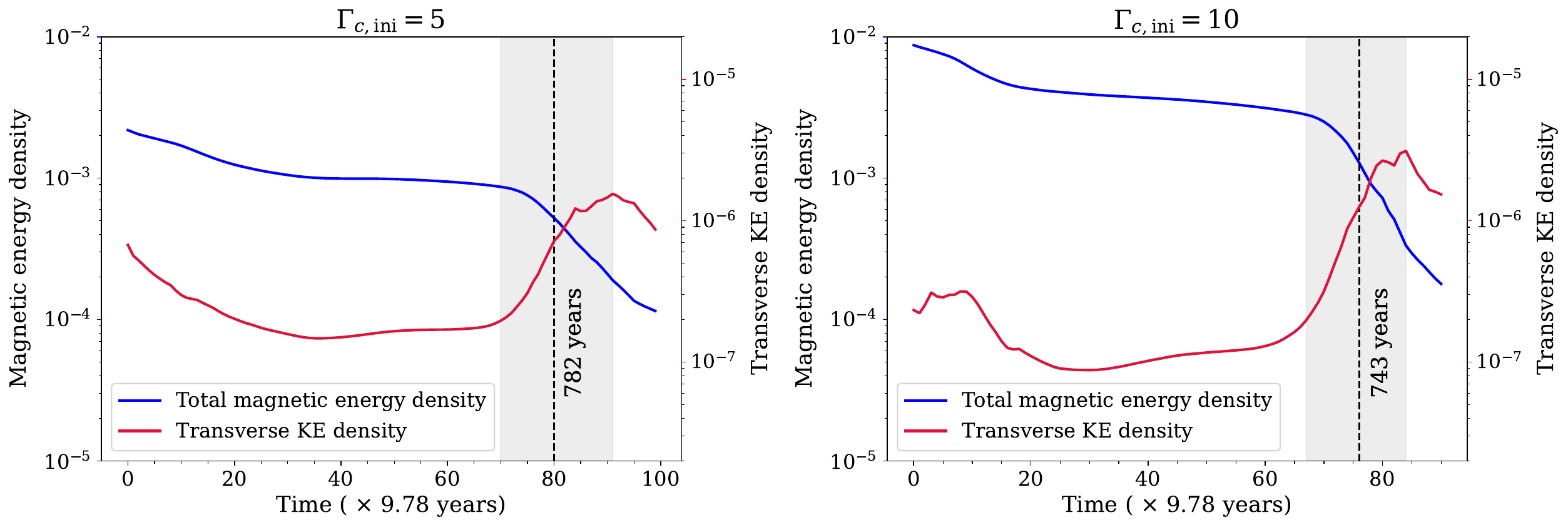}
    \caption{\small Evolution of total magnetic energy density and the transverse (\textit{xy}-plane) kinetic energy density for the jets with $\Gamma_{\rm c, ini} = 5$ (\textit{left panel}) and $\Gamma_{\rm c, ini} = 10$ (\textit{right panel}), respectively.
     The linear phase of kink instability formation is shaded gray and the \textit{}{dashed black line} represents the optimal time slice chosen for the multi-zone analysis ($t_{\rm s1}=782$\, years for the $\Gamma_{\rm c, ini} = 5$ jet and $t_{\rm s2}=743$\, years for the $\Gamma_{\rm c, ini} = 10$ jet). There are two vertical axes in each plot, the total magnetic energy density (\textit{blue solid line}) is associated with the vertical axis on the left side of the plots  while the vertical axis on the right side is associated with the transverse kinetic energy density (\textit{red solid line}) for each plot.
     }
    \label{energetics}
\end{figure*}

In Figure~\ref{jetplots}, we present the temporal evolution of the density, thermal pressure, and magnetic pressure for the $\Gamma_{\rm c,ini} = 5$ jet. 
The kink instability, characterized by a non-axisymmetric helical displacement of the jet body, grows when the magnetic pressure locally exceeds the 
stabilizing effect of magnetic tension. 

With the onset of the kink mode in Figure~\ref{jetplots}, 
we observe that regions of elevated density correlate with enhancements in thermal pressure. We interpret this as indicative of shock formation, 
which is marked by discontinuities in density, pressure and also the local flow speed. 
Notably, these regions also correspond to amplified magnetic pressure. Since we operate in the non-resistive, ideal MHD regime in our relativistic simulations, matter is expected to be more concentrated in such regions, simply due to flux-freezing. 

Kink-induced distortions can trigger the formation of multiple shocks and regions of intense magnetic compression, 
which in a resistive MHD scenario could serve as ideal sites for magnetic reconnection \citep{Striani_2016, Medina-Torrejon_2021}. 
As evident from the figure, while such regions are present across the jet, they are more concentrated near the sites where prominent helical
displacements occur, consequently triggered by the formation of a jet kink. We observe a qualitatively similar morphological evolution for the $\Gamma_{\rm c,ini} = 10$ jet, which for brevity we omit from the main 
text of the paper.

\begin{figure*}[t]
    \centering
        \includegraphics[width=0.475\textwidth]{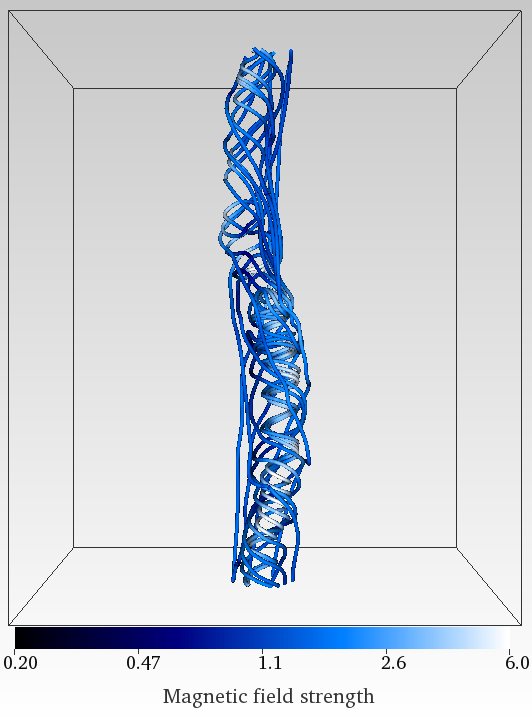}
        \includegraphics[width=0.475\textwidth]{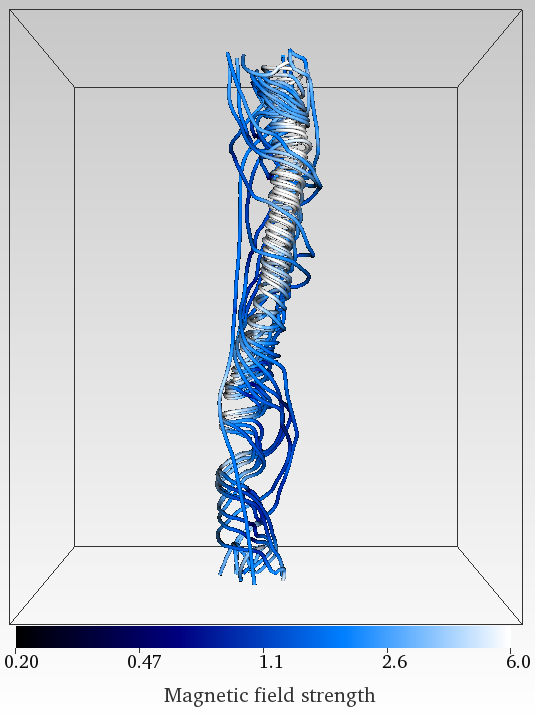}
    \caption{\small Magnetic field structure for the two jet simulations with different initial jet speed.    
    Shown are the field lines for the jet with $\Gamma_{\rm c,ini} = 5.0$ at $t_{\rm s1} =782$ years (\textit{left panel}) 
    and for the jet with $\Gamma_{\rm c,ini} = 10.0$ at $t_{\rm s2} = 743$ years (\textit{right panel}). 
    The color bar represents the magnetic field strength (in code units).
      }
    \label{Bline_plot}
\end{figure*}

 In Figure~\ref{energetics} we illustrate how the formation of the instability plays a significant role in governing the evolution of the different energy components of the jet over time. 
We have calculated the volume-averaged components of energy density for the jet in the lab frame (see Equation 10 in \citealt{Dubey_2023}).
The formation of the kink instability advances through linear and non-linear phases. 
In the linear phase, small perturbations to the equilibrium state of the plasma grow over time 
resulting in the formation of a helical kink instability, marked by a rise in the transverse(\textit{xy}-plane) kinetic energy of the jet. 
The system transitions into the non-linear phase as the transverse kinetic energy declines, characterized by a significant transverse deformation of the overall jet structure. 
Simulation studies such as \citet{Mizuno_2009,Mizuno_2011, Mizuno_2012, ONeill2012, Acharya2021} give significant insights on these phases of kink instability growth.

During the linear growth phase, the perturbation causes a reconfiguration of the initially ordered toroidal magnetic field within the jet, 
causing spatial variations in the field strength across the jet. 
This introduces imbalances in magnetic tension which drives the plasma into a helical displacement pattern. The growing instability causes the magnetic energy of the jet to decline as the field distorts further from its initial configuration, weakening the toroidal component locally.
The associated transverse motion of the plasma during 
the development of the kink leads to an increase in transverse kinetic energy. To highlight the linear phase for each jet, we have shaded the corresponding region gray in Figure~\ref{energetics}. 

A portion of the magnetic energy lost by the jet during kink formation, is fed into the transverse kinetic energy channel and the thermal energy of the jet, while the remaining part is lost to the ambient medium surrounding the jet. 
We suggest that this possibly happens due to the shocks produced in the ambient medium due to a combination of kink induced deformations (see Figure~\ref{jetplots}) and jet rotation.

Figure~\ref{Bline_plot} presents 3D visualizations of the magnetic field lines at the chosen simulation times for each jet. The structure of the magnetic field lines clearly illustrates how the bending of the toroidal field component distorts the body of the jet, leading to kink formation. We see that there exists a higher concentration of magnetic field lines at the location of the kink distortions, which is also reflected in the respective density, thermal pressure, and magnetic pressure distribution.

In this work, we are interested in exploring whether intrinsic shocks produced during the linear growth phase of the kink instability, can generate significant neutrino emissions. As the jet progresses into the later stages of kink evolution, we see that the particle spectra across the zones show significant exponential cooling. Hence we select a simulation time near the midpoint of the linear phase to perform the multi-zone analysis (black dashed lines in Fig.~\ref{energetics}).  This allows us to study the interplay of both shock-driven particle acceleration as well as radiative cooling without the latter suppressing the imprints of shock acceleration on the particle spectra. 
%-------------------------------------------------------------------------------------------------------------------------
\subsection{Zonal analysis}
\label{sec_Zonal}
\begin{figure*}[t]
    \centering
    \includegraphics[width=0.7\linewidth]{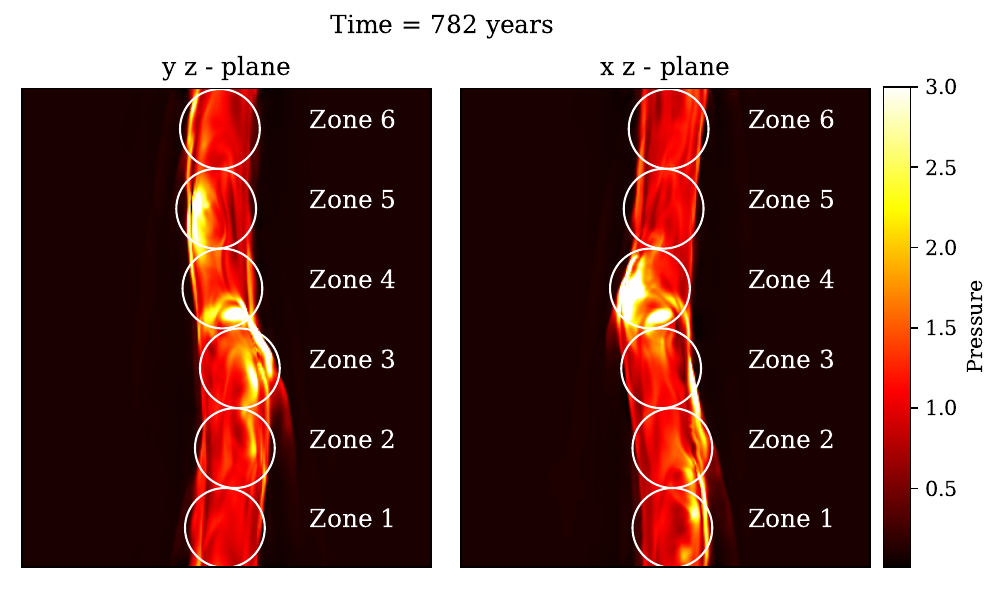}
    \caption{\small Slices of the $yz$-plane (\textit{Left}) and the $xz$-plane (\textit{Right}) with their color representing the thermal pressure (see color bar) 
    for the jet with $\Gamma_{\rm c,ini} = 5.0$ at time $t_{\rm s1} = 782$ years. 
    The respective spherical zones are marked by white circles.
    }
    \label{g5zones}
\end{figure*}

In Figure~\ref{g5zones}, we show the thermal pressure for the selected time instance $t_{\rm s1} = 782$ years for $\Gamma_{\rm c,ini} = 5$. 
The thermal pressure for both \textit{yz} and \textit{xz}-slices of the jet have been presented to illustrate the rationale behind the zone selection. 
As evident from the figure, we have segregated the jet column into six spherical zones. 
The centers of the spherical zones have been strategically positioned, such that the spheres encompass regions of elevated pressure which serve as signatures of shocks, generated due to the structural deformities caused by the formation of kink instability.
Regions with a higher concentration of shocks are located near the kinks due to the pronounced deformation of magnetic field lines in those regions.
This results in abrupt changes in plasma parameters, such as density, pressure and velocity leading to the formation of localized areas with increased shock activity. 

\begin{figure*}
    \centering
    \includegraphics[width=\linewidth]{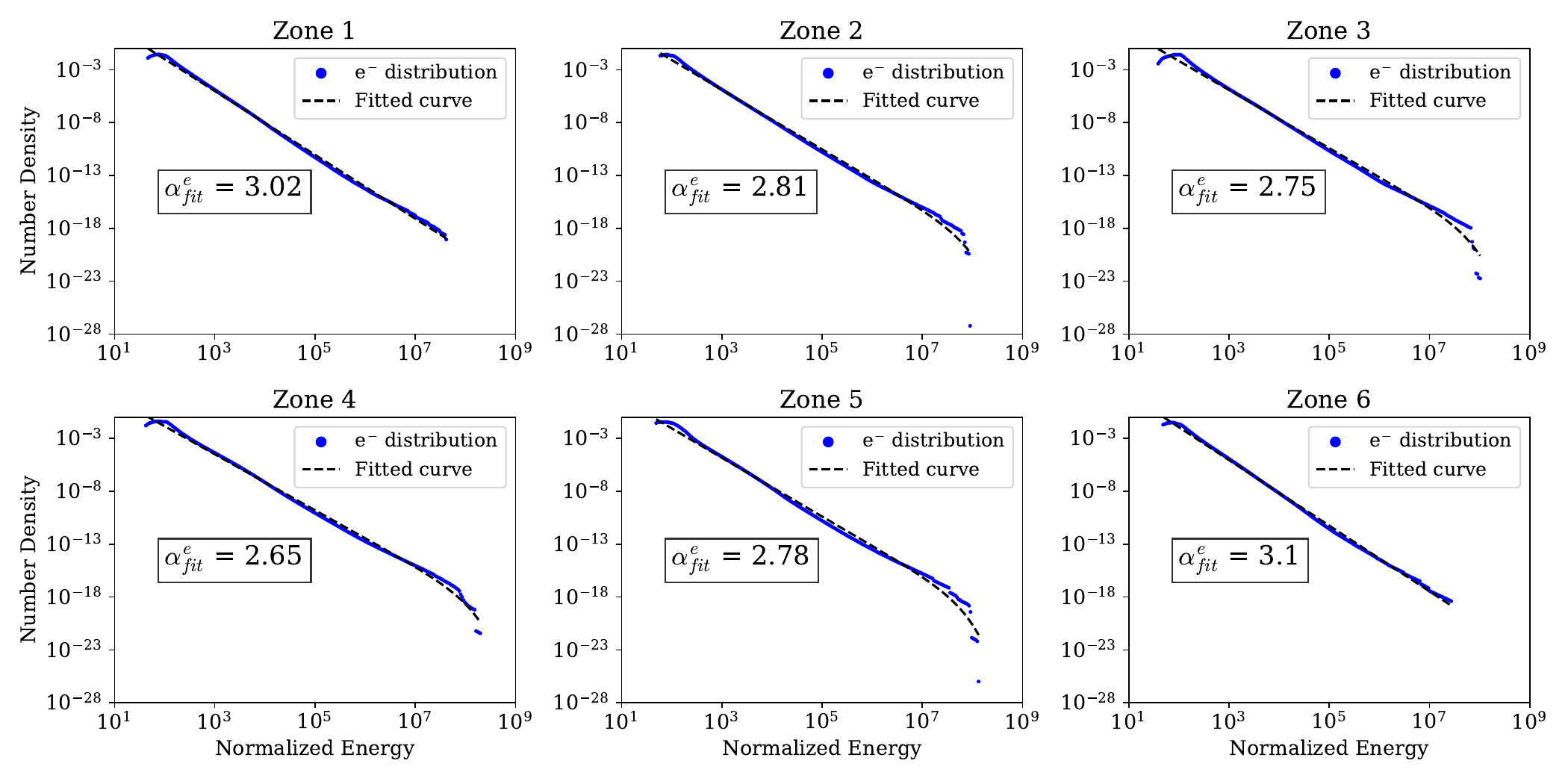}
    \caption{\small The respective electron distributions and the fitted PL/PLEC curve for the six zones of the jet with $\Gamma_{\rm c,ini} = 5$ at time $t_{\rm s1} = 782$ years. }
    \label{g5edist}
\end{figure*}

Subsequently, we present the electron distribution for all six zones for the jet with $\Gamma_{\rm c,ini} = 5$ in Figure~\ref{g5edist}. 
The electron distribution reveals that regions with elevated pressure are associated with a flatter electron power-law distribution. 
Additionally, these regions also exhibit an exponential cutoff in their power-law spectra.
This possibly arises from heightened radiative cooling activity induced by mechanisms like synchrotron and inverse Compton cooling in regions 
containing intense shocks and enhanced magnetic field variations. 

For the jet with $\Gamma_{\rm c,ini} = 5$, in Figure~\ref{g5zones} we can see that Zone 4 stands out among all the zones, as it encloses one of the helical kinks 
along with the largest high pressure region in the jet possibly arising due to shock induced plasma compression. 
Consequently, the electron distribution for this zone as depicted in Figure~\ref{g5edist}, shows the flattest power-law spectrum in the jet with an index 
of $\alpha^{\rm e}_{\rm fit} =2.65$.

\begin{figure}[t]
    \centering
    \includegraphics[width=1\linewidth]{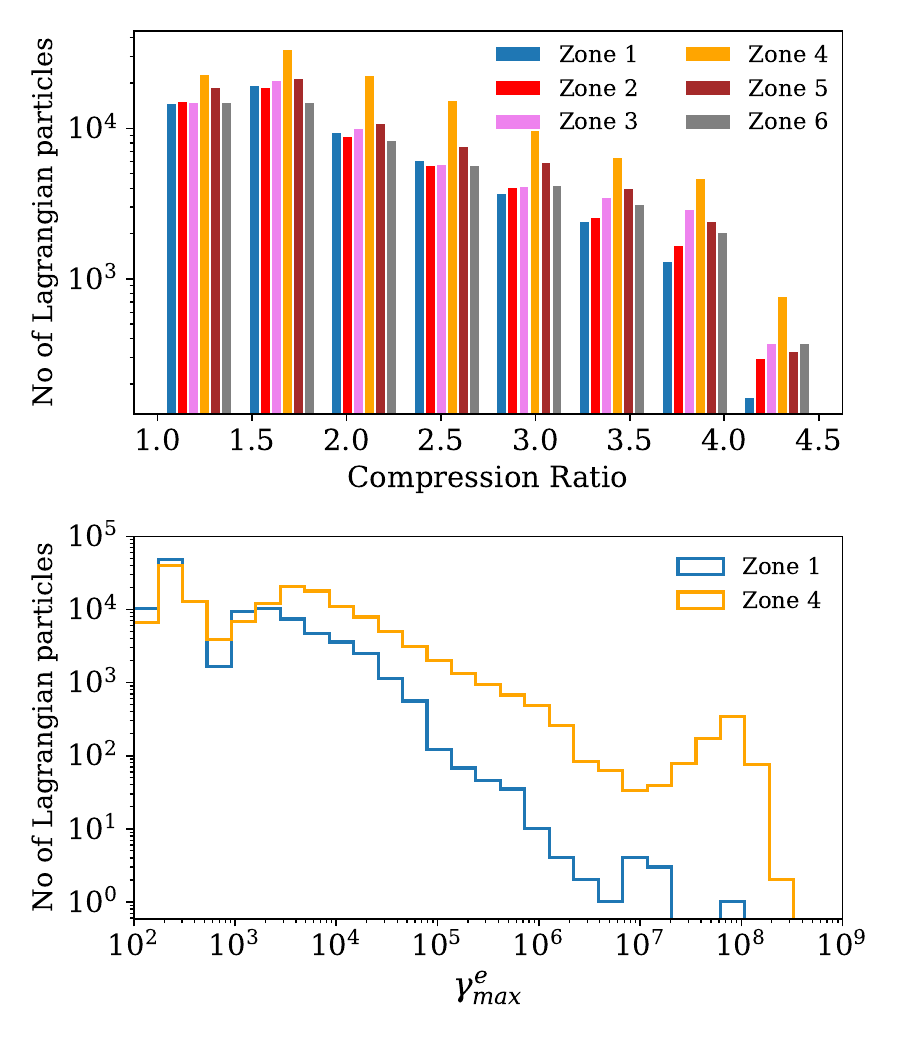}
    \caption{\small Histogram for the compression ratio of the last encountered shock for the Lagrangian particles inside the six zones of the jet with 
    $\Gamma_{\rm c,ini} = 5$ (\textit{top panel}).
    Step histogram for the $\gamma_{\rm max}$ of each Lagrangian particle specifically in Zone 1 and Zone 4 of the jet with $\Gamma_{\rm c,ini} = 5$ (\textit{bottom panel}).
    }
    \label{g5_hist}
\end{figure}

To understand how the shock strength associated with these dynamical signatures influences particle acceleration, in Figure~\ref{g5_hist} (\textit{top panel}) we present a histogram for the compression ratio of the last encountered shock for the electron Lagrangian particles within the six spherical zones of the $\Gamma_{c,ini} = 5$ jet. We see that Zone 4,  which features both a helical kink and a very prominent high-pressure region, consistently shows a higher number of macro-particles that have traversed through shocks with larger compression ratios. This suggests that such regions act as hotspots for the formation of strong shocks. Motivated by these results, we plot a step histogram (\textit{bottom panel} of Figure~\ref{g5_hist}) for the $\gamma_{max}$ of each Lagrangian particle within Zone 1 and Zone 4  - representing the zones with the lowest and highest number of Lagrangian particles undergoing high-compression-ratio shocks, respectively. We observe that Zone 4 indeed has more particles getting accelerated to higher energies compared to Zone 1, as the latter lacks a prominent high pressure region and has a much steeper electron distribution (see Figure~\ref{g5edist}).

\begin{figure*}[t]
    \centering
    \includegraphics[width=0.7\linewidth]{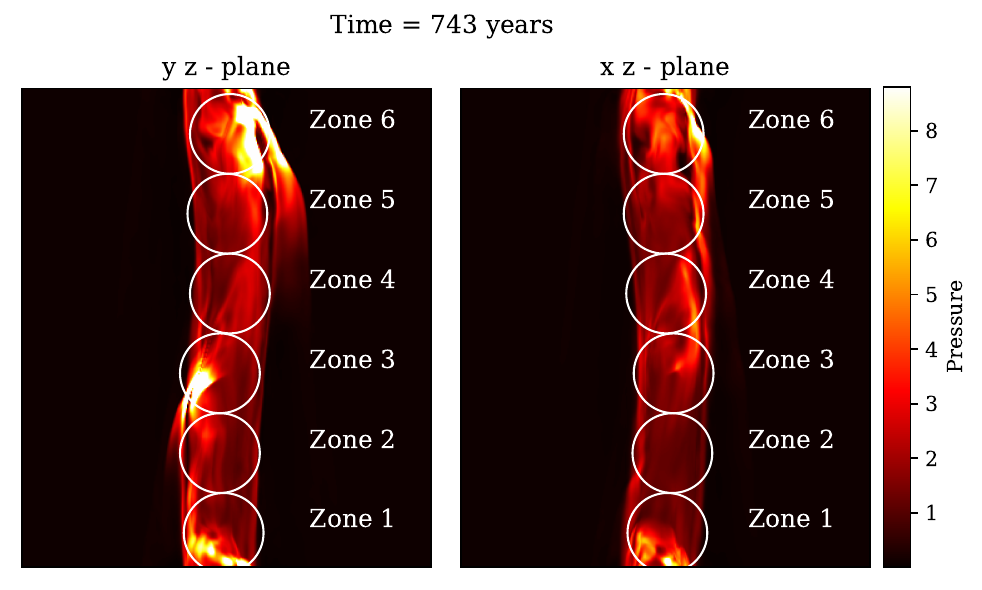}
   \caption{\small Slices of the $yz$-plane (\textit{Left}) and the $xz$-plane (\textit{Right}) with their color representing the thermal pressure (see color bar)
    for the jet with $\Gamma_{\rm c,ini} = 10$ at time $t_{\rm s2} = 743$ years. 
    }
    \label{g10zones}
\end{figure*}

Figure~\ref{g10zones} illustrates the thermal pressure for the selected time instance $t_{\rm s2} = 743$ years for the $\Gamma_{\rm c,ini} = 10$ jet. Here again we have divided the jet into six spherical zones. As evident from the figure, we observe that Zones 1, 3 and 6 in the \textit{yz}-plane show indications of high pressure regions which are characteristic signatures of shocks. The jet column also shows bending at these locations which is consistent with the $\Gamma_{c,ini} = 5$ jet simulations. While in the \textit{xz}-plane the shock signatures are observed in Zones 1 and 6.

\begin{figure*}[t]
    \centering
    \includegraphics[width=\linewidth]{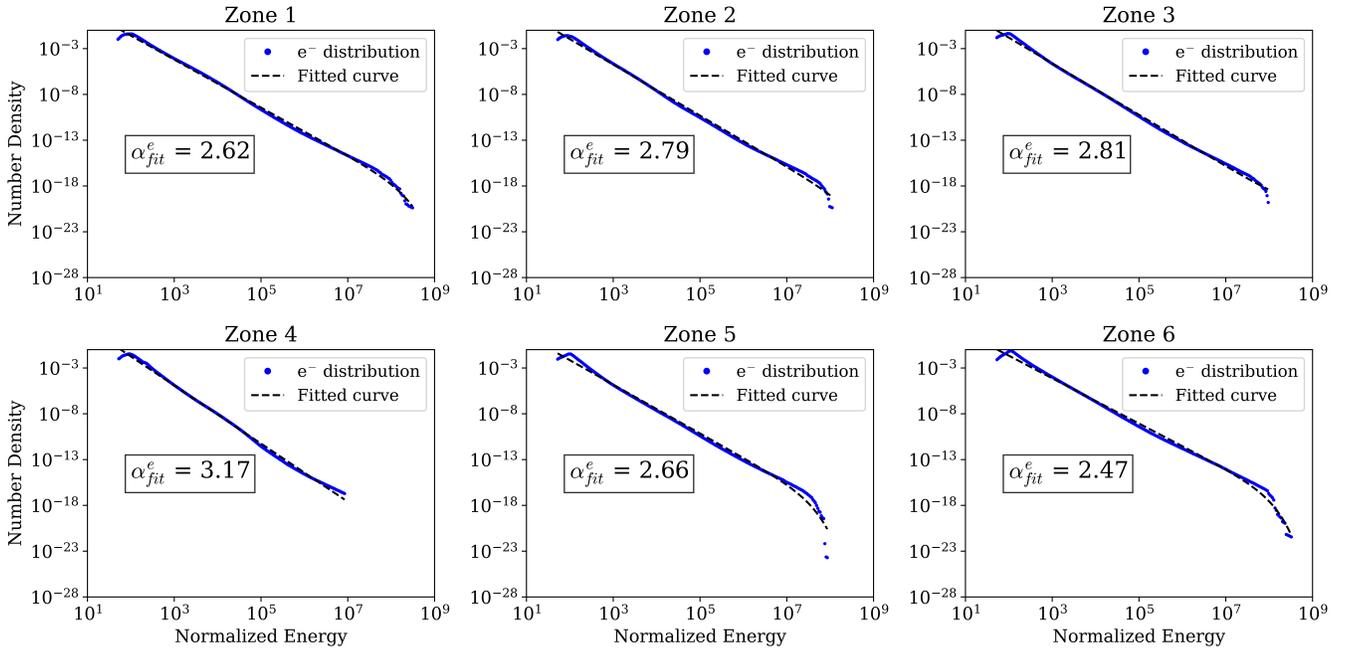}
    \caption{\small The respective electron distributions and the fitted PL/PLEC curve for the six zones of the jet with $\Gamma_{\rm c,ini} = 10$ at time $t_{\rm s2} = 743$ years.}
    \label{g10edist}
\end{figure*}

Figure~\ref{g10edist} depicts the respective electron distributions for all six zones of the $\Gamma_{c,ini} = 10$ jet. 
A pattern similar to the zonal analysis of the $\Gamma_{\rm c,ini}=5$ jet, can be observed for the $\Gamma_{\rm c,ini} = 10$ jet simulation, 
where Zone 6 consisting of the highest pressure region, 
corresponds to the flattest electron power-law distribution with an index of $\alpha^{\rm e}_{\rm fit} =2.47$. 

These results indicate that varying particle distributions emerge from different emission zones, as the jet undergoes instability dynamics. Additionally, higher pressure regions are associated with flatter particle distributions, likely due to heightened shock dynamics resulting in more efficient acceleration processes. 
Additionally, such regions exhibit exponential cooling in higher energies, potentially arising from amplified radiative cooling that modifies the underlying electron energy distribution in those zones.

%-----------------------------------------------------------------------
\subsection{Multi-zone SED}
\begin{figure*}[t]
    \centering
    % --- top row: two panels ---
    \begin{minipage}{0.49\textwidth}
        \centering
        \includegraphics[width=\linewidth]{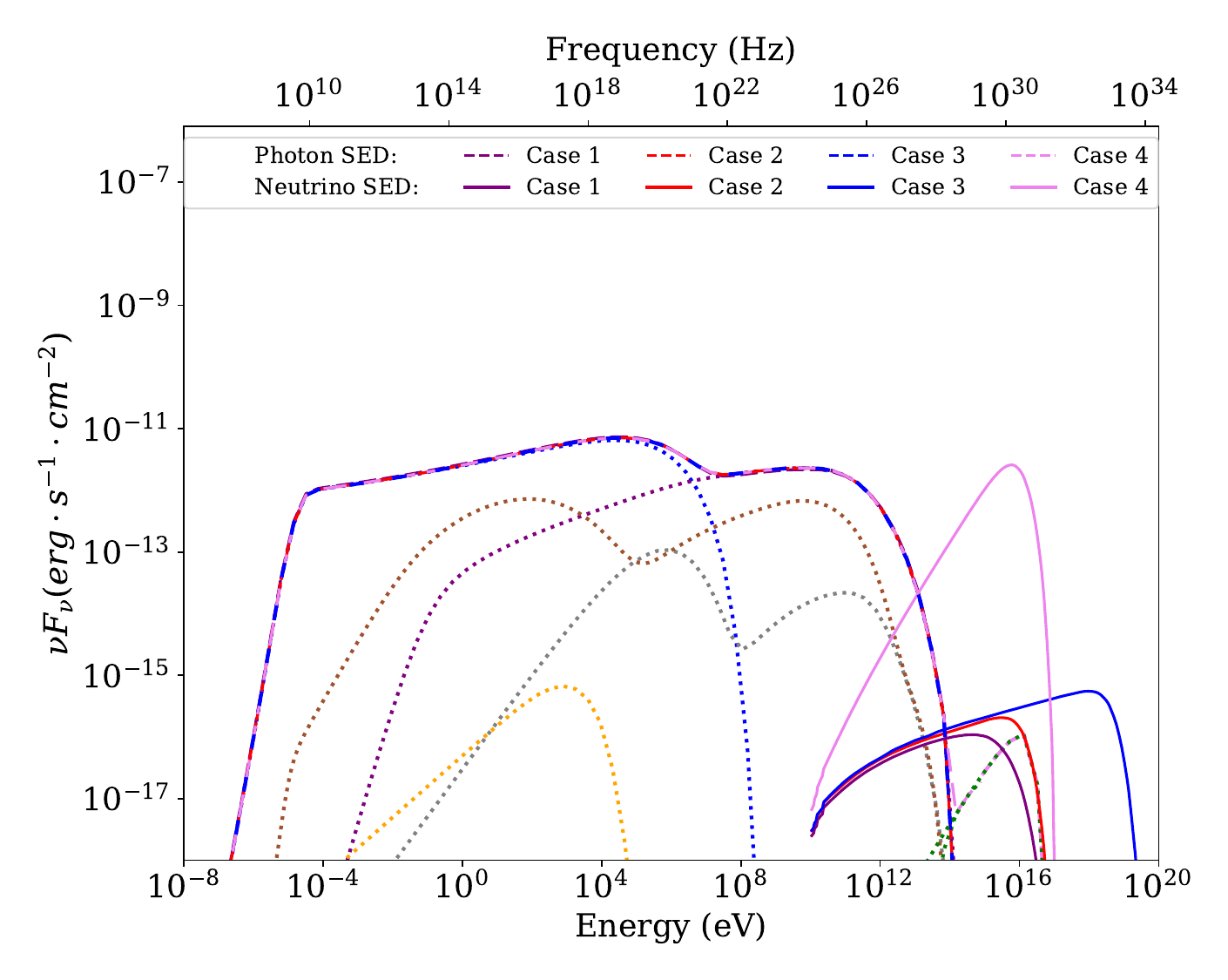}
    \end{minipage}\hfill
    \begin{minipage}{0.49\textwidth}
        \centering
        \includegraphics[width=\linewidth]{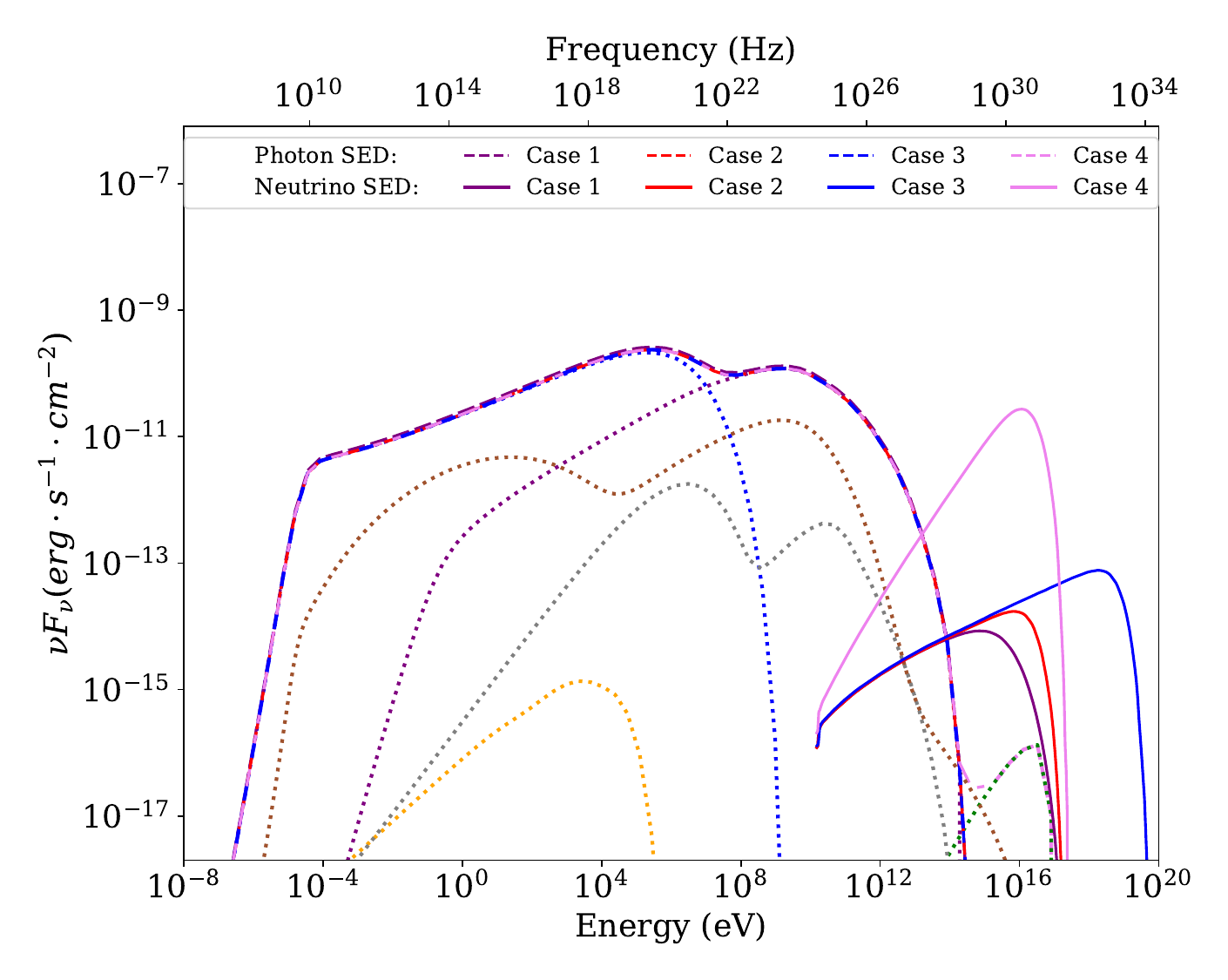}
    \end{minipage}

    % --- bottom row: legend spanning full text width ---
    \par\vspace{0.5em}
    \begin{minipage}{0.6\textwidth}
        \centering
        \includegraphics[width=\textwidth]{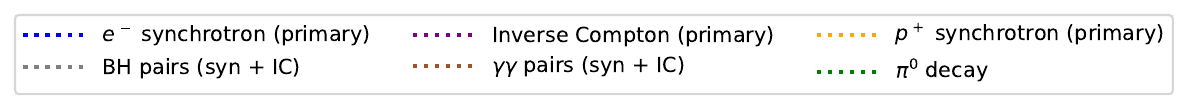}
    \end{minipage}

    \caption{\small The Multi-zone photon (\textit{dashed lines}) and neutrino (\textit{solid lines}) SED for all four cases for the proton distribution for the jets with $\Gamma_{\rm c,ini} = 5.0$ (left), and $\Gamma_{\rm c,ini} = 10.0$ (right), respectively. The multi-zone photon SED components for Case 4 arising from different processes have been depicted using \textit{dotted lines} (each component sums up the contributions from different zones).}
    \label{MZ}
\end{figure*}

 We present the multi-zone SEDs for all four proton distribution cases considered (see Table~\ref{tab_proton}) for the two jet simulations in Figure~\ref{MZ}.
These SEDs are obtained by combining the line-of-sight integrated contributions from individual zonal SEDs across the six emission zones. 
In this figure, we depict both the multi-wavelength photon flux as well as the combined muon neutrino and anti-neutrino $(\nu_\mu + \bar{\nu}_\mu)$ flux in the observer frame. This has been calculated as one-third of the total neutrino population generated at the source to take into account neutrino oscillations.
For the SED modeling, we have assumed a redshift of $z=0.3365$, identical to the TXS 0506+056 source, which gives us a luminosity distance of $d_{\rm L}=1762$\,Mpc. 
Additionally, we assume the jet to be a blazar, with the jet axis making an angle of $\theta_{\rm los} = 1\degree$ with our line of sight.

For both jet simulations, we find that a clear trend emerges for the neutrino flux for the four proton distribution cases. Compared to Case 1, we observe an increase in the multi-zone neutrino flux for both jets in Case 2.
The Case 3 setup not only exhibits a noticeable enhancement in neutrino flux, but also displays  flux distributed over a broader energy range, due to more higher energy protons being available to participate in the photo-meson interaction owing to the Hillas criteria. 
For Case 4, we see a significant increase in the neutrino flux compared to the other cases, which arises from the power-law of the proton distribution being assumed to be very flat with an index of $\alpha^{\rm p} = 2.0$. 
  We also observe that the flux for both photon and neutrino flux is a few orders of magnitude higher for the jet with $\Gamma_{\rm c,ini} = 10$, 
  as compared to the jet with $\Gamma_{\rm c,ini} = 5$, across all cases.
This is primarily due to higher average Doppler factors in certain zones arising from a higher initial bulk Lorentz factor. 

For both jets, the multi-wavelength photon flux features two peaks - the relatively lower energy peak arising due to synchrotron emissions and the high-energy component resulting from inverse Compton scattering. 

Figure~\ref{MZ} also depicts the prominent components for Case 4 of the multi-zone photon SED (\textit{dotted lines}), for both the simulated jets. The SED components are calculated by combining the contributions from each zone in the jet.
Different zones contribute across the frequency spectrum to varying extents, shaping the multi-zone photon flux, with certain zones dominating in specific frequency bands. Interestingly, the corresponding total multi-zone photon flux remains almost identical across all four proton distribution cases for each jet simulations. 
Only notable difference arises at high frequencies, where the $\pi^{0}$-decay photons contribute to the total photon flux, with Case 4 exhibiting comparatively higher $\pi^{0}$-decay flux. We also observe that the primary lepton synchrotron and inverse Compton components of the photon SED dominate over the hadronic counterparts. This clearly reflects the consistently low values of the underlying proton-to-electron number density ratio $\eta_{\rm zone}$ across all six zones for our jet simulations.

 The average power for the RMHD jets excluding the rest mass energy \citep[as described in][]{Mckinney, English2016, Mukherjee2020, UPRETI2024146} at initial time, is estimated to be $\approx 4 \times 10^{45}$\, erg s$^{-1}$ for the $\Gamma_{\rm c,ini} = 5$ jet, and $\approx 2 \times 10^{46}$\, erg s$^{-1}$ for the $\Gamma_{\rm c,ini} = 10$ jet. 

%-----------------------
\subsection{Impact of external photon fields}
\label{sec_EC}
With an aim to study the impact of photon fields external to the jet as targets for inverse Compton emission, we conducted additional simulations for the $\Gamma_{c,ini} = 5$ jet. We refer to the simulation run considering the BLR photon field as \textit{Ref\_g5\_BLR}, and the run considering the DT photon field as \textit{Ref\_g5\_DT}. For the lepto-hadronic analysis we consider only Case 4 for the proton distribution where we take $\alpha^{\rm p} = 2.0$ (see Table~\ref{tab_proton}).

\begin{figure}
    \centering
    \includegraphics[width=\linewidth]{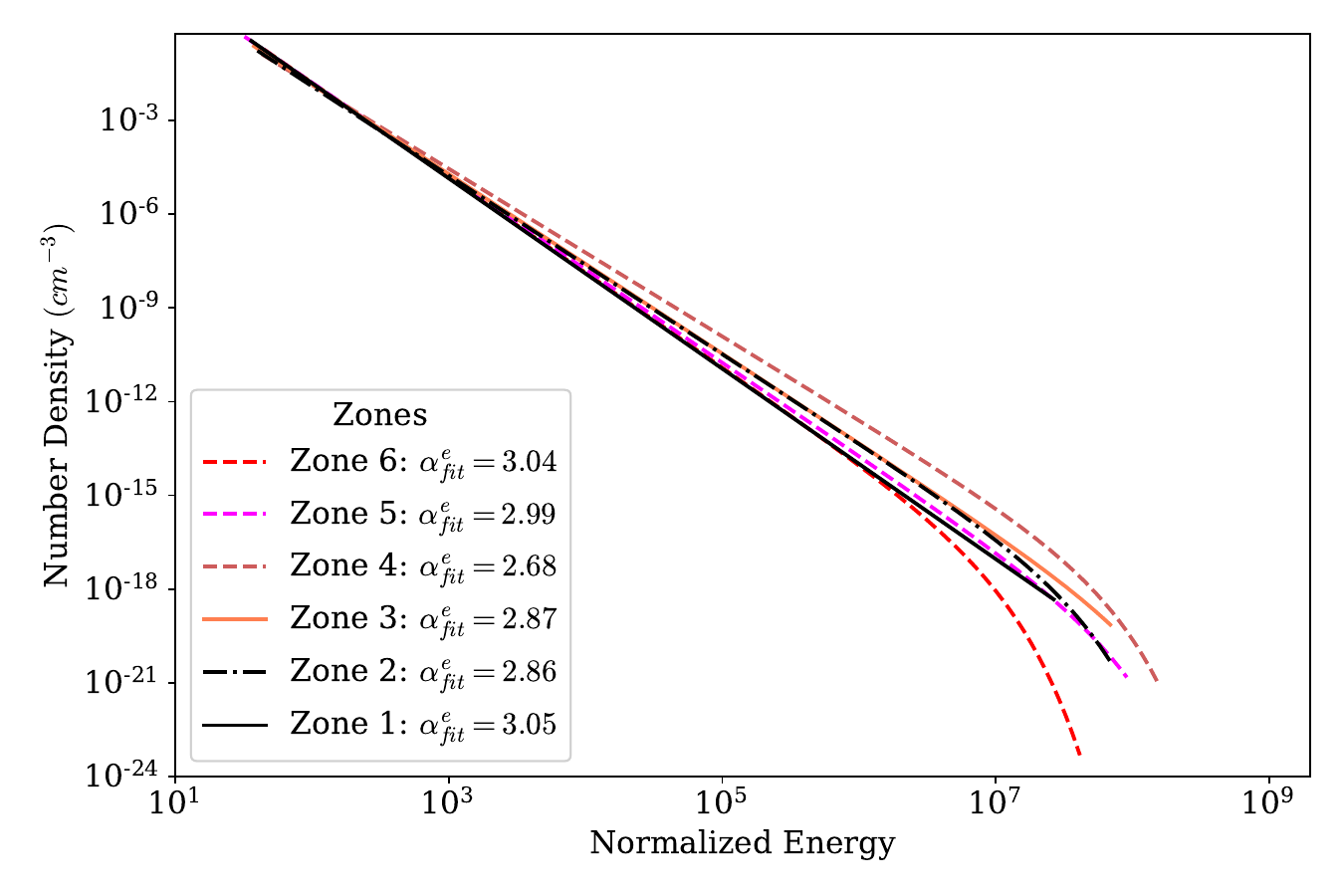}
    \caption{Electron distribution fitted curves for the 6 spherical zones of the \textit{Ref\_g5\_BLR} jet.}
    \label{edist_EC}
\end{figure}

\begin{figure}
    \centering
    \includegraphics[width=\linewidth]{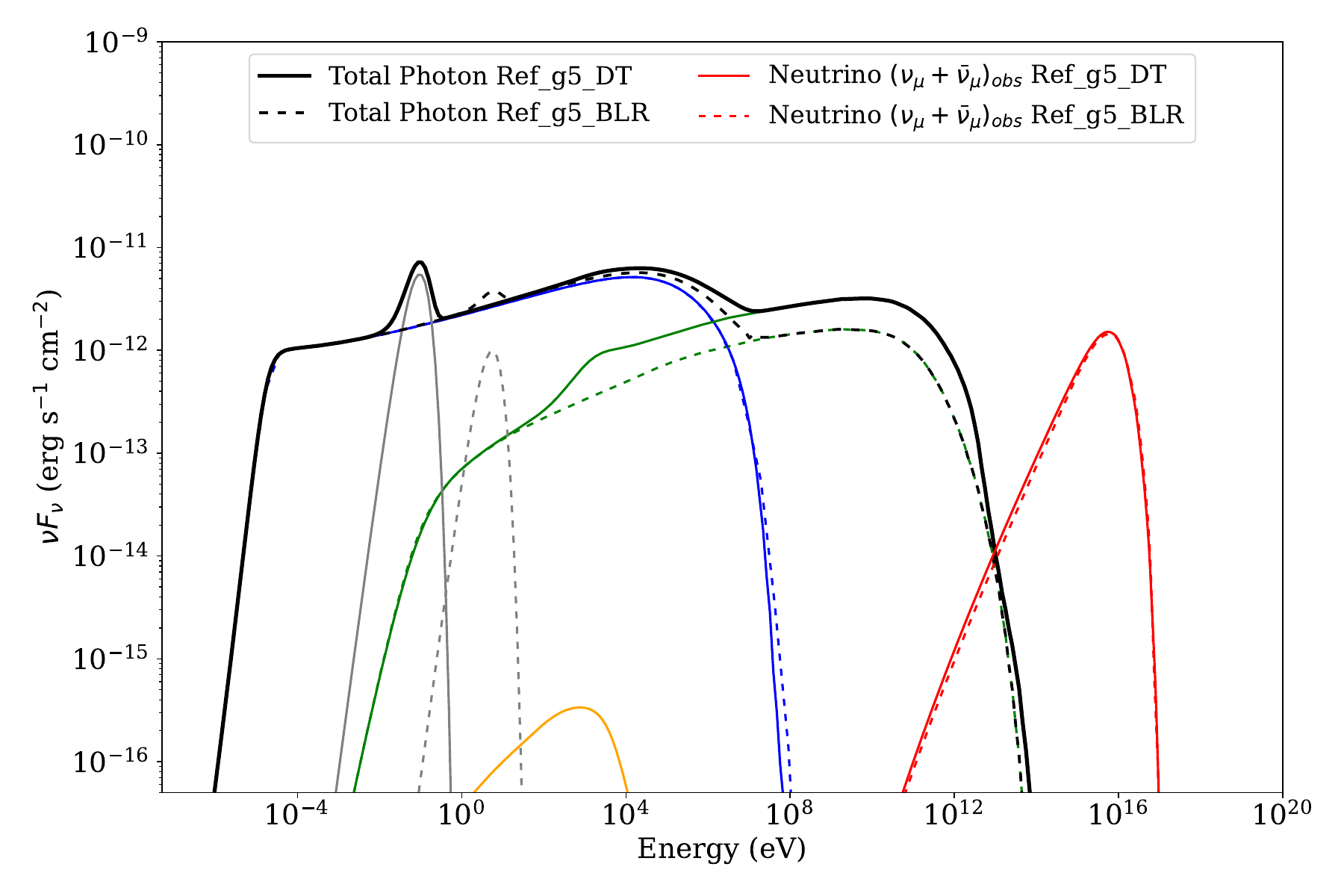}
    \caption{The multi-zone photon and neutrino SED for \textit{Ref\_g5\_BLR} and \textit{Ref\_g5\_DT} jets. The multi-zone SED components depicted includes: electron synchrotron emission (\textit{blue lines}), proton synchrotron emission (\textit{orange lines}), inverse Compton emissions (\textit{green lines}), and the external photon field (\textit{orange lines}). The \textit{solid lines} are for \textit{Ref\_g5\_DT} and \textit{dashed lines} are for \textit{Ref\_g5\_BLR}.}
    \label{MZ_EC}
\end{figure}

The zonal segmentation for the \textit{Ref\_g5\_BLR} and \textit{Ref\_g5\_DT} jets are done at the time $t_{\rm s1} = 782$ years, 
similar to its $\Gamma_{\rm c,ini}=5$ counterpart without EC (Section \ref{sec_Zonal}).
In Figure \ref{edist_EC}, we show the PL/PLEC fits for the electron distribution in each of the six spherical zones for \textit{Ref\_g5\_BLR}. 
Due to the distance of our simulated jet segments from the external photon fields, as well as the presence of active DSA during the linear kink phase causing acceleration to be dominant at the selected time, 
the consideration of EC results in small to moderate modifications in the zonal electron distributions compared to the non-EC $\Gamma_{\rm c,ini}=5$ jet simulations.
Specifically, we find that $\gamma^{\rm e}_{\max}$ relatively decreases by $\approx 6.9 - 12 \%$ 
in certain zones for \textit{Ref\_g5\_BLR} and by $\approx 7.2 - 25\%$ in certain zones for \textit{Ref\_g5\_DT}.
The power-law indices for the EC simulations also exhibit small variations compared to the zones of the non-EC jet simulation.

 In Figure \ref{MZ_EC} we compare the multi-zone photon and neutrino SEDs, combining contributions from each zone for the \textit{Ref\_g5\_BLR} and \textit{Ref\_g5\_DT} jets.
 For both simulations, the figure also depicts the major spectral components of the SED (combining all zonal contributions), which includes electron synchrotron emission, 
proton synchrotron emission, and inverse Compton scattering. Additionally, we have depicted the respective external photon fields for both the EC simulations. 

 Due to the distance-dependent dilution, the BLR photon field energy density becomes too weak to substantially modify the inverse Compton or neutrino components for pc scale distances. Hence, the resulting SED for \textit{Ref\_g5\_BLR} closely resembles that of the $\Gamma_{\rm c,ini} = 5$ jet without EC, in terms of the flux across different frequencies. While for \textit{Ref\_g5\_DT}, we observe relatively higher IC emissions, by a factor of $2 - 80$ across different frequencies, compared to \textit{Ref\_g5\_BLR}. This is expected due to the relatively higher co-moving frame energy density provided by the DT field in the emission zones. This trend is consistent with previous studies \citep[e.g.][]{Ghisellini_Tavecchio_2009, Tavecchio2013}, which showed that the DT is the dominant radiation field at the distances of $\gtrsim\, 10$\, pc from the central engine which is relevant to our study. 
We also find modest enhancements of the total photon flux  around the peak frequencies of each of the external photon fields. However, the neutrino SED remains similar for both \textit{Ref\_g5\_BLR} and \textit{Ref\_g5\_DT}. 

\section{Bridging synthetic emission signatures with observations} 
A major objective of this study is to explore the intriguing relationship between jet composition, the onset of jet instabilities,
and the emission features of AGN jets, using a computational framework for lepto-hadronic multi-zone modeling that we have developed as part of this work. 
We have carried out the multi-zone lepto-hadronic analysis as described in Section~\ref{MZ_sec} and have derived the resultant synthetic photon SEDs and neutrino fluxes for a kink unstable 3D magnetized jet column. 
Essentially, the adopted approach greatly reduces the otherwise extensive free parameter space that would be required for multi-zone modeling. 
The study further aims to pave the way for connecting RMHD simulations of AGN jets with its multi-wavelength and neutrino emission signatures. Here, we present a discussion of the findings from our RMHD jet simulations and its synthetic SEDs alongside the observations from the well known blazar TXS 0506+056 to gain deeper insights into the underlying physics of such parsec-scale blazar jets. 

\subsection{Insights from synthetic multi-zone photon flux}
 Our jet simulations do exhibit some key similarities to features of the blazar TXS 0506+056, as indicated by VLBI studies such as \cite{Li_2020} and \cite{Ros_2020}, 
 which describe the source as a parsec-scale relativistic, magnetized jet showing evidence of jet bending.
 This resembles our simulations of parsec-scale sections of relativistic, magnetized jets undergoing helical kink instabilities that result in prominent structural bending at these scales(see Figures \ref{jetplots}, \ref{g5zones}, and \ref{g10zones}). 
 
 While there are studies such as \citet{Britzen:2019} that suggest that such structural bending can arise from two-jet cosmic collisions, there are also simulation based studies like \citet{Tchekhovskoy2016} and \citet{Barniol2017} that show kink instabilities causing large scale bending in jets.

Despite some of these similarities between the TXS 0506+056 blazar and our jet simulations, there are some distinctions, notably the magnetic fields in our simulations are weaker than those inferred for TXS 0506+056 in \citet{Li_2020}. 
Additionally, the particle distribution parameters in blazar jets are poorly constrained and highly model-dependent, often relying on best-fit values within the framework of simplified one-zone models. 
These uncertainties, combined with limited knowledge of key physical parameters that influence the emission, inevitably introduce differences
between our RMHD jet and the actual conditions in TXS 0506+056. 

As such, we refrain from a direct quantitative comparison with the observational data for TXS 0506+056. 
Nonetheless, a rough qualitative comparison between the synthetic multi-zone SED for the simulated $\Gamma_{c,\mathrm{ini}} = 10$ jet and observations of 
TXS 0506+056 during its electromagnetic flares \citep{IceCube_MMA2018} reveals that their photon fluxes are of similar orders of magnitude in the optical and gamma-ray bands, and partially so in the radio band. 
However, the synchrotron component of the multi-zone SED from our simulated jet peaks at a significantly higher energy than that observed in 
TXS 0506+056.
Since the synchrotron photon energy scales as $\propto \gamma^2 B$, the variation in the synchrotron peak between the synthetic multi-zone SED for the $\Gamma_{c,\mathrm{ini}} = 10$ jet and the observed SED for TXS 0506+056 likely arises due to the potential differences in the magnetic fields as well as the differences in the maximum particle energies achieved in the jet. In particular, during the linear growth phase of the kink instability in our jet simulation, particles are accelerated to very high energies in each zone, with maximum energies reaching $\gamma_{\max} \sim 10^8$ which may substantially contribute to shifting the synchrotron peak to higher energies in our synthetic SEDs. 

\subsection{Comparing synthetic neutrino emissions with observations for TXS 0506+056}
 In the subsequent discussion, we provide a qualitative comparison between the synthetic multi-zone neutrino fluxes generated across different cases of our jet simulations and the observed neutrino flux from TXS 0506+056.

\begin{figure}[h]
    \centering
    \includegraphics[width=\linewidth]{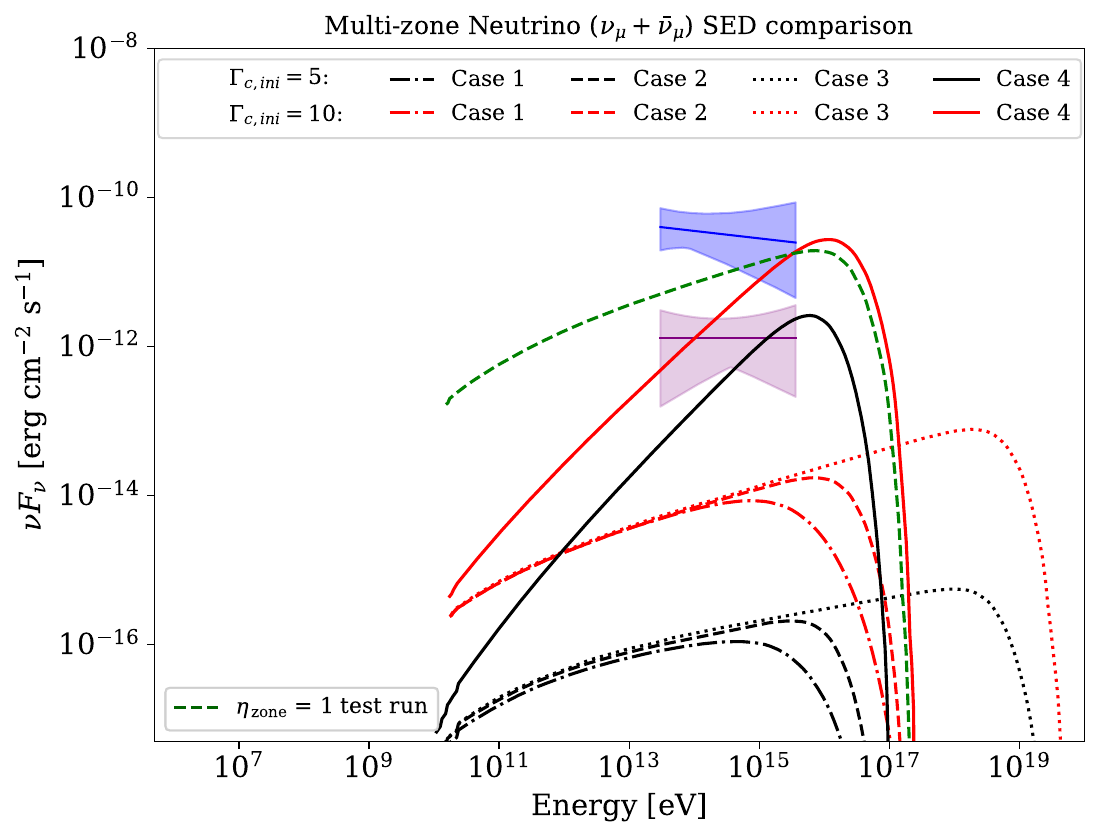}
    \caption{\small Multi-zone muon neutrino SEDs for the four proton distribution cases of $\Gamma_{\rm c,ini} = 5$ (\textit{black curves}) and $10$ (\textit{red curves}) jets. The \textit{dashed green curve} represents the multi-zone neutrino flux for the test simulation with $\eta_{\rm zone} = 1$. The blue "bow-tie" represents the best-fit muon neutrino flux (\textit{blue line}) during the 110 day flaring state of TXS 0506+056 obtained by IceCube with its $68\, \%$ confidence region (\textit{blue shaded region}), The pink bow-tie depicts IceCube's best fit muon neutrino flux (\textit{purple line}) from the 9.5 year data of TXS 0506+056 with the its the $68\, \%$ confidence region (\textit{pink shaded region}) \citep{icecube2018dataset,IceCube2018}. 
    }
    \label{Neu_SED}
\end{figure}

In Figure~\ref{Neu_SED}, we only use the IceCube data for TXS 0506+056 as a benchmark to explore whether the multi-zone muon neutrino fluxes derived from our simulated jets are of comparable magnitude and lie within an observationally relevant range, without attempting a strict quantitative fit (discussed in detail later).To understand the observational implications of these synthetic neutrino fluxes, we complement this figure by performing a calorimetric analysis that gives us a more quantitative estimate of the expected number of events that such fluxes may possibly generate in the IceCube Neutrino Observatory over different observational time periods. For the analysis, we use
\begin{equation}
    n^{\rm exp}_{E_{\nu}} = A_{\rm eff} (E_{\nu})\, \frac{\mathcal{F}^{\nu}}{E_{\nu}}\,  \Delta T,
    \label{eq_n_exp}
\end{equation}
 where $\mathcal{F}^{\nu}$ is the integrated neutrino flux, $A_{\rm eff} (E_{\nu})$ represents the effective area of the detector for a neutrino of energy $E_{\nu}$ arriving from a particular direction, and $\Delta T$ is the observational time interval. This estimate has been previously used in \citet{Krauss2014, 2016Kadler_nu, Kreter_2020}. These studies adopt an upper limit for the integrated neutrino flux by using the observed gamma-ray flux. Whereas, for our analysis, we directly utilize the synthetic neutrino flux derived from the multi-zone modeling of our simulated jets to estimate $n^{\rm exp}_{E_{\nu}}$.
 
 As specified in Table~\ref{tab_proton}, for the multiple emission zone analysis of our jet simulations, we have taken 4 cases of assumptions for the proton distribution parameters. In Table~\ref{tab_n_exp}, for both of our simulated jets we present the expected number of neutrino events  $n^{\rm exp}_{E_{\nu,\rm peak}}$ for these cases, associated with the characteristic neutrino energy $E_{\nu,\rm peak}$. We define $E_{\nu,\rm peak}$ as the energy at which the neutrino flux for each case peaks. 
We calculate $\mathcal{F}^{\nu}$ by integrating the multi-zone muon neutrino $(\nu_\mu + \bar{\nu}_\mu)$ flux over the simulated neutrino energy range. 
The $n^{\rm exp}_{E_{\nu,\rm peak}}$ values are computed over three observational time intervals, $\Delta T=1,3,9.5$ years, respectively.

As discussed earlier, to generate the synthetic photon and neutrino emissions for our simulated jets, we have assumed the redshift and hence the luminosity distance for the jets to be identical to that of TXS 0506+056. 
Subsequently, considering the same sky position as TXS 0506+056 reasonably provides a basis to use the $A_{\rm eff}$ data provided in IceCube's 10 year data release for TXS 0506+056 \citep{icecube2018dataset} as representative effective areas for the $n^{\rm exp}_{E_{\nu, \rm peak}}$ calculations using the synthetic neutrino emissions from our jet simulations.
We have specifically used the IC86a (May 2011- May 2012), IC86b (May 2012 - May 2015),and IC86c (May 2015 - Oct 2017) $A_{\rm eff}$ data provided in the data release to get a range for $n^{\rm exp}_{E_\nu, \rm peak}$.

\begin{deluxetable*}{c cccc c cccc}
\tablecaption{Expected neutrino event counts for jets with $\Gamma_{\rm c,ini} = 5$ and 10, for $\Delta T = 1$, 3, 9.5 years. \label{tab_n_exp}}
\tablehead{
\colhead{$\Delta T$} & \multicolumn{4}{c}{$n^{\rm exp}_{E_{\nu,\rm peak}}$ (For $\Gamma_{\rm c,ini} = 5.0$)} && \multicolumn{4}{c}{$n^{\rm exp}_{E_{\nu,\rm peak}}$ (For $\Gamma_{\rm c,ini} = 10.0$)} \\
\cline{2-5} \cline{7-10}
\colhead{(years)} & \colhead{Case 1} & \colhead{Case 2} & \colhead{Case 3} & \colhead{Case 4} && \colhead{Case 1} & \colhead{Case 2} & \colhead{Case 3} & \colhead{Case 4}\\
 \colhead{} & \colhead{$(\times 10^{-3})$} & \colhead{$(\times 10^{-3})$} & \colhead{$(\times 10^{-5})$} & \colhead{} && \colhead{$(\times 10^{-2})$} & \colhead{$(\times 10^{-2})$} & \colhead{$(\times 10^{-3})$} & \colhead{}
}
\startdata
1   & 0.09–0.12 & 0.07–0.09 & 0.14–0.21 & 0.24–0.30 && 0.56–0.72 & 0.29–0.38 & 0.13–0.19 & 1.85–2.51 \\
3   & 0.28–0.36 & 0.20–0.26 & 0.41–0.63 & 0.72–0.90 && 1.68–2.15 & 0.89–1.15 & 0.38–0.57 & 5.54–7.52 \\
9.5 & 0.88–1.14 & 0.65–0.82 & 1.30–1.99 & 2.29–2.84 && 5.31–6.82 & 2.82–3.66 & 1.19–1.82 & 17.53–23.80 \\
\enddata
\tablecomments{Here $n^{\rm exp}_{E_{\nu,\rm peak}}$ denotes the expected number of neutrino events at the energy where the multi-zone neutrino flux peaks($E_{\nu,\rm peak}$), computed for four proton distribution cases. At the respective $E_{\nu,\rm peak}$ for each case, the IC86a, IC86b, and IC86c $A_{\rm eff}$ data from the TXS 0506+056 10-year IceCube data release \citep{icecube2018dataset} are used as representative effective areas to determine the range for each entry in the table.}
\end{deluxetable*}

Since for Case 1, we consider the proton distribution to have the same functional form (PL or PLEC) as the electron distribution in the zone, while for Case 2 the proton distribution is strictly a power-law with no exponential cutoff from cooling, we observe a relatively higher peak multi-zone neutrino flux in Case 2. Subsequently we also see a higher $n^{\rm exp}_{E_{\nu,\rm peak}}$ for Case 2 in both $\Gamma_{\rm c,ini} = 5$ and $10$ jets. On the other hand, Case 3 which again assumes a power-law with no cooling cutoff along with a higher particle $\gamma_{\rm max}$ specified by the Hillas criteria, despite having a higher peak neutrino flux than the first two cases, has a smaller $n^{\rm exp}_{E_{\nu,\rm peak}}$ as it gets scaled down by a higher $E_{\nu,peak}$. 

For both jets, the expected number of events $(n^{\rm exp}_{E_{\nu,\rm peak}})$ remains below one for Cases 1, 2, and 3 across the aforementioned time intervals $\Delta T$, implying that these fluxes may not be detectable by the current IceCube detector. However, a detector like IceCube Gen2 with a larger instrumented volume may offer enhanced sensitivity to such lower fluxes. 

Only Case 4, where we assume a very flat power-law distribution for protons with an index of $\alpha^{\rm p} = 2.0$ has a neutrino flux high enough to yield $n^{\rm exp}_{E_{\nu,\rm peak}}\geq 1$ for both jets within the observational time periods $\Delta T$ considered. Hence, suggesting that only this particular case produces a flux that may have a reasonable possibility of detection by the current IceCube detector. 

This is further illustrated in Figure~\ref{Neu_SED}, where we have depicted the multi-zone muon neutrino flux for all the proton distribution cases corresponding to both the simulated jets. For qualitative comparison, we overlay the IceCube observational estimates from TXS 0506+056 in the form of bow-tie plots representing the best-fit muon neutrino flux and associated $68\, \%$ confidence intervals for (a) the 2014–2015 flare period (from Fig. 3A in \cite{IceCube2018}), depicted by the blue-shaded bow-tie, and (b) the 9.5-year archival excess (from Fig. 4A in \cite{IceCube2018}), represented by the pink-shaded bow-tie. 

Case 4 for both jet configurations yields synthetic fluxes that are comparable to or exceed the archival excess, while Case 4 specifically for the $\Gamma_{\rm c,\mathrm{ini}} = 10$ jet reaches flux levels consistent with the 2014-2015 flare scenario. 
Neutrino flux for the other cases are a few orders of magnitude lower than the observational estimates for TXS 0506+056. 

In the synthetic SEDs that we derived for the jet simulations with no external photon field considered, 
we find a mismatch in the neutrino emission in the sub-PeV range as well as the neutrino spectral shape compared to the observations.
Although, the flux levels in the PeV range for the favoured Case 4 scenario ($\Gamma_{\rm c,ini} = 10$ jet) do lie within the $68\, \%$ confidence region of the observed neutrino flux for the TXS 0506+056 flare. 

In fact, according to literature studies \citep{Reimer_2019}, reproducing the observed spectral shape for TXS 0506+056, requires a photon field (co-moving or stationary) in X-ray frequencies such that the differential photon spectrum roughly scales as $dN^{\rm ph}/dE^{\rm ph} \propto (E^{\rm ph})^{-1}$ in the relevant frequency range.  
The relevant photon energy range to produce sub-PeV range neutrinos, for the $\delta^{\rm zone}_{\rm D}$ in our model, lies in soft X-rays.  
While we do have photons in the soft X-ray range, they scale as $dN^{\rm ph}/dE^{\rm ph} \propto (E^{ph})^{-2.8}$ instead of $\propto (E^{ph})^{-1}$ 
as found by \citet{Reimer_2019} in the relevant range. 
This could be the reason for the sub-PeV range mismatch in our neutrino spectrum compared to the observations of TXS 0506+056. 

As noted in Section~\ref{sec_EC}, even with the presence of external stationary photon fields in the AGN frame, one of which is the BLR field peaking around UV to very soft X-ray frequencies, due to the distance of the modeled jet section from the BLR this external field remains sub-dominant to the internal co-moving photon field. While the DT field does modify the IC emissions, we do not see much of an impact on the neutrino emissions. 
Hence the neutrino spectral shape doesn't change much in our synthetic SEDs even with the presence of external photon fields. 

This leads us to conclude that reproducing the spectral shape of the neutrino SED that corresponds to the TXS 0506+056 flare specially in the sub-PeV frequencies, likely requires an emission region placed much closer to the external photon fields considered in our simulations.
Additionally, a photon field at X-ray energies, as suggested by \citet{Reimer_2019, Winter_2019}, may be necessary to reproduce the exact spectral shape. However, we emphasize that our study explores the potential of kink-driven shocks within jets, in producing neutrino emissions comparable to a parsec-scale neutrino-emitting blazar source, instead of an attempt to replicate the exact emissions.

We see from the zonal electron distribution results, that the synergy between kink instability, internal shocks, and diffusive shock acceleration did not produce an electron distribution with a power-law index as flat as $\alpha^{\rm e} = 2.0$ in our simulations.
This indicates that the shocks formed in a RMHD scenario within our jets may not be strong enough to generate particularly flat power-law particle spectra. 
 
Moreover, as suggested by \citet{Dubey_2023, Dubey_2024}, even in the presence of weaker shocks, the cumulative number of shocks encountered by a particle may also play a significant role in shaping the slope of the particle distributions within the jet.
 
 Hence, Case 4 is indeed an extreme scenario but there is merit to such an assumption given lepto-hadronic simulation based studies modeling TXS 0506+056 like \cite{Gao_2019,Petropoulou_2020} have shown that a power-law with $\alpha^{\rm p} = 2.0$ for protons proves to be promising for reproducing the observed neutrino flux levels for the blazar source. 
 A similar hard spectra proton distribution with $\alpha^{\rm p} = 2.0$, has also been used to successfully recreate the observed neutrino emission signatures from NGC 1068\citep{Das_2024}, as well as to simulate neutrino emission from proto-magnetar jets \citep{Mukul2023}.

In lepto-hadronic models, the jet composition and the subsequent photon and neutrino emissions, depend significantly on the proton-to-electron number density ratio $\eta_{\rm zone}$. 
A larger $\eta_{\rm zone}$ reflects a greater proton content in the jet, naturally giving rise to an enhanced neutrino production. 

In all 4 cases for the multi-zone analysis of our jet simulations, $\eta_{\rm zone}$ is quite low, lying in the range of $\sim 0.004 - 0.010$ across different zones with the mean of the zonal electron number densities being $<n^{\rm e}_{\rm zone}> \sim 2.7$\,cm$^{-3}$ for the $\Gamma_{\rm c,ini} = 5$ jet. While $\eta_{\rm zone} \sim 0.0007 - 0.013$ and $<n^{\rm e}_{\rm zone}> \sim 3.6$\,cm$^{-3}$ for the $\Gamma_{\rm c,ini} = 10$ jet. The values for these parameters are estimated with the constraints specified in Section~\ref{Approx_MZ}. 
 
Since only the extreme case (Case 4) produces $n^{\rm exp}_{E_{\nu,\rm peak}} \geq 1$ for both jets, this suggests that for our  multi-zone neutrino fluxes to have a reasonable chance of detection at the IceCube detector, we either require a flat power-law distribution for protons like in Case 4 from our low $\eta_{\rm zone}$ simulations, or alternatively have a higher proton density within the jet. 
 
To confirm the above assertion, we did a test simulation with our multi-zone framework using $\eta_{\rm zone} = 1$ i.e. assuming equipartition between the proton and electron number densities across all zones. 
For this test run, we used the parameters derived for Case 2 from the RMHD simulation for the jet with $\Gamma_{\rm c,ini} = 10$, and performed the lepto-hadronic analysis.

 The neutrino flux generated by this test run is represented by the green dashed curve in Figure~\ref{Neu_SED}. 
 We find that increasing the proton-to-electron ratio $\eta_{\rm zone}$ in order to achieve an equipartition in proton and electron number densities is sufficient to produce a neutrino flux comparable to the case with the very flat power-law distribution for protons. The resultant neutrino flux from the test simulation is also comparable to the TXS 0506+056 flux during the 2014-2015 flare. 
 This proves that just shock acceleration that does not produce a very flat particle spectra in our jets, can still produce a significant neutrino flux, comparable to the TXS 0506+056 blazar, given the jet has a higher number density of protons.

\subsection{Origin of neutrinos: Intrinsic or extrinsic?}
There are both intrinsic and extrinsic pathways to achieve the aforementioned conditions favorable to substantial neutrino production. Reconnection within the jet as well as the extensive shocks generated by jet-environment interaction could be viable mechanisms to accelerate a significant amount of protons to high energies, possibly generating proton distributions like Case 4 even with small $\eta_{\rm zone}$ values. 
Jet–environment interactions can manifest in several scenarios, such as the interaction of the jet with broad-line region clouds \citep{Palacio}, 
jets interacting with disk winds or stellar winds \citep{Barkov2012}, or on certain occasions, via jet–jet interactions which \citet{Britzen:2019} hypothesize to be the possible origin for the 2014-2015 neutrino flare from TXS 0506+056.

Apart from producing significant dynamical consequences resulting in particle acceleration hence possibly generating a flat power-law proton 
distribution, such interactions can cause entrainment of proton-rich matter from the surroundings into the jet thereby enhancing the proton density 
within the jet. The proton distribution within each zone may then be defined as a combination of the intrinsic and entrained components:
\begin{align}
N^{\rm p}(\gamma) &= N^{\rm p}_{\rm int}(\gamma) + N^{\rm p}_{\rm ent}(\gamma) \nonumber \\
            &= \frac{\eta_{\rm zone} \rho^{\rm NT}_{\rm zone}}{(m_{\rm e} + \eta_{\rm zone} m_{\rm p})} f_{\rm int} (\gamma) + N^{\rm p}_{\rm ent} (\gamma)
\label{eq_int_ext}
\end{align}

where $N^{\rm p}_{\rm int}(\gamma)$ and $N^{\rm p}_{\rm ent}(\gamma)$ are the intrinsic and entrained proton populations with $f_{\rm int} (\gamma)$ being the distribution function for the intrinsic population. This additional component for entrained protons within the jet along with the associated dynamics during the jet-environment interactions can create limited periods of flaring activity in neutrino flux similar to what is observed from sources like TXS 0506+056. \\

To break the dichotomy between the different pathways for possible proton enrichment in the jet, we need to incorporate an actual population of protons into our simulations and study how the dynamical evolution of the jet impacts the proton distribution instead of taking phenomenological approximations for the same. Overcoming this limitation by building a module to incorporate proton macro-particles into our RMHD simulations will be a central focus of our upcoming work, as we move toward developing a more self-consistent and physically complete multi-zone framework for AGN jets and apply it to study different intrinsic and extrinsic scenarios.

%---------------------------------------------------------------------------------
\section{Conclusions} 
The particle composition of AGN jets remains an open question since long, posing a fundamental challenge to our understanding of 
the dynamics of these jets and their emission mechanisms. 
Most of the emission zone modeling for AGN jets is typically confined to a single-zone approximation, 
which, however, fails to capture how a complex jet morphology and dynamics will affect the resultant multi-messenger emission.

In this work, we have followed a new, more general approach.
We have first performed 3D RMHD simulations of parsec-scale AGN jet sections using the PLUTO code.
The jet dynamics is then coupled to a multi-zone framework that utilizes a lepto-hadronic code \citep{Katu2020} for modeling the interactions
and emissions from individual zones. 
Our framework produces synthetic spectral energy distributions of photons and neutrinos combining the contribution from many jet emission
zones. 
The multi-zone framework presented in this work is also tested and validated with established one-zone codes that are widely used in the 
community (see Appendix~\ref{Appendix}).

Our framework utilizes RMHD simulations to derive quantitatively the parameters that are required for the multi-zone modeling, 
thus circumventing the challenge posed by the large number of free parameters for such a model while improving upon certain simplified approximations for one-zone models. 
The key insights from our study can be summarized as follows.

\begin{itemize}
\item Our 3D RMHD simulations show that jet instabilities generate complex magnetic field configurations, naturally facilitating the formation of
different emission regions across the jet. 
The zonal analysis of the jet simulations reveal that during the onset of instabilities, varying particle distributions emerge across these zones, 
translating to significant differences in their individual zonal SEDs. Emissions from different zones contribute in varying extents across different frequency bands. This highlights that single zone approximations are insufficient to capture a realistic picture of the jet.

\item Our results clearly validate that proper modeling of AGN jets require a multi-zone approach - one that captures the jet dynamics capable of creating multiple emission regions across the jet, each of which can have a significant impact on the resultant multi-zone photon and neutrino SEDs. 

\item
We find that a single-zone model yields the best-fit parameters under the assumption that majority of the jet's emissions arise from a compact, homogeneous emission zone. 
However, when this assumption fails, which our simulation results and supporting studies suggest to be a likely possibility, this failure can lead to a highly biased interpretation of the underlying jet physics. 
We therefore conclude that a multi-zone framework is the way to go in this regard, as it can accommodate both a
single zone scenario 
where the contribution from only one of the many zones dominate, as well as scenarios where multiple zones significantly contribute to 
the overall emission.

\item There can be both intrinsic and extrinsic pathways for hadron enrichment in jets. 
The present study focuses on exploring the neutrino emissions produced through the intrinsic pathway.
The zonal proton-to-electron number density ratio in our lepto-hadronic analysis are constrained to consistently low values. 
Under such conditions, producing a possibly detectable neutrino flux for a detector like IceCube requires an underlying proton population which is accelerated to a flat power-law having a power-law index $\sim 2.0$.
However, we observe in our model that intrinsic shocks generated due to the onset of kink instabilities alone, are not strong enough to produce such flat particle power-law indices for the parameters of our simulations.

\item We suggest that other intrinsic mechanisms, such as magnetic reconnection within the jet, which can efficiently accelerate particles, may need to be considered to achieve the flat proton power-law spectra required to produce significant neutrino emissions. 
Further, shocks generated due to other intrinsic processes (apart from the kink instability) could aid in accelerating particles to high energies.
This needs to be investigated in further studies.

\item
 Alternatively, we find that increasing the value of the proton-to-electron number density ratio for the zones in the jet, even with just shock acceleration dominating, can significantly enhance the neutrino flux. It can generate flux levels comparable to the relatively flat proton power-law scenarios of our low proton-to-electron number density ratio jets. A probable pathway to achieve a higher proton number density within the jet is to account for jet-environment interactions, through which proton-rich matter from the surroundings may be entrained into the jet, a pathway we will investigate in our future work.

\item
 Including external Compton scattering from the BLR photon field in our simulation produces only minor changes in the electron distribution and in the resulting photon and neutrino SEDs. Comparatively, including the dusty torus photon field, yields $2-80$ times higher inverse Compton emissions across a range of frequencies, highlighting the dominance of the dusty torus photon field over the BLR field when emission zones are located at parsec-scale distances. 

 \item
 In terms of potential limitations of our modeling, we note that the proton spectra are phenomenologically assumed from the leptonic spectra that is evolved as described in Section~\ref{Hybrid_module}.
 Furthermore, the leptonic spectral evolution for each macro-particle co-moving with the jet flow is evolved without taking into account cooling due to SSC while
 the one-zone lepto-hadronic code used in the present study incorporates SSC process to obtain the multi-zone SED introducing a potential inconsistency. However, this discrepancy will only have a secondary impact on the scenario that is favorable for significant neutrino production.
 In future studies, effects due to SSC on the macro-particle spectra evolution would be incorporated with RMHD simulations.

\end{itemize}

The aim of this study was essentially to develop a tool to model and test out how different dynamical features in a jet can affect the multi-messenger emissions 
generated by the jet material. 
In a future approach, it would be essential to include the evolution of hadrons as well in order to be able to apply proton spectra that are self-consistently 
derived from the jet dynamics as an enhancement to our current multi-zone framework.
This would further allow us to disentangle and also to quantify the role of the number density of non-thermal protons as compared to non-thermal electrons in the jet. 
Such a model development, whereby leptons and hadrons are evolved along with the dynamics, will require substantial coding effort and will be taken up in a subsequent study.

\section*{Acknowledgments}
We thank an anonymous referee for their extremely valuable suggestions that have helped to improve our paper considerably.
 We thank Andrea Mignone and collaborators for developing and providing the PLUTO code \citep{Mignone_2007}, which was used for our RMHD simulations. 
We also acknowledge the authors of the \texttt{Katu} code \citep{Katu2020}, used for radiative and neutrino emission modeling. 
We also acknowledge the authors of the new PyPLUTO code \citep{Mattia2025} which has been extensively used for our work.
B.V. would like to acknowledge the support from the Max Planck Partner group Award established at IIT Indore. 
H.B. is funded by the DST INSPIRE Fellowship and acknowledges its support towards the Ph.D. H.B. also thanks Mukul Bhattacharya for the very helpful discussions.
C.F. and B.V. acknowledges the support of the Deutsche Forschungsgemeinschaft (DFG, German Research Foundation), project number 443220636, via the Research Unit FOR\,5195.
All the simulations have been performed using the MPCDF computing clusters \textit{Raven} and \textit{Vera} of the Max Planck Society
and the facilities provided at IIT Indore. 

\bibliographystyle{aasjournal}
\bibliography{refs}

@article{IceCube:2022der,
    author = "Abbasi, R. and others",
    collaboration = "IceCube",
    title = "{Evidence for neutrino emission from the nearby active galaxy NGC 1068}",
    eprint = "2211.09972",
    archivePrefix = "arXiv",
    primaryClass = "astro-ph.HE",
    doi = "10.1126/science.abg3395",
    journal = "Science",
    volume = "378",
    number = "6619",
    pages = "538--543",
    year = "2022"
}

@article{IceCube2018,
   title={Neutrino emission from the direction of the blazar TXS 0506+056 prior to the IceCube-170922A alert},
   volume={361},
   ISSN={1095-9203},
   url={http://dx.doi.org/10.1126/science.aat2890},
   DOI={10.1126/science.aat2890},
   number={6398},
   journal={Science},
   publisher={American Association for the Advancement of Science (AAAS)},
   author={Aartsen, Mark and others},
   year={2018},
   month=jul, pages={147–151} }

@article{Vaidya_2018,
	doi = {10.3847/1538-4357/aadd17},
  
	url = {https://doi.org/10.3847%2F1538-4357%2Faadd17},
  
	year = 2018,
	month = {oct},
  
	publisher = {American Astronomical Society},
  
	volume = {865},
  
	number = {2},
  
	pages = {144},
  
	author = {Bhargav Vaidya and Andrea Mignone and Gianluigi Bodo and Paola Rossi and Silvano Massaglia},
  
	title = {A Particle Module for the {PLUTO} Code. {II}. Hybrid Framework for Modeling Nonthermal Emission from Relativistic Magnetized Flows},
  
	journal = {The Astrophysical Journal}
}

@article{Abdo_2010,
	doi = {10.1088/0004-637x/716/1/30},
  
	url = {https://doi.org/10.1088%2F0004-637x%2F716%2F1%2F30},
  
	year = 2010,
	month = {may},
  
	publisher = {American Astronomical Society},
  
	volume = {716},
  
	number = {1},
  
	pages = {30--70},
  
	author = {A. A. Abdo and others},
  
	title = {{THE} {SPECTRAL} {ENERGY} {DISTRIBUTION} {OF} $\less$i$\greater${FERMI}$\less$/i$\greater${BRIGHT} {BLAZARS}},
  
	journal = {The Astrophysical Journal}
}

@article{Mignone_2007,
	doi = {10.1086/513316},
  
	url = {https://doi.org/10.1086%2F513316},
  
	year = 2007,
	month = {may},
  
	publisher = {American Astronomical Society},
  
	volume = {170},
  
	number = {1},
  
	pages = {228--242},
  
	author = {A. Mignone and G. Bodo and S. Massaglia and T. Matsakos and O. Tesileanu and C. Zanni and A. Ferrari},
  
	title = {{PLUTO}: A Numerical Code for Computational Astrophysics},
  
	journal = {The Astrophysical Journal Supplement Series}
}

@ARTICLE{Mastichiadis1995,
       author = {{Mastichiadis}, A. and {Kirk}, J.~G.},
        title = "{Self-consistent particle acceleration in active galactic nuclei.}",
      journal = {\aap},
         year = 1995,
        month = mar,
       volume = {295},
        pages = {613},
       adsurl = {https://ui.adsabs.harvard.edu/abs/1995A&A...295..613M}
}

@article{Petropoulou:2014lja,
    author = "Petropoulou, Maria and Giannios, Dimitrios and Dimitrakoudis, Stavros",
    title = "{Implications of a PeV neutrino spectral cutoff in GRB models}",
    eprint = "1405.2091",
    archivePrefix = "arXiv",
    primaryClass = "astro-ph.HE",
    doi = "10.1093/mnras/stu1757",
    journal = "Mon. Not. Roy. Astron. Soc.",
    volume = "445",
    number = "1",
    pages = "570--580",
    year = "2014"
}

@article{Bottcher_2013,
doi = {10.1088/0004-637X/768/1/54},
url = {https://dx.doi.org/10.1088/0004-637X/768/1/54},
year = {2013},
month = {apr},
publisher = {The American Astronomical Society},
volume = {768},
number = {1},
pages = {54},
author = {M. Böttcher and A. Reimer and K. Sweeney and A. Prakash},
title = {LEPTONIC AND HADRONIC MODELING OF FERMI-DETECTED BLAZARS},
journal = {The Astrophysical Journal}
}

@article{10.1093/mnras/stu2691,
    author = {Cerruti, M. and Zech, A. and Boisson, C. and Inoue, S.},
    title = "{A hadronic origin for ultra-high-frequency-peaked BL Lac objects}",
    journal = {Monthly Notices of the Royal Astronomical Society},
    volume = {448},
    number = {1},
    pages = {910-927},
    year = {2015},
    month = {02},
    issn = {0035-8711},
    doi = {10.1093/mnras/stu2691},
    url = {https://doi.org/10.1093/mnras/stu2691},
    eprint = {https://academic.oup.com/mnras/article-pdf/448/1/910/9379400/stu2691.pdf},
}

@article{Walker_2018,
	doi = {10.3847/1538-4357/aaafcc},
  
	url = {https://doi.org/10.3847%2F1538-4357%2Faaafcc},
  
	year = 2018,
	month = {mar},
  
	publisher = {American Astronomical Society},
  
	volume = {855},
  
	number = {2},
  
	pages = {128},
  
	author = {R. Craig Walker and Philip E. Hardee and Frederick B. Davies and Chun Ly and William Junor},
  
	title = {The Structure and Dynamics of the Subparsec Jet in M87 Based on 50 {VLBA} Observations over 17 Years at 43 {GHz}
},
  
	journal = {The Astrophysical Journal}
}

@article{Blandford_2019,
   author = "Blandford, Roger and Meier, David and Readhead, Anthony",
   title = "Relativistic Jets from Active Galactic Nuclei", 
   journal= "Annual Review of Astronomy and Astrophysics",
   year = "2019",
   volume = "57",
   number = "Volume 57, 2019",
   pages = "467-509",
   doi = "https://doi.org/10.1146/annurev-astro-081817-051948",
   url = "https://www.annualreviews.org/content/journals/10.1146/annurev-astro-081817-051948",
   publisher = "Annual Reviews",
   issn = "1545-4282",
   type = "Journal Article", 
  }

@article{Acharya_2023,
	author = {Acharya, Sriyasriti and Vaidya, Bhargav and Kalpa Dihingia, Indu and Agarwal, Sushmita and Shukla, Amit},
	title = {A numerical study on the role of instabilities on multi-wavelength emission signatures of blazar jets},
	DOI= "10.1051/0004-6361/202244256",
	url= "https://doi.org/10.1051/0004-6361/202244256",
	journal = {A\&A},
	year = 2023,
	volume = 671,
	pages = "A161",
}

@article{Bodo_2019,
   title={Linear stability analysis of magnetized relativistic rotating jets},
   volume={485},
   ISSN={1365-2966},
   url={http://dx.doi.org/10.1093/mnras/stz591},
   DOI={10.1093/mnras/stz591},
   number={2},
   journal={Monthly Notices of the Royal Astronomical Society},
   publisher={Oxford University Press (OUP)},
   author={Bodo, G and Mamatsashvili, G and Rossi, P and Mignone, A},
   year={2019},
   month=mar, pages={2909–2921} }

@article{Fiorillo_2024,
doi = {10.3847/2041-8213/ad192b},
url = {https://dx.doi.org/10.3847/2041-8213/ad192b},
year = {2024},
month = {jan},
publisher = {The American Astronomical Society},
volume = {961},
number = {1},
pages = {L14},
author = {Damiano F. G. Fiorillo and Maria Petropoulou and Luca Comisso and Enrico Peretti and Lorenzo Sironi},
title = {TeV Neutrinos and Hard X-Rays from Relativistic Reconnection in the Corona of NGC 1068},
journal = {The Astrophysical Journal Letters}
}

@article{Giannios,
    author = {Giannios, Dimitrios},
    title = "{UHECRs from magnetic reconnection in relativistic jets}",
    journal = {Monthly Notices of the Royal Astronomical Society: Letters},
    volume = {408},
    number = {1},
    pages = {L46-L50},
    year = {2010},
    month = {10},
    issn = {1745-3925},
    doi = {10.1111/j.1745-3933.2010.00925.x},
    url = {https://doi.org/10.1111/j.1745-3933.2010.00925.x},
    eprint = {https://academic.oup.com/mnrasl/article-pdf/408/1/L46/54672079/mnrasl\_408\_1\_l46.pdf},
}

@article{Mbarek_2021,
doi = {10.3847/1538-4357/ac1da8},
url = {https://dx.doi.org/10.3847/1538-4357/ac1da8},
year = {2021},
month = {nov},
publisher = {The American Astronomical Society},
volume = {921},
number = {1},
pages = {85},
author = {Rostom Mbarek and Damiano Caprioli},
title = {Espresso and Stochastic Acceleration of Ultra-high-energy Cosmic Rays in Relativistic Jets},
journal = {The Astrophysical Journal}
}

@article{Xue_2021,
   title={A Two-zone Blazar Radiation Model for “Orphan” Neutrino Flares},
   volume={906},
   ISSN={1538-4357},
   url={http://dx.doi.org/10.3847/1538-4357/abc886},
   DOI={10.3847/1538-4357/abc886},
   number={1},
   journal={The Astrophysical Journal},
   publisher={American Astronomical Society},
   author={Xue, Rui and Liu, Ruo-Yu and Wang, Ze-Rui and Ding, Nan and Wang, Xiang-Yu},
   year={2021},
   month=jan, pages={51} }

@article{Aguilar_Ruiz_2023,
   title={Evidence of a lepto-hadronic two-zone emission in flare states},
   volume={83},
   ISSN={1434-6052},
   url={http://dx.doi.org/10.1140/epjc/s10052-023-11523-w},
   DOI={10.1140/epjc/s10052-023-11523-w},
   number={4},
   journal={The European Physical Journal C},
   publisher={Springer Science and Business Media LLC},
   author={Aguilar-Ruiz, E. and Fraija, N. and Galván-Gámez, A.},
   year={2023},
   month=apr }

@article{Rossi2008,
author = {Rossi, P. and Mignone, A. and Bodo, G. and Massaglia, S. and Ferrari, Attilio},
year = {2008},
month = {06},
pages = {},
title = {Formation of dynamical structures in relativistic jets: The FRI case},
volume = {488},
journal = {Astronomy and Astrophysics},
doi = {10.1051/0004-6361:200809687}
}

@article{Katu2020,
   title={A Bayesian approach to modelling multimessenger emission from blazars using lepto-hadronic kinetic equations},
   volume={500},
   ISSN={1365-2966},
   url={http://dx.doi.org/10.1093/mnras/staa3163},
   DOI={10.1093/mnras/staa3163},
   number={3},
   journal={Monthly Notices of the Royal Astronomical Society},
   publisher={Oxford University Press (OUP)},
   author={Jiménez-Fernández, Bruno and van Eerten, Hendrik Jan},
   year={2020},
   month=nov, pages={3613–3630} }

@ARTICLE{Mckinney,
       author = {{Mignone}, A. and {McKinney}, Jonathan C.},
        title = "{Equation of state in relativistic magnetohydrodynamics: variable versus constant adiabatic index}",
      journal = {\mnras},
         year = 2007,
        month = jul,
       volume = {378},
       number = {3},
        pages = {1118-1130},
          doi = {10.1111/j.1365-2966.2007.11849.x},
archivePrefix = {arXiv},
       eprint = {0704.1679},
 primaryClass = {astro-ph},
       adsurl = {https://ui.adsabs.harvard.edu/abs/2007MNRAS.378.1118M}
}

@article{Kreter_2020,
   title={On the Detection Potential of Blazar Flares for Current Neutrino Telescopes},
   volume={902},
   ISSN={1538-4357},
   url={http://dx.doi.org/10.3847/1538-4357/abb5b1},
   DOI={10.3847/1538-4357/abb5b1},
   number={2},
   journal={The Astrophysical Journal},
   publisher={American Astronomical Society},
   author={Kreter, M. and Kadler, M. and Krauß, F. and Mannheim, K. and Buson, S. and Ojha, R. and Wilms, J. and Böttcher, M.},
   year={2020},
   month=oct, pages={133} }

@article{Britzen:2019,
    author = {Britzen, S. and Fendt, C. and B\"ottcher, M. and Zaja\v{c}ek, M. and Jaron, F. and Pashchenko, I. N. and Araudo, A. and Karas, V. and Kurtanidze, O.},
    title = "{A cosmic collider: Was the IceCube neutrino generated in a precessing jet-jet interaction in TXS 0506+056?}",
    doi = "10.1051/0004-6361/201935422",
    journal = "Astron. Astrophys.",
    volume = "630",
    pages = "A103",
    year = "2019",
    note = "[Erratum: Astron.Astrophys. 632, C3 (2019)]"
}

@ARTICLE{Barkov2012,
       author = {{Barkov}, M.~V. and {Aharonian}, F.~A. and {Bogovalov}, S.~V. and {Kelner}, S.~R. and {Khangulyan}, D.},
        title = "{Rapid TeV Variability in Blazars as a Result of Jet-Star Interaction}",
      journal = {\apj},
         year = 2012,
        month = apr,
       volume = {749},
       number = {2},
          eid = {119},
        pages = {119},
          doi = {10.1088/0004-637X/749/2/119},
archivePrefix = {arXiv},
       eprint = {1012.1787},
 primaryClass = {astro-ph.HE},
       adsurl = {https://ui.adsabs.harvard.edu/abs/2012ApJ...749..119B}
}

@article{ Palacio,
	author = {{del Palacio, S.} and {Bosch-Ramon, V.} and {Romero, G. E.}},
	title = {Gamma rays from jets interacting with BLR clouds in blazars},
	DOI= "10.1051/0004-6361/201834231",
	url= "https://doi.org/10.1051/0004-6361/201834231",
	journal = {A\&A},
	year = 2019,
	volume = 623,
	pages = "A101",
}

@article{Petropoulou_2020,
doi = {10.3847/1538-4357/ab76d0},
url = {https://dx.doi.org/10.3847/1538-4357/ab76d0},
year = {2020},
month = {mar},
publisher = {The American Astronomical Society},
volume = {891},
number = {2},
pages = {115},
author = {Petropoulou, Maria and Murase, Kohta and Santander, Marcos and Buson, Sara and Tohuvavohu, Aaron and Kawamuro, Taiki and Vasilopoulos, Georgios and Negoro, Hiroshi and Ueda, Yoshihiro and Siegel, Michael H. and Keivani, Azadeh and Kawai, Nobuyuki and Mastichiadis, Apostolos and Dimitrakoudis, Stavros},
title = {Multi-epoch Modeling of TXS 0506+056 and Implications for Long-term High-energy Neutrino Emission},
journal = {The Astrophysical Journal},

}

@ARTICLE{2016Kadler_nu,
       author = {{Kadler}, M. and {Krau{\ss}}, F. and {Mannheim}, K. and {Ojha}, R. and {M{\"u}ller}, C. and {Schulz}, R. and {Anton}, G. and {Baumgartner}, W. and {Beuchert}, T. and {Buson}, S. and {Carpenter}, B. and {Eberl}, T. and {Edwards}, P.~G. and {Eisenacher Glawion}, D. and {Els{\"a}sser}, D. and {Gehrels}, N. and {Gr{\"a}fe}, C. and {Gulyaev}, S. and {Hase}, H. and {Horiuchi}, S. and {James}, C.~W. and {Kappes}, A. and {Kappes}, A. and {Katz}, U. and {Kreikenbohm}, A. and {Kreter}, M. and {Kreykenbohm}, I. and {Langejahn}, M. and {Leiter}, K. and {Litzinger}, E. and {Longo}, F. and {Lovell}, J.~E.~J. and {McEnery}, J. and {Natusch}, T. and {Phillips}, C. and {Pl{\"o}tz}, C. and {Quick}, J. and {Ros}, E. and {Stecker}, F.~W. and {Steinbring}, T. and {Stevens}, J. and {Thompson}, D.~J. and {Tr{\"u}stedt}, J. and {Tzioumis}, A.~K. and {Weston}, S. and {Wilms}, J. and {Zensus}, J.~A.},
        title = "{Coincidence of a high-fluence blazar outburst with a PeV-energy neutrino event}",
      journal = {Nature Physics},
         year = 2016,
        month = aug,
       volume = {12},
       number = {8},
        pages = {807-814},
          doi = {10.1038/nphys3715},
archivePrefix = {arXiv},
       eprint = {1602.02012},
 primaryClass = {astro-ph.HE},
       adsurl = {https://ui.adsabs.harvard.edu/abs/2016NatPh..12..807K}
}

@article{Garrappa_2019,
doi = {10.3847/1538-4357/ab2ada},
url = {https://dx.doi.org/10.3847/1538-4357/ab2ada},
year = {2019},
month = {jul},
publisher = {The American Astronomical Society},
volume = {880},
number = {2},
pages = {103},
author = {Garrappa, S. and Buson, S. and Franckowiak, A. and Fermi-LAT collaboration and Shappee, B. J. and Beacom, J. F. and Dong, S. and Holoien, T. W.-S. and Kochanek, C. S. and Prieto, J. L. and Stanek, K. Z. and Thompson, T. A. and ASAS-SN collaboration and Aartsen, M. G. and Ackermann, M. and Adams, J. and Aguilar, J. A. and Ahlers, M. and Ahrens, M. and Alispach, C. and Andeen, K. and Anderson, T. and Ansseau, I. and Anton, G. and Argüelles, C. and Auffenberg, J. and Axani, S. and Backes, P. and Bagherpour, H. and Bai, X. and Barbano, A. and Barwick, S. W. and Baum, V. and Bay, R. and Beatty, J. J. and Becker, K.-H. and Tjus, J. Becker and BenZvi, S. and Berley, D. and Bernardini, E. and Besson, D. Z. and Binder, G. and Bindig, D. and Blaufuss, E. and Blot, S. and Bohm, C. and Börner, M. and Böser, S. and Botner, O. and Bourbeau, E. and Bourbeau, J. and Bradascio, F. and Braun, J. and Bretz, H.-P. and Bron, S. and Brostean-Kaiser, J. and Burgman, A. and Busse, R. S. and Carver, T. and Chen, C. and Cheung, E. and Chirkin, D. and Clark, K. and Classen, L. and Collin, G. H. and Conrad, J. M. and Coppin, P. and Correa, P. and Cowen, D. F. and Cross, R. and Dave, P. and de André, J. P. A. M. and Clercq, C. De and DeLaunay, J. J. and Dembinski, H. and Deoskar, K. and Ridder, S. De and Desiati, P. and Vries, K. D. de and Wasseige, G. de and With, M. de and DeYoung, T. and Diaz, A. and Díaz-Vélez, J. C. and Dujmovic, H. and Dunkman, M. and Dvorak, E. and Eberhardt, B. and Ehrhardt, T. and Eller, P. and Evenson, P. A. and Fahey, S. and Fazely, A. R. and Felde, J. and Filimonov, K. and Finley, C. and Franckowiak, A. and Friedman, E. and Fritz, A. and Gaisser, T. K. and Gallagher, J. and Ganster, E. and Garrappa, S. and Gerhardt, L. and Ghorbani, K. and Glauch, T. and Glüsenkamp, T. and Goldschmidt, A. and Gonzalez, J. G. and Grant, D. and Griffith, Z. and Günder, M. and Gündüz, M. and Haack, C. and Hallgren, A. and Halve, L. and Halzen, F. and Hanson, K. and Hebecker, D. and Heereman, D. and Helbing, K. and Hellauer, R. and Henningsen, F. and Hickford, S. and Hignight, J. and Hill, G. C. and Hoffman, K. D. and Hoffmann, R. and Hoinka, T. and Hokanson-Fasig, B. and Hoshina, K. and Huang, F. and Huber, M. and Hultqvist, K. and Hünnefeld, M. and Hussain, R. and In, S. and Iovine, N. and Ishihara, A. and Jacobi, E. and Japaridze, G. S. and Jeong, M. and Jero, K. and Jones, B. J. P. and Kang, W. and Kappes, A. and Kappesser, D. and Karg, T. and Karl, M. and Karle, A. and Katz, U. and Kauer, M. and Keivani, A. and Kelley, J. L. and Kheirandish, A. and Kim, J. and Kintscher, T. and Kiryluk, J. and Kittler, T. and Klein, S. R. and Koirala, R. and Kolanoski, H. and Köpke, L. and Kopper, C. and Kopper, S. and Koskinen, D. J. and Kowalski, M. and Krings, K. and Krückl, G. and Kulacz, N. and Kunwar, S. and Kurahashi, N. and Kyriacou, A. and Labare, M. and Lanfranchi, J. L. and Larson, M. J. and Lauber, F. and Lazar, J. P. and Leonard, K. and Leuermann, M. and Liu, Q. R. and Lohfink, E. and Mariscal, C. J. Lozano and Lu, L. and Lucarelli, F. and Lünemann, J. and Luszczak, W. and Madsen, J. and Maggi, G. and Mahn, K. B. M. and Makino, Y. and Mallot, K. and Mancina, S. and Mariş, I. C. and Maruyama, R. and Mase, K. and Maunu, R. and Meagher, K. and Medici, M. and Medina, A. and Meier, M. and Meighen-Berger, S. and Menne, T. and Merino, G. and Meures, T. and Miarecki, S. and Micallef, J. and Momenté, G. and Montaruli, T. and Moore, R. W. and Moulai, M. and Nagai, R. and Nahnhauer, R. and Nakarmi, P. and Naumann, U. and Neer, G. and Niederhausen, H. and Nowicki, S. C. and Nygren, D. R. and Pollmann, A. Obertacke and Olivas, A. and O’Murchadha, A. and O’Sullivan, E. and Palczewski, T. and Pandya, H. and Pankova, D. V. and Park, N. and Peiffer, P. and de los Heros, C. Pérez and Pieloth, D. and Pinat, E. and Pizzuto, A. and Plum, M. and Price, P. B. and Przybylski, G. T. and Raab, C. and Raissi, A. and Rameez, M. and Rauch, L. and Rawlins, K. and Rea, I. C. and Reimann, R. and Relethford, B. and Renzi, G. and Resconi, E. and Rhode, W. and Richman, M. and Robertson, S. and Rongen, M. and Rott, C. and Ruhe, T. and Ryckbosch, D. and Rysewyk, D. and Safa, I. and Herrera, S. E. Sanchez and Sandrock, A. and Sandroos, J. and Santander, M. and Sarkar, S. and Sarkar, S. and Satalecka, K. and Schaufel, M. and Schlunder, P. and Schmidt, T. and Schneider, A. and Schneider, J. and Schumacher, L. and Sclafani, S. and Seckel, D. and Seunarine, S. and Silva, M. and Snihur, R. and Soedingrekso, J. and Soldin, D. and Song, M. and Spiczak, G. M. and Spiering, C. and Stachurska, J. and Stamatikos, M. and Stanev, T. and Stasik, A. and Stein, R. and Stettner, J. and Steuer, A. and Stezelberger, T. and Stokstad, R. G. and Stößl, A. and Strotjohann, N. L. and Stuttard, T. and Sullivan, G. W. and Sutherland, M. and Taboada, I. and Tenholt, F. and Ter-Antonyan, S. and Terliuk, A. and Tilav, S. and Tomankova, L. and Tönnis, C. and Toscano, S. and Tosi, D. and Tselengidou, M. and Tung, C. F. and Turcati, A. and Turcotte, R. and Turley, C. F. and Ty, B. and Unger, E. and Elorrieta, M. A. Unland and Usner, M. and Vandenbroucke, J. and Driessche, W. Van and Eijk, D. van and Eijndhoven, N. van and Vanheule, S. and Santen, J. van and Vraeghe, M. and Walck, C. and Wallace, A. and Wallraff, M. and Wandkowsky, N. and Watson, T. B. and Weaver, C. and Weiss, M. J. and Weldert, J. and Wendt, C. and Werthebach, J. and Westerhoff, S. and Whelan, B. J. and Whitehorn, N. and Wiebe, K. and Wiebusch, C. H. and Wille, L. and Williams, D. R. and Wills, L. and Wolf, M. and Wood, J. and Wood, T. R. and Woschnagg, K. and Wrede, G. and Xu, D. L. and Xu, X. W. and Xu, Y. and Yanez, J. P. and Yodh, G. and Yoshida, S. and Yuan, T. and IceCube Collaboration},
title = {Investigation of Two Fermi-LAT Gamma-Ray Blazars Coincident with High-energy Neutrinos Detected by IceCube},
journal = {The Astrophysical Journal},
}

@article{Sahakyan_2022,
    author = {Sahakyan, N and Giommi, P and Padovani, P and Petropoulou, M and Bégué, D and Boccardi, B and Gasparyan, S},
    title = {A multimessenger study of the blazar PKS 0735+178: a new major neutrino source candidate},
    journal = {Monthly Notices of the Royal Astronomical Society},
    volume = {519},
    number = {1},
    pages = {1396-1408},
    year = {2022},
    month = {12},
    issn = {0035-8711},
    doi = {10.1093/mnras/stac3607},
    url = {https://doi.org/10.1093/mnras/stac3607},
    eprint = {https://academic.oup.com/mnras/article-pdf/519/1/1396/48411871/stac3607.pdf},
}

@article{Omeliukh:2024kgk,
    author = "Omeliukh, A. and others",
    title = "{Multi-epoch leptohadronic modeling of neutrino source candidate blazar PKS 0735+178}",
    eprint = "2409.04165",
    archivePrefix = "arXiv",
    primaryClass = "astro-ph.HE",
    doi = "10.1051/0004-6361/202452143",
    journal = "Astron. Astrophys.",
    volume = "695",
    pages = "A266",
    year = "2025"
}

@article{Lister_2009,
doi = {10.1088/0004-6256/138/6/1874},
url = {https://dx.doi.org/10.1088/0004-6256/138/6/1874},
year = {2009},
month = {nov},
publisher = {The American Astronomical Society},
volume = {138},
number = {6},
pages = {1874},
author = {Lister, M. L. and Cohen, M. H. and Homan, D. C. and Kadler, M. and Kellermann, K. I. and Kovalev, Y. Y. and Ros, E. and Savolainen, T. and Zensus, J. A.},
title = {MOJAVE: MONITORING OF JETS IN ACTIVE GALACTIC NUCLEI WITH VLBA EXPERIMENTS. VI. KINEMATICS ANALYSIS OF A COMPLETE SAMPLE OF BLAZAR JETS},
journal = {The Astronomical Journal},
}

@article{HOVATTA2019101541,
title = {Relativistic Jets of Blazars},
journal = {New Astronomy Reviews},
volume = {87},
pages = {101541},
year = {2019},
issn = {1387-6473},
doi = {https://doi.org/10.1016/j.newar.2020.101541},
url = {https://www.sciencedirect.com/science/article/pii/S138764732030018X},
author = {Talvikki Hovatta and Elina Lindfors},
}

@article{Li_2020,
doi = {10.3847/1538-4357/ab8f9f},
url = {https://dx.doi.org/10.3847/1538-4357/ab8f9f},
year = {2020},
month = {jun},
publisher = {The American Astronomical Society},
volume = {896},
number = {1},
pages = {63},
author = {Li, Xiaofeng and An, Tao and Mohan, Prashanth and Giroletti, Marcello},
title = {The Parsec-scale Jet of the Neutrino-emitting Blazar TXS 0506+056},
journal = {The Astrophysical Journal},
}

@article{ Ros_2020,
	author = {{Ros, E.} and {Kadler, M.} and {Perucho, M.} and {Boccardi, B.} and {Cao, H.-M.} and {Giroletti, M.} and {Krauß, F.} and {Ojha, R.}},
	title = {Apparent superluminal core expansion and limb brightening in the candidate neutrino blazar TXS 0506+056},
	DOI= "10.1051/0004-6361/201937206",
	url= "https://doi.org/10.1051/0004-6361/201937206",
	journal = {A\&A},
	year = 2020,
	volume = 633,
	pages = "L1",
}

@article{Striani_2016,
    author = {Striani, E. and Mignone, A. and Vaidya, B. and Bodo, G. and Ferrari, A.},
    title = {MHD simulations of three-dimensional resistive reconnection in a cylindrical plasma column},
    journal = {Monthly Notices of the Royal Astronomical Society},
    volume = {462},
    number = {3},
    pages = {2970-2979},
    year = {2016},
    month = {07},
    issn = {0035-8711},
    doi = {10.1093/mnras/stw1848},
    url = {https://doi.org/10.1093/mnras/stw1848},
    eprint = {https://academic.oup.com/mnras/article-pdf/462/3/2970/8012453/stw1848.pdf},
}

@article{Medina-Torrejon_2021,
doi = {10.3847/1538-4357/abd6c2},
url = {https://dx.doi.org/10.3847/1538-4357/abd6c2},
year = {2021},
month = {feb},
publisher = {The American Astronomical Society},
volume = {908},
number = {2},
pages = {193},
author = {Medina-Torrejón, Tania E. and de Gouveia Dal Pino, Elisabete M. and Kadowaki, Luis H. S. and Kowal, Grzegorz and Singh, Chandra B. and Mizuno, Yosuke},
title = {Particle Acceleration by Relativistic Magnetic Reconnection Driven by Kink Instability Turbulence in Poynting Flux–Dominated Jets},
journal = {The Astrophysical Journal},
}

@article{URRY1999159,
title = {Multiwavelength properties of blazars},
journal = {Astroparticle Physics},
volume = {11},
number = {1},
pages = {159-167},
year = {1999},
note = {TeV Astrophysics of Extragalactic Sources},
issn = {0927-6505},
doi = {https://doi.org/10.1016/S0927-6505(99)00043-2},
url = {https://www.sciencedirect.com/science/article/pii/S0927650599000432},
author = {C.M. Urry},
}

@article{DILTZ201463,
title = {Time dependent leptonic modeling of Fermi II processes in the jets of flat spectrum radio quasars},
journal = {Journal of High Energy Astrophysics},
volume = {1-2},
pages = {63-70},
year = {2014},
issn = {2214-4048},
doi = {https://doi.org/10.1016/j.jheap.2014.04.001},
url = {https://www.sciencedirect.com/science/article/pii/S2214404814000044},
author = {C. Diltz and M. Böttcher},
}

@article{Murase2014,
  title = {Diffuse neutrino intensity from the inner jets of active galactic nuclei: Impacts of external photon fields and the blazar sequence},
  author = {Murase, Kohta and Inoue, Yoshiyuki and Dermer, Charles D.},
  journal = {Phys. Rev. D},
  volume = {90},
  issue = {2},
  pages = {023007},
  numpages = {18},
  year = {2014},
  month = {Jul},
  publisher = {American Physical Society},
  doi = {10.1103/PhysRevD.90.023007},
  url = {https://link.aps.org/doi/10.1103/PhysRevD.90.023007}
}

@article{Petropoulou2015,
    author = {Petropoulou, M. and Dimitrakoudis, S. and Padovani, P. and Mastichiadis, A. and Resconi, E.},
    title = {Photohadronic origin of \$\\boldsymbol \{\\gamma \}\$-ray BL Lac emission: implications for IceCube neutrinos},
    journal = {Monthly Notices of the Royal Astronomical Society},
    volume = {448},
    number = {3},
    pages = {2412-2429},
    year = {2015},
    month = {03},
    issn = {0035-8711},
    doi = {10.1093/mnras/stv179},
    url = {https://doi.org/10.1093/mnras/stv179},
    eprint = {https://academic.oup.com/mnras/article-pdf/448/3/2412/6008935/stv179.pdf},
}

@article{IceCube_flavorOsc_2015,
  title = {Flavor Ratio of Astrophysical Neutrinos above 35 TeV in IceCube},
  author = {Aartsen, M. G. and Ackermann, M. and Adams, J. and Aguilar, J. A. and Ahlers, M. and Ahrens, M. and Altmann, D. and Anderson, T. and Arguelles, C. and Arlen, T. C. and Auffenberg, J. and Bai, X. and Barwick, S. W. and Baum, V. and Bay, R. and Beatty, J. J. and Becker Tjus, J. and Becker, K.-H. and BenZvi, S. and Berghaus, P. and Berley, D. and Bernardini, E. and Bernhard, A. and Besson, D. Z. and Binder, G. and Bindig, D. and Bissok, M. and Blaufuss, E. and Blumenthal, J. and Boersma, D. J. and Bohm, C. and Bos, F. and Bose, D. and B\"oser, S. and Botner, O. and Brayeur, L. and Bretz, H.-P. and Brown, A. M. and Buzinsky, N. and Casey, J. and Casier, M. and Cheung, E. and Chirkin, D. and Christov, A. and Christy, B. and Clark, K. and Classen, L. and Clevermann, F. and Coenders, S. and Cowen, D. F. and Cruz Silva, A. H. and Daughhetee, J. and Davis, J. C. and Day, M. and de Andr\'e, J. P. A. M. and De Clercq, C. and Dembinski, H. and De Ridder, S. and Desiati, P. and de Vries, K. D. and de With, M. and DeYoung, T. and D\'{\i}az-V\'elez, J. C. and Dumm, J. P. and Dunkman, M. and Eagan, R. and Eberhardt, B. and Ehrhardt, T. and Eichmann, B. and Eisch, J. and Euler, S. and Evenson, P. A. and Fadiran, O. and Fazely, A. R. and Fedynitch, A. and Feintzeig, J. and Felde, J. and Filimonov, K. and Finley, C. and Fischer-Wasels, T. and Flis, S. and Frantzen, K. and Fuchs, T. and Gaisser, T. K. and Gaior, R. and Gallagher, J. and Gerhardt, L. and Gier, D. and Gladstone, L. and Gl\"usenkamp, T. and Goldschmidt, A. and Golup, G. and Gonzalez, J. G. and Goodman, J. A. and G\'ora, D. and Grant, D. and Gretskov, P. and Groh, J. C. and Gro\ss{}, A. and Ha, C. and Haack, C. and Haj Ismail, A. and Hallen, P. and Hallgren, A. and Halzen, F. and Hanson, K. and Hebecker, D. and Heereman, D. and Heinen, D. and Helbing, K. and Hellauer, R. and Hellwig, D. and Hickford, S. and Hill, G. C. and Hoffman, K. D. and Hoffmann, R. and Homeier, A. and Hoshina, K. and Huang, F. and Huelsnitz, W. and Hulth, P. O. and Hultqvist, K. and Ishihara, A. and Jacobi, E. and Jacobsen, J. and Japaridze, G. S. and Jero, K. and Jurkovic, M. and Kaminsky, B. and Kappes, A. and Karg, T. and Karle, A. and Kauer, M. and Keivani, A. and Kelley, J. L. and Kheirandish, A. and Kiryluk, J. and Kl\"as, J. and Klein, S. R. and K\"ohne, J.-H. and Kohnen, G. and Kolanoski, H. and Koob, A. and K\"opke, L. and Kopper, C. and Kopper, S. and Koskinen, D. J. and Kowalski, M. and Kriesten, A. and Krings, K. and Kroll, G. and Kroll, M. and Kunnen, J. and Kurahashi, N. and Kuwabara, T. and Labare, M. and Lanfranchi, J. L. and Larsen, D. T. and Larson, M. J. and Lesiak-Bzdak, M. and Leuermann, M. and L\"unemann, J. and Madsen, J. and Maggi, G. and Maruyama, R. and Mase, K. and Matis, H. S. and Maunu, R. and McNally, F. and Meagher, K. and Medici, M. and Meli, A. and Meures, T. and Miarecki, S. and Middell, E. and Middlemas, E. and Milke, N. and Miller, J. and Mohrmann, L. and Montaruli, T. and Morse, R. and Nahnhauer, R. and Naumann, U. and Niederhausen, H. and Nowicki, S. C. and Nygren, D. R. and Obertacke, A. and Olivas, A. and Omairat, A. and O'Murchadha, A. and Palczewski, T. and Paul, L. and Penek, \"O. and Pepper, J. A. and P\'erez de los Heros, C. and Pfendner, C. and Pieloth, D. and Pinat, E. and Posselt, J. and Price, P. B. and Przybylski, G. T. and P\"utz, J. and Quinnan, M. and R\"adel, L. and Rameez, M. and Rawlins, K. and Redl, P. and Rees, I. and Reimann, R. and Relich, M. and Resconi, E. and Rhode, W. and Richman, M. and Riedel, B. and Robertson, S. and Rodrigues, J. P. and Rongen, M. and Rott, C. and Ruhe, T. and Ruzybayev, B. and Ryckbosch, D. and Saba, S. M. and Sander, H.-G. and Sandroos, J. and Santander, M. and Sarkar, S. and Schatto, K. and Scheriau, F. and Schmidt, T. and Schmitz, M. and Schoenen, S. and Sch\"oneberg, S. and Sch\"onwald, A. and Schukraft, A. and Schulte, L. and Schulz, O. and Seckel, D. and Sestayo, Y. and Seunarine, S. and Shanidze, R. and Smith, M. W. E. and Soldin, D. and Spiczak, G. M. and Spiering, C. and Stamatikos, M. and Stanev, T. and Stanisha, N. A. and Stasik, A. and Stezelberger, T. and Stokstad, R. G. and St\"o\ss{}l, A. and Strahler, E. A. and Str\"om, R. and Strotjohann, N. L. and Sullivan, G. W. and Taavola, H. and Taboada, I. and Tamburro, A. and Ter-Antonyan, S. and Terliuk, A. and Te\ifmmode \check{s}\else \v{s}\fi{}i\ifmmode \acute{c}\else \'{c}\fi{}, G. and Tilav, S. and Toale, P. A. and Tobin, M. N. and Tosi, D. and Tselengidou, M. and Unger, E. and Usner, M. and Vallecorsa, S. and van Eijndhoven, N. and Vandenbroucke, J. and van Santen, J. and Vanheule, S. and Vehring, M. and Voge, M. and Vraeghe, M. and Walck, C. and Wallraff, M. and Weaver, Ch. and Wellons, M. and Wendt, C. and Westerhoff, S. and Whelan, B. J. and Whitehorn, N. and Wichary, C. and Wiebe, K. and Wiebusch, C. H. and Williams, D. R. and Wissing, H. and Wolf, M. and Wood, T. R. and Woschnagg, K. and Xu, D. L. and Xu, X. W. and Xu, Y. and Yanez, J. P. and Yodh, G. and Yoshida, S. and Zarzhitsky, P. and Ziemann, J. and Zoll, M.},
  collaboration = {IceCube Collaboration},
  journal = {Phys. Rev. Lett.},
  volume = {114},
  issue = {17},
  pages = {171102},
  numpages = {8},
  year = {2015},
  month = {Apr},
  publisher = {American Physical Society},
  doi = {10.1103/PhysRevLett.114.171102},
  url = {https://link.aps.org/doi/10.1103/PhysRevLett.114.171102}
}

@article{Ball_2018,
doi = {10.3847/1538-4357/aac820},
url = {https://dx.doi.org/10.3847/1538-4357/aac820},
year = {2018},
month = {jul},
publisher = {The American Astronomical Society},
volume = {862},
number = {1},
pages = {80},
author = {Ball, David and Sironi, Lorenzo and Özel, Feryal},
title = {Electron and Proton Acceleration in Trans-relativistic Magnetic Reconnection: Dependence on Plasma Beta and Magnetization},
journal = {The Astrophysical Journal},
}

@article{Das_2024,
doi = {10.3847/1538-4357/ad5a04},
url = {https://dx.doi.org/10.3847/1538-4357/ad5a04},
year = {2024},
month = {aug},
publisher = {The American Astronomical Society},
volume = {972},
number = {1},
pages = {44},
author = {Das, Abhishek and Zhang, B. Theodore and Murase, Kohta},
title = {Revealing the Production Mechanism of High-energy Neutrinos from NGC 1068},
journal = {The Astrophysical Journal},
}

@article{Mukul2023,
    author = {Bhattacharya, Mukul and Carpio, Jose A and Murase, Kohta and Horiuchi, Shunsaku},
    title = {High-energy neutrino emission from magnetized jets of rapidly rotating protomagnetars},
    journal = {Monthly Notices of the Royal Astronomical Society},
    volume = {521},
    number = {2},
    pages = {2391-2407},
    year = {2023},
    month = {02},
    issn = {0035-8711},
    doi = {10.1093/mnras/stad494},
    url = {https://doi.org/10.1093/mnras/stad494},
    eprint = {https://academic.oup.com/mnras/article-pdf/521/2/2391/49552300/stad494.pdf},
}

@article{ Nigro2022,
	author = {{Nigro, C.} and {Sitarek, J.} and {Gliwny, P.} and {Sanchez, D.} and {Tramacere, A.} and {Craig, M.}},
	title = {agnpy: An open-source python package modelling the radiative processes of jetted active galactic nuclei},
	DOI= "10.1051/0004-6361/202142000",
	url= "https://doi.org/10.1051/0004-6361/202142000",
	journal = {A\&A},
	year = 2022,
	volume = 660,
	pages = "A18",
}

@BOOK{DermerMenon,
       author = {{Dermer}, Charles D. and {Menon}, Govind},
        title = "{High Energy Radiation from Black Holes: Gamma Rays, Cosmic Rays, and Neutrinos}",
         year = 2009,
       adsurl = {https://ui.adsabs.harvard.edu/abs/2009herb.book.....D}
}

@ARTICLE{Aharonian2010,
       author = {{Aharonian}, F.~A. and {Kelner}, S.~R. and {Prosekin}, A. Yu.},
        title = "{Angular, spectral, and time distributions of highest energy protons and associated secondary gamma rays and neutrinos propagating through extragalactic magnetic and radiation fields}",
      journal = {\prd},
         year = 2010,
        month = aug,
       volume = {82},
       number = {4},
          eid = {043002},
        pages = {043002},
          doi = {10.1103/PhysRevD.82.043002},
archivePrefix = {arXiv},
       eprint = {1006.1045},
 primaryClass = {astro-ph.HE},
       adsurl = {https://ui.adsabs.harvard.edu/abs/2010PhRvD..82d3002A}
}

@ARTICLE{Trama1,
       author = {{Tramacere}, A. and {Massaro}, E. and {Taylor}, A.~M.},
        title = "{Stochastic Acceleration and the Evolution of Spectral Distributions in Synchro-Self-Compton Sources: A Self-consistent Modeling of Blazars' Flares}",
      journal = {\apj},
         year = 2011,
        month = oct,
       volume = {739},
       number = {2},
          eid = {66},
        pages = {66},
          doi = {10.1088/0004-637X/739/2/66},
archivePrefix = {arXiv},
       eprint = {1107.1879},
 primaryClass = {astro-ph.HE},
       adsurl = {https://ui.adsabs.harvard.edu/abs/2011ApJ...739...66T}
}

@ARTICLE{Trama2,
       author = {{Tramacere}, A. and {Giommi}, P. and {Perri}, M. and {Verrecchia}, F. and {Tosti}, G.},
        title = "{Swift observations of the very intense flaring activity of Mrk 421 during 2006. I. Phenomenological picture of electron acceleration and predictions for MeV/GeV emission}",
      journal = {\aap},
         year = 2009,
        month = jul,
       volume = {501},
       number = {3},
        pages = {879-898},
          doi = {10.1051/0004-6361/200810865},
archivePrefix = {arXiv},
       eprint = {0901.4124},
 primaryClass = {astro-ph.HE},
       adsurl = {https://ui.adsabs.harvard.edu/abs/2009A&A...501..879T}
}

@ARTICLE{Trama3,
       author = {{Massaro}, E. and {Tramacere}, A. and {Perri}, M. and {Giommi}, P. and {Tosti}, G.},
        title = "{Log-parabolic spectra and particle acceleration in blazars. III. SSC emission in the TeV band from Mkn501}",
      journal = {\aap},
         year = 2006,
        month = mar,
       volume = {448},
       number = {3},
        pages = {861-871},
          doi = {10.1051/0004-6361:20053644},
archivePrefix = {arXiv},
       eprint = {astro-ph/0511673},
 primaryClass = {astro-ph},
       adsurl = {https://ui.adsabs.harvard.edu/abs/2006A&A...448..861M}
}

@ARTICLE{Dedner2002,
       author = {{Dedner}, A. and {Kemm}, F. and {Kr{\"o}ner}, D. and {Munz}, C. -D. and {Schnitzer}, T. and {Wesenberg}, M.},
        title = "{Hyperbolic Divergence Cleaning for the MHD Equations}",
      journal = {Journal of Computational Physics},
         year = 2002,
        month = jan,
       volume = {175},
       number = {2},
        pages = {645-673},
          doi = {10.1006/jcph.2001.6961},
       adsurl = {https://ui.adsabs.harvard.edu/abs/2002JCoPh.175..645D}
}

@article{Allakhverdyan2023,
    author = {Allakhverdyan, V A and Avrorin, A D and Avrorin, A V and Aynutdinov, V M and Bardačová, Z and Belolaptikov, I A and Bondarev, E A and Borina, I V and Budnev, N M and Chadymov, V A and Chepurnov, A S and Dik, V Y and Domogatsky, G V and Doroshenko, A A and Dvornický, R and Dyachok, A N and Dzhilkibaev, Zh-A M and Eckerová, E and Elzhov, T V and Fajt, L and Fomin, V N and Gafarov, A R and Golubkov, K V and Gorshkov, N S and Gress, T I and Kebkal, K G and Kharuk, I and Khramov, E V and Kolbin, M M and Koligaev, S O and Konischev, K V and Korobchenko, A V and Koshechkin, A P and Kozhin, V A and Kruglov, M V and Kulepov, V F and Lemeshev, Y E and Milenin, M B and Mirgazov, R R and Naumov, D V and Nikolaev, A S and Petukhov, D P and Pliskovsky, E N and Rozanov, M I and Ryabov, E V and Safronov, G B and Seitova, D and Shaybonov, B A and Shelepov, M D and Shilkin, S D and Shirokov, E V and Šimkovic, F and Sirenko, A E and Skurikhin, A V and Solovjev, A G and Sorokovikov, M N and Štekl, I and Stromakov, A P and Suvorova, O V and Tabolenko, V A and Ulzutuev, B B and Yablokova, Y V and Zaborov, D N and Zavyalov, S I and Zvezdov, D Y and (Baikal-GVD Collaboration) and Erkenov, A K and Kosogorov, N A and Kovalev, Yu A and Kovalev, Y Y and Plavin, A V and Popkov, A V and Pushkarev, A B and Semikoz, D V and Sotnikova, Y V and Troitsky, S V},
    title = {High-energy neutrino-induced cascade from the direction of the flaring radio blazar TXS 0506+056 observed by Baikal-GVD in 2021},
    journal = {Monthly Notices of the Royal Astronomical Society},
    volume = {527},
    number = {3},
    pages = {8784-8792},
    year = {2023},
    month = {11},
    issn = {0035-8711},
    doi = {10.1093/mnras/stad3653},
    url = {https://doi.org/10.1093/mnras/stad3653},
    eprint = {https://academic.oup.com/mnras/article-pdf/527/3/8784/54766239/stad3653.pdf},
}

@article{Dubey_2023,
doi = {10.3847/1538-4357/ace0bf},
url = {https://dx.doi.org/10.3847/1538-4357/ace0bf},
year = {2023},
month = {jul},
publisher = {The American Astronomical Society},
volume = {952},
number = {1},
pages = {1},
author = {Dubey, Ravi Pratap and Fendt, Christian and Vaidya, Bhargav},
title = {Particles in Relativistic MHD Jets. I. Role of Jet Dynamics in Particle Acceleration},
journal = {The Astrophysical Journal},
}

@article{Dubey_2024,
doi = {10.3847/1538-4357/ad8135},
url = {https://dx.doi.org/10.3847/1538-4357/ad8135},
year = {2024},
month = {nov},
publisher = {The American Astronomical Society},
volume = {976},
number = {1},
pages = {144},
author = {Dubey, Ravi Pratap and Fendt, Christian and Vaidya, Bhargav},
title = {Particles in Relativistic Magnetohydrodynamic Jets. II. Bridging Jet Dynamics with Multi–wave band Nonthermal Emission Signatures},
journal = {The Astrophysical Journal},
}

@misc{icecube2018dataset,
  doi          ={10.21234/B4QG92},
  url          ={https://doi.org/10.21234/B4QG92},
  author       = {{IceCube Collaboration}},
  year         = {2018},
  title        = {IceCube data from 2008 to 2017 related to analysis of TXS 0506+056},
  note         = {Dataset},
}

@article{Mizuno_2009,
doi = {10.1088/0004-637X/700/1/684},
url = {https://dx.doi.org/10.1088/0004-637X/700/1/684},
year = {2009},
month = {jul},
publisher = {The American Astronomical Society},
volume = {700},
number = {1},
pages = {684},
author = {Mizuno, Yosuke and Lyubarsky, Yuri and Nishikawa, Ken-Ichi and Hardee, Philip E.},
title = {THREE-DIMENSIONAL RELATIVISTIC MAGNETOHYDRODYNAMIC SIMULATIONS OF CURRENT-DRIVEN INSTABILITY. I. INSTABILITY OF A STATIC COLUMN},
journal = {The Astrophysical Journal},
}

@article{Mizuno_2011,
doi = {10.1088/0004-637X/734/1/19},
url = {https://dx.doi.org/10.1088/0004-637X/734/1/19},
year = {2011},
month = {may},
publisher = {The American Astronomical Society},
volume = {734},
number = {1},
pages = {19},
author = {Mizuno, Yosuke and Hardee, Philip E. and Nishikawa, Ken-Ichi},
title = {THREE-DIMENSIONAL RELATIVISTIC MAGNETOHYDRODYNAMIC SIMULATIONS OF CURRENT-DRIVEN INSTABILITY WITH A SUB-ALFVÉNIC JET: TEMPORAL PROPERTIES},
journal = {The Astrophysical Journal},
}

@article{Mizuno_2012,
doi = {10.1088/0004-637X/757/1/16},
url = {https://dx.doi.org/10.1088/0004-637X/757/1/16},
year = {2012},
month = {aug},
publisher = {The American Astronomical Society},
volume = {757},
number = {1},
pages = {16},
author = {Mizuno, Yosuke and Lyubarsky, Yuri and Nishikawa, Ken-Ichi and Hardee, Philip E.},
title = {THREE-DIMENSIONAL RELATIVISTIC MAGNETOHYDRODYNAMIC SIMULATIONS OF CURRENT-DRIVEN INSTABILITY. III. ROTATING RELATIVISTIC JETS},
journal = {The Astrophysical Journal},
}

@article{Acharya2021,
    author = {Acharya, Sriyasriti and Borse, Nikhil S and Vaidya, Bhargav},
    title = {Numerical analysis of long-term variability of AGN jets through RMHD simulations},
    journal = {Monthly Notices of the Royal Astronomical Society},
    volume = {506},
    number = {2},
    pages = {1862-1878},
    year = {2021},
    month = {06},
    issn = {0035-8711},
    doi = {10.1093/mnras/stab1775},
    url = {https://doi.org/10.1093/mnras/stab1775},
    eprint = {https://academic.oup.com/mnras/article-pdf/506/2/1862/39067883/stab1775.pdf},
}

@article{ONeill2012,
    author = {O’Neill, Sean M. and Beckwith, Kris and Begelman, Mitchell C.},
    title = {Local simulations of instabilities in relativistic jets – I. Morphology and energetics of the current-driven instability},
    journal = {Monthly Notices of the Royal Astronomical Society},
    volume = {422},
    number = {2},
    pages = {1436-1452},
    year = {2012},
    month = {04},
    issn = {0035-8711},
    doi = {10.1111/j.1365-2966.2012.20721.x},
    url = {https://doi.org/10.1111/j.1365-2966.2012.20721.x},
    eprint = {https://academic.oup.com/mnras/article-pdf/422/2/1436/3489233/mnras0422-1436.pdf},
}

@article{Tchekhovskoy2016,
    author = {Tchekhovskoy, Alexander and Bromberg, Omer},
    title = {Three-dimensional relativistic MHD simulations of active galactic nuclei jets: magnetic kink instability and Fanaroff–Riley dichotomy},
    journal = {Monthly Notices of the Royal Astronomical Society: Letters},
    volume = {461},
    number = {1},
    pages = {L46-L50},
    year = {2016},
    month = {04},
    issn = {1745-3925},
    doi = {10.1093/mnrasl/slw064},
    url = {https://doi.org/10.1093/mnrasl/slw064},
    eprint = {https://academic.oup.com/mnrasl/article-pdf/461/1/L46/56943188/mnrasl\_461\_1\_l46.pdf},
}

@article{Barniol2017,
    author = {Barniol Duran, Rodolfo and Tchekhovskoy, Alexander and Giannios, Dimitrios},
    title = {Simulations of AGN jets: magnetic kink instability versus conical shocks},
    journal = {Monthly Notices of the Royal Astronomical Society},
    volume = {469},
    number = {4},
    pages = {4957-4978},
    year = {2017},
    month = {05},
    issn = {0035-8711},
    doi = {10.1093/mnras/stx1165},
    url = {https://doi.org/10.1093/mnras/stx1165},
    eprint = {https://academic.oup.com/mnras/article-pdf/469/4/4957/17822952/stx1165.pdf},
}

@ARTICLE{Plavin2020,
       author = {{Plavin}, Alexander and {Kovalev}, Yuri Y. and {Kovalev}, Yuri A. and {Troitsky}, Sergey},
        title = "{Observational Evidence for the Origin of High-energy Neutrinos in Parsec-scale Nuclei of Radio-bright Active Galaxies}",
      journal = {\apj},
         year = 2020,
        month = may,
       volume = {894},
       number = {2},
          eid = {101},
        pages = {101},
          doi = {10.3847/1538-4357/ab86bd},
archivePrefix = {arXiv},
       eprint = {2001.00930},
 primaryClass = {astro-ph.HE},
       adsurl = {https://ui.adsabs.harvard.edu/abs/2020ApJ...894..101P}
}

@article{Plavin2021,
doi = {10.3847/1538-4357/abceb8},
url = {https://dx.doi.org/10.3847/1538-4357/abceb8},
year = {2021},
month = {feb},
publisher = {The American Astronomical Society},
volume = {908},
number = {2},
pages = {157},
author = {Plavin, A. V. and Kovalev, Y. Y. and Kovalev, Yu. A. and Troitsky, S. V.},
title = {Directional Association of TeV to PeV Astrophysical Neutrinos with Radio Blazars},
journal = {The Astrophysical Journal},
}

@article{Plavin2023,
    author = {Plavin, A V and Kovalev, Y Y and Kovalev, Yu A and Troitsky, S V},
    title = {Growing evidence for high-energy neutrinos originating in radio blazars},
    journal = {Monthly Notices of the Royal Astronomical Society},
    volume = {523},
    number = {2},
    pages = {1799-1808},
    year = {2023},
    month = {05},
    issn = {0035-8711},
    doi = {10.1093/mnras/stad1467},
    url = {https://doi.org/10.1093/mnras/stad1467},
    eprint = {https://academic.oup.com/mnras/article-pdf/523/2/1799/50503386/stad1467.pdf},
}

@article{Lister_2021,
doi = {10.3847/1538-4357/ac230f},
url = {https://dx.doi.org/10.3847/1538-4357/ac230f},
year = {2021},
month = {dec},
publisher = {The American Astronomical Society},
volume = {923},
number = {1},
pages = {30},
author = {Lister, M. L. and Homan, D. C. and Kellermann, K. I. and Kovalev, Y. Y. and Pushkarev, A. B. and Ros, E. and Savolainen, T.},
title = {Monitoring Of Jets in Active Galactic Nuclei with VLBA Experiments. XVIII. Kinematics and Inner Jet Evolution of Bright Radio-loud Active Galaxies},
journal = {The Astrophysical Journal},
}

@Article{Bottcher_2019,
AUTHOR = {B\"ottcher, Markus},
TITLE = {Progress in Multi-Wavelength and Multi-Messenger Observations of Blazars and Theoretical Challenges},
JOURNAL = {Galaxies},
VOLUME = {7},
YEAR = {2019},
NUMBER = {1},
ARTICLE-NUMBER = {20},
URL = {https://www.mdpi.com/2075-4434/7/1/20},
ISSN = {2075-4434},
DOI = {10.3390/galaxies7010020}
}

@article{Krauss2014,
  author = "Krauß, F. and Kadler, M. and Mannheim, K. and Schulz, R. and Trüstedt, J. and Wilms, J. and Ojha, R. and Ros, E. and Anton, G. and Baumgartner, W. and Beuchert, T. and Blanchard, J. and Bürkel, C. and Carpenter, B. and Eberl, T. and Edwards, P. G. and Eisenacher, D. and Elsässer, D. and Fehn, K. and Fritsch, U. and Gehrels, N. and Gräfe, C. and Großberger, C. and Hase, H. and Horiuchi, S. and James, C. and Kappes, A. and Katz, U. and Kreikenbohm, A. and Kreykenbohm, I. and Langejahn, M. and Leiter, K. and Litzinger, E. and Lovell, J. E. J. and Müller, C. and Phillips, C. and Plötz, C. and Quick, J. and Steinbring, T. and Stevens, J. and Thompson, D. J. and Tzioumis, A. K.",
  title = "TANAMI blazars in the IceCube PeV-neutrino fields",
  journal = "A\&A",
  volume = "566",
  pages = "L7",
  year = "2014",
  doi = "10.1051/0004-6361/201424219",
  url = "https://doi.org/10.1051/0004-6361/201424219"
}

@article{Gao_2017,
doi = {10.3847/1538-4357/aa7754},
url = {https://dx.doi.org/10.3847/1538-4357/aa7754},
year = {2017},
month = {jul},
publisher = {The American Astronomical Society},
volume = {843},
number = {2},
pages = {109},
author = {Gao, Shan and Pohl, Martin and Winter, Walter},
title = {On the Direct Correlation between Gamma-Rays and PeV Neutrinos from Blazars},
journal = {The Astrophysical Journal},
}

@article{ Del_zanna2003,
	author = {Del Zanna, L. and Bucciantini, N. and Londrillo, P.},
	title = {An efficient shock-capturing central-type scheme  
 for multidimensional relativistic flows - II. Magnetohydrodynamics},
	DOI= "10.1051/0004-6361:20021641",
	url= "https://doi.org/10.1051/0004-6361:20021641",
	journal = {A\&A},
	year = 2003,
	volume = 400,
	number = 2,
	pages = "397-413",
}

@article{MIGNONE2014784,
title = {High-order conservative reconstruction schemes for finite volume methods in cylindrical and spherical coordinates},
journal = {Journal of Computational Physics},
volume = {270},
pages = {784-814},
year = {2014},
issn = {0021-9991},
doi = {https://doi.org/10.1016/j.jcp.2014.04.001},
url = {https://www.sciencedirect.com/science/article/pii/S0021999114002538},
author = {A. Mignone},
}

@article{
IceCube_MMA2018,
author = {{The IceCube Collaboration} and Fermi-LAT and MAGIC and AGILE and ASAS-SN and HAWC and H.E.S.S. and INTEGRAL and Kanata and Kiso and Kapteyn and Liverpool Telescope and Subaru and Swift/NuSTAR and VERITAS and VLA/17B-403 teams and Mark Aartsen  and Markus Ackermann  and Jenni Adams  and Juan Antonio Aguilar  and Markus Ahlers  and Maryon Ahrens  and Imen Al Samarai  and David Altmann  and Karen Andeen  and Tyler Anderson  and Isabelle Ansseau  and Gisela Anton  and Carlos Argüelles  and Jan Auffenberg  and Spencer Axani  and Hadis Bagherpour  and Xinhua Bai  and Jared Barron  and Steve Barwick  and Volker Baum  and Ryan Bay  and James Beatty  and Karl Heinz Becker  and Julia Tjus  and Segev BenZvi  and David Berley  and Elisa Bernardini  and David Besson  and Gary Binder  and Daniel Bindig  and Erik Blaufuss  and Summer Blot  and Christian Bohm  and Mathis Boerner  and Fabian Bos  and Sebastian Boeser  and Olga Botner  and Etienne Bourbeau  and James Bourbeau  and Federica Bradascio  and Jim Braun  and Martin Brenzke  and Hans-Peter Bretz  and Stephanie Bron  and Jannes Brostean-Kaiser  and Alexander Burgman  and Raffaela Busse  and Tessa Carver  and Edward Cheng  and Dmitry Chirkin  and Asen Christov  and Ken Clark  and Lew Classen  and S. Coenders  and Gabriel Collin  and Janet Conrad  and Paul Coppin  and Pablo Correa  and Doug Cowen  and Robert Cross  and Pranav Dave  and Melanie Day  and Joao Pedro A M de Andre  and Catherine De Clercq  and James Delaunay  and Hans Dembinski  and Sam DeRidder  and Paolo Desiati  and Krijn de Vries  and Gwenhael DeWasseige  and Meike DeWith  and Ty DeYoung  and Juan Carlos Díaz-Vélez  and Vincenzo Di Lorenzo  and Hrvoje Dujmovic  and Jonathan Dumm  and Matt Dunkman  and Emily Dvorak  and Benjamin Eberhardt  and Thomas Ehrhardt  and Bjorn Eichmann  and Philipp Eller  and Paul Evenson  and Sam Fahey  and Ali Fazely  and John Felde  and Kirill Filimonov  and Chad Finley  and Samuel Flis  and Anna Franckowiak  and Elizabeth Friedman  and Alexander Fritz  and Tom Gaisser  and Jay Gallagher  and Lisa Gerhardt  and Kevin Ghorbani  and Theo Glauch  and Thorsten Gluesenkamp  and Azriel Goldschmidt  and Javier Gonzalez  and Darren Grant  and Zachary Griffith  and Christian Haack  and Allan Hallgren  and Francis Halzen  and Kael Hanson  and Dustin Hebecker  and David Heereman  and Klaus Helbing  and Robert Hellauer  and Stephanie Hickford  and Joshua Hignight  and Gary Hill  and Kara Hoffman  and Ruth Hoffmann  and Tobias Hoinka  and Benjamin Hokanson-Fasig  and Kotoyo Hoshina  and Feifei Huang  and Matthias Huber  and Klas Hultqvist  and Mirco Huennefeld  and Raamis Hussain  and Seongjin In  and Nadège Iovine  and Aya Ishihara  and Emanuel Jacobi  and George Japaridze  and Minjin Jeong  and Kyle Jero  and Benjamin Jones  and Piotr Kalaczynski  and Woosik Kang  and Alexander Kappes  and David Kappesser  and Timo Karg  and Albrecht Karle  and Uli Katz  and Matt Kauer  and Azadeh Keivani  and John Kelley  and Ali Kheirandish  and JongHyun Kim  and Myoungchul Kim  and Thomas Kintscher  and Joanna Kiryluk  and Thomas Kittler  and Spencer Klein  and Ramesh Koirala  and Hermann Kolanoski  and Lutz Koepke  and Claudio Kopper  and Sandro Kopper  and Jan Paul Koschinsky  and Jason Koskinen  and Marek Kowalski  and Kai Krings  and Mike Kroll  and Gerald Krueckl  and Samridha Kunwar  and Naoko Kurahashi Neilson  and Takao Kuwabara  and Alexander Kyriacou  and Mathieu Labare  and Justin Lanfranchi  and Michael Larson  and Frederik Lauber  and Kayla Leonard  and Mariola Lesiak-Bzdak  and Martin Leuermann  and Qinrui Liu  and Cristian Jesús Lozano Mariscal  and Lu Lu  and Jan Luenemann  and William Luszczak  and James Madsen  and Giuliano Maggi  and Kendall Mahn  and Sarah Mancina  and Reina Maruyama  and Keiichi Mase  and Ryan Maunu  and Kevin Meagher  and Morten Medici  and Maximilian Meier  and Thorben Menne  and Gonzalo Merino  and Thomas Meures  and Sandy Miarecki  and Jessie Micallef  and Giulio Momente  and Teresa Montaruli  and Roger Moore  and Robert Morse  and Marjon Moulai  and Rolf Nahnhauer  and Prabandha Nakarmi  and Uwe Naumann  and Garrett Neer  and Hans Niederhausen  and Sarah Nowicki  and Dave Nygren  and Anna Pollmann  and Alex Olivas  and Aongus Ó Murchadha  and Erin O'Sullivan  and Tomasz Palczewski  and Hershal Pandya  and Daria Pankova  and Peter Peiffer  and James Pepper  and Carlos de los Heros  and Damian Pieloth  and Elisa Pinat  and Matthias Plum  and Buford Price  and Gerald Przybylski  and Christoph Raab  and Leif Raedel  and Mohamed Rameez  and Ludwig Rauch  and Katherine Rawlins  and Immacolata Carmen Rea  and Rene Reimann  and Ben Relethford  and Matt Relich  and Elisa Resconi  and Wolfgang Rhode  and Mike Richman  and Sally Robertson  and Martin Rongen  and Carsten Rott  and Tim Ruhe  and Dirk Ryckbosch  and Devyn Rysewyk  and Ibrahim Safa  and Tobias Saelzer  and Sebastian Sanchez  and Alexander Sandrock  and Joakim Sandroos  and Marcos Santander  and Sourav Sarkar  and Subir Sarkar  and Konstancja Satalecka  and Philipp Schlunder  and Torsten Schmidt  and Austin Schneider  and Sebastian Schoenen  and Sebastian Schoneberg  and Lisa Schumacher  and Stephen Sclanfani  and Dave Seckel  and Suruj Seunarine  and Jan Soedingrekso  and Dennis Soldin  and Ming Song  and Glenn Spiczak  and Christian Spiering  and Juliana Stachurska  and Michael Stamatikos  and Todor Stanev  and Alexander Stasik  and Robert Stein  and Joeran Stettner  and Anna Steuer  and Thorsten Stezelberger  and Robert Stokstad  and Achim Stoessl  and Nora Linn Strotjohann  and Thomas Stuttard  and Greg Sullivan  and Michael Sutherland  and Ignacio Taboada  and Joulien Tatar  and Frederik Tenholt  and Samvel Ter-Antonyan  and Andrii Terliuk  and Serap Tilav  and Pat Toale  and Moriah Tobin  and Christoph Toennis  and Simona Toscano  and Delia Tosi  and Maria Tselengidou  and ChunFai Tung  and Andrea Turcati  and Colin Turley  and Bunheng Ty  and Lisa Unger  and Marcel Usner  and Ward Van Driessche  and Daan Van Eijk  and Nick van Eijndhoven  and Justin Vandenbroucke  and Sander Vanheule  and Jakob van Santen  and Eric Vogel  and Matthias Vraeghe  and Christian Walck  and Alexander Wallace  and Marius Wallraff  and Frank Wandler  and Nancy Wandkowsky  and Aatif Waza  and Chris Weaver  and Matthew Weiss  and Chris Wendt  and Johannes Werthebach  and Stefan Westerhoff  and Ben Whelan  and Nathan Whitehorn  and Klaus Wiebe  and Christopher Wiebusch  and Logan Wille  and Dawn Williams  and Lizz Wills  and Martin Wolf  and Joshua Wood  and Tania Wood  and Kurt Woschnagg  and Donglian Xu  and Xianwu Xu  and Yiqian Xu  and Juan Pablo Yanez  and Gaurang Yodh  and Shigeru Yoshida  and Tianlu Yuan  and Soheila Abdollahi  and Marco Ajello  and Roberto Angioni  and Luca Baldini  and Jean Ballet  and Guido Barbiellini  and Denis Bastieri  and Keith Bechtol  and Ronaldo Bellazzini  and Bijan Berenji  and Elisabetta Bissaldi  and Roger Blandford  and Raffaella Bonino  and Eugenio Bottacini  and Johan Bregeon  and Philippe Bruel  and Rolf Büehler  and Toby Burnett  and Eric Burns  and Sara Buson  and Rob Cameron  and Regina Caputo  and Patrizia A. Caraveo  and Elisabetta Cavazzuti  and Eric Charles  and Sina Chen  and Teddy Cheung  and James Chiang  and Graziano Chiaro  and Stefano Ciprini  and Johann Cohen-Tanugi  and Jan Conrad  and Denise Costantin  and Sara Cutini  and Filippo D'Ammando  and Francesco de Palma  and Seth Digel  and Niccolò Di Lalla  and Mattia Di Mauro  and Leonardo Di Venere  and Alberto Domínguez  and Cecilia Favuzzi  and Anna Franckowiak  and Yasushi Fukazawa  and Stefan Funk  and Piergiorgio Fusco  and Fabio Gargano  and Dario Gasparrini  and Nico Giglietto  and Matteo Giomi  and Paolo Giommi  and Francesco Giordano  and Marcello Giroletti  and Thomas Glanzman  and David Green  and Isabelle Grenier  and Marie-Hélène Grondin  and Sylvain Guiriec  and Alice Harding  and Masaaki Hayashida  and Liz Hays  and John Hewitt  and Deirdre Horan  and Guölaugur Jóhannesson  and Matthias Kadler  and Shiki Kensei  and Daniel Kocevski  and Felicia Krauss  and Michael Kreter  and Michael Kuss  and Giovanni La Mura  and Stefan Larsson  and Luca Latronico  and Marianne Lemoine-Goumard  and Jian Li  and Francesco Longo  and Francesco Loparco  and Michael Lovellette  and Pasquale Lubrano  and Jeffrey Magill  and Simone Maldera  and Dmitry Malyshev  and Alberto Manfreda  and Mario Nicola Mazziotta  and Julie McEnery  and Manuel Meyer  and Peter Michelson  and Tsunefumi Mizuno  and Maria Elena Monzani  and Aldo Morselli  and Igor Moskalenko  and Michela Negro  and Eric Nuss  and Roopesh Ojha  and Nicola Omodei  and Monica Orienti  and Elena Orlando  and Michele Palatiello  and Vaidehi Paliya  and Jeremy Perkins  and Massimo Persic  and Melissa Pesce-Rollins  and Frederic Piron  and Troy Porter  and Giacomo Principe  and Silvia Rainò  and Riccardo Rando  and Bindu Rani  and Massimiliano Razzano  and Soebur Razzaque  and Anita Reimer  and Olaf Reimer  and Nicolas Renault-Tinacci  and Steve Ritz  and Leon Rochester  and Pablo Saz Parkinson  and Carmelo Sgrò  and Eric J. Siskind  and Gloria Spandre  and Paolo Spinelli  and Dan Suson  and Hiro Tajima  and Mitsunari Takahashi  and Yasuyuki Tanaka  and Jana Thayer  and David J. Thompson  and Luigi Tibaldo  and Diego F. Torres  and Eleonora Torresi  and Gino Tosti  and Eleonora Troja  and Janeth Verónica Valverde  and Giacomo Vianello  and Matthew Vogel  and Kent Wood  and Matthew Wood  and Gabrijela Zaharijas  and Max Ludwig Ahnen  and Stefano Ansoldi  and Lucio Angelo Antonelli  and Cornelia Arcaro  and Dominik Baack  and Ana Babić  and Biswajit Banerjee  and Priyadarshini Bangale  and Ulisses Barres de Almeida  and Juan Abel Barrio  and Josefa Becerra González  and Wlodek Bednarek  and Elisa Bernardini  and Alessio Berti  and Wrijupan Bhattacharyya  and Adrian Biland  and Oscar Blanch  and Giacomo Bonnoli  and Roberto Carosi  and Alessandro Carosi  and Giovanni Ceribella  and Anshu Chatterjee  and Sidika Merve Colak  and Pierre Colin  and Eduardo Colombo  and Jose Luis Contreras  and Juan Cortina  and Stefano Covino  and Paolo Cumani  and Paolo Da Vela  and Francesco Dazzi  and Alessandro De Angelis  and Barbara De Lotto  and Manuel Delfino  and Jordi Delgado  and Federico Di Pierro  and Alberto Domínguez  and Dijana Dominis Prester  and Daniela Dorner  and Michele Doro  and Sabrina Einecke  and Dominik Elsaesser  and Vandad Fallah Ramazani  and Alba Fernández-Barral  and David Fidalgo  and Luca Foffano  and Konstantin Pfrang  and Maria Victoria Fonseca  and Lluis Font  and Christian Fruck  and Daniel Galindo  and Stefano Gallozzi  and Ramon J. García López  and Markus Garczarczyk  and Markus Gaug  and Paola Giammaria  and Nikola Godinović  and Dariusz Gora  and Daniel Guberman  and Daniela Hadasch  and Alexander Hahn  and Tarek Hassan  and Masaaki Hayashida  and Javier Herrera  and Juergen Hose  and Dario Hrupec  and Susumu Inoue  and Kazuma Ishio  and Yusuke Konno  and Hidetoshi Kubo  and Junko Kushida  and Damira Lelas  and Elina Lindfors  and Saverio Lombardi  and Francesco Longo  and Marcos López  and Camilla Maggio  and Pratik Majumdar  and Martin Makariev  and Galina Maneva  and Marina Manganaro  and Karl Mannheim  and Laura Maraschi  and Mosé Mariotti  and Manel Martínez  and Shu Masuda  and Daniel Mazin  and Milen Minev  and Jose Miguel Miranda  and Razmik Mirzoyan  and Abelardo Moralejo  and Victoria Moreno  and Elena Moretti  and Tsutomu Nagayoshi  and Vitaly Neustroev  and Andrzej Niedzwiecki  and Mireia Nievas Rosillo  and Cosimo Nigro  and Kari Nilsson  and Daniele Ninci  and Kyoshi Nishijima  and Koji Noda  and Leyre Nogués  and Simona Paiano  and Joaquim Palacio  and David Paneque  and Riccardo Paoletti  and Josep M. Paredes  and Giovanna Pedaletti  and Michele Peresano  and Massimo Persic  and Pier Giorgio Prada Moroni  and Elisa Prandini  and Ivica Puljak  and Jezabel Rodriguez  and Ignasi Reichardt  and Wolfgang Rhode  and Marc Ribó  and Javier Rico  and Chiara Righi  and Andrea Rugliancich  and Takayuki Saito  and Konstancja Satalecka  and Thomas Schweizer  and Julian Sitarek  and Iva Šnidarić  and Dorota Sobczynska  and Antonio Stamerra  and Marcel Strzys  and Tihomir Surić  and Mitsunari Takahashi  and Fabrizio Tavecchio  and Petar Temnikov  and Tomislav Terzić  and Masahiro Teshima  and Nuria Torres-Albà  and Aldo Treves  and Shimpei Tsujimoto  and Gaia Vanzo  and Monica Vazquez Acosta  and Ievgen Vovk  and John E. Ward  and Martin Will  and Darko Zarić  and Alberto Franceschini  and Fabrizio Lucarelli  and Marco Tavani  and Giovanni Piano  and Imma Donnarumma  and Carlotta Pittori  and Francesco Verrecchia  and Guido Barbiellini  and Andrea Bulgarelli  and Patrizia Caraveo  and Paolo Walter Cattaneo  and Sergio Colafrancesco  and Enrico Costa  and Guido Di Cocco  and Attilio Ferrari  and Fulvio Gianotti  and Andrea Giuliani  and Paolo Lipari  and Sandro Mereghetti  and Aldo Morselli  and Luigi Pacciani  and Franco Paoletti  and Nicolò Parmiggiani  and Alberto Pellizzoni  and Piergiorgio Picozza  and Maura Pilia  and Andrea Rappoldi  and Alessio Trois  and Stefano Vercellone  and Valerio Vittorini  and Andrea Albert  and Ruben Alfaro  and César Álvarez  and Roberto Arceo  and Juan Carlos Arteaga Velázquez  and Daniel Omar Avila Rojas  and Hugo Alberto Ayala Solares  and Ana Delia Becerril  and Ernesto Belmont-Moreno  and Abel Bernal  and Karen S. Caballero Mora  and Tomás Capistrán Rojas  and Alberto Carramiñana  and Sabrina Casanova  and Mario Alberto Castillo Maldonado  and Umberto Cotti  and Jorge Cotzomi  and Sara Coutiño de León  and Cederik León De León Acuña  and Eduardo De la Fuente  and Raquel Diaz Hernandez  and Simone Dichiara  and Brenda Dingus  and Michael DuVernois  and Juan Carlos Díaz Velez  and Robert Ellsworth  and Kristi Engel  and Daniel W. Fiorino  and Henrike Fleischhack  and Nissim Illich Fraija  and José Andrés García González  and Fernando Garfias  and Maria Magdalena González  and Adiv González Muñoz  and Jordan A. Goodman  and Zigfried Hampel-Arias  and J. Patrick Harding  and Sergio Hernandez Cadena  and Binita Hona  and Filiberto Hueyotl-Zahuantitla  and Michelle Hui  and Petra Hüntemeyer  and Arturo Iriarte  and Armelle Jardin-Blicq  and Vikas Joshi  and Sarah Kaufmann  and Gerd J. Kunde  and Alejandro Lara  and Robert Lauer  and William Lee  and Dirk Lennarz  and Hermes León Vargas  and Jim Linnemann  and Anna Lia Longinotti  and Gilgamesh Luis-Raya  and Rene Luna-García  and Kelly Malone  and Samuel Stephens Marinelli  and Oscar Martinez  and Israel Martinez Castellanos  and Humberto Martínez Huerta  and Jesús Martínez Castro  and John Matthews  and Pedro Miranda-Romagnoli  and Eduardo Moreno Barbosa  and Miguel Mostafa  and Amid Nayerhoda  and Lukas Nellen  and Michael Newbold  and Mehr Un Nisa  and Roberto Noriega-Papaqui  and Rodrigo Pelayo  and John Pretz  and Eucario Gonzalo Pérez Pérez  and Zhixiang Ren  and Chang Dong Rho  and Colas Rivière  and Daniel Rosa González  and Matthew Rosenberg  and Edna Ruiz-Velasco  and Enrique Ruiz-Velasco  and Francisco Salesa Greus  and Andres Sandoval  and Michael Schneider  and Harm Schoorlemmer  and Gus Sinnis  and Andrew James Smith  and Wayne Springer  and Pooja Surajbali  and Omar Tibolla  and Kirsten Tollefson  and Ibrahim Torres  and Luis Villaseñor  and Thomas Weisgarber  and Felix Werner  and Tolga Yapici  and Gaurang Yodh  and Arnulfo Zepeda  and Hao Zhou  and Juan de Dios Álvarez Romero  and H. Abdalla  and E. O. Angüner  and C. Armand  and M. Backes  and Y. Becherini  and D. Berge  and M. Böttcher  and C. Boisson  and J. Bolmont  and S. Bonnefoy  and P. Bordas  and F. Brun  and M. Büchele  and T. Bulik  and S. Caroff  and A. Carosi  and S. Casanova  and M. Cerruti  and N. Chakraborty  and S. Chandra  and A. Chen  and S. Colafrancesco  and I. D. Davids  and C. Deil  and J. Devin  and A. Djannati-Ataï  and K. Egberts  and G. Emery  and S. Eschbach  and A. Fiasson  and G. Fontaine  and S. Funk  and M. Füßling  and Y. A. Gallant  and F. Gaté  and G. Giavitto  and D. Glawion  and J. F. Glicenstein  and D. Gottschall  and M.-H. Grondin  and M. Haupt  and G. Henri  and J. A. Hinton  and C. Hoischen  and T. L. Holch  and D. Huber  and M. Jamrozy  and D. Jankowsky  and F. Jankowsky  and L. Jouvin  and I. Jung-Richardt  and D. Kerszberg  and B. Khélifi  and J. King  and S. Klepser  and W. Kluźniak  and Nu. Komin  and M. Kraus  and J. Lefaucheur  and A. Lemière  and M. Lemoine-Goumard  and J.-P. Lenain  and E. Leser  and T. Lohse  and R. López-Coto  and M. Lorentz  and I. Lypova  and V. Marandon  and G. Guillem Martí-Devesa  and G. Maurin  and A.M.W. Mitchell  and R. Moderski  and M. Mohamed  and L. Mohrmann  and E. Moulin  and T. Murach  and M. de Naurois  and F. Niederwanger  and J. Niemiec  and L. Oakes  and P. O'Brien  and S. Ohm  and M. Ostrowski  and I. Oya  and M. Panter  and R. D. Parsons  and C. Perennes  and Q. Piel  and S. Pita  and V. Poireau  and A. Priyana Noel  and H. Prokoph  and G. Pühlhofer  and A. Quirrenbach  and S. Raab  and R. Rauth  and M. Renaud  and F. Rieger  and L. Rinchiuso  and C. Romoli  and G. Rowell  and B. Rudak  and D. A. Sanchez  and M. Sasaki  and R. Schlickeiser  and F. Schüssler  and A. Schulz  and U. Schwanke  and M. Seglar-Arroyo  and N. Shafi  and R. Simoni  and H. Sol  and C. Stegmann  and C. Steppa  and T. Tavernier  and A. M. Taylor  and D. Tiziani  and C. Trichard  and M. Tsirou  and C. van Eldik  and C. van Rensburg  and B. van Soelen  and J. Veh  and P. Vincent  and F. Voisin  and S. J. Wagner  and R. M. Wagner  and A. Wierzcholska  and R. Zanin  and A. A. Zdziarski  and A. Zech  and A. Ziegler  and J. Zorn  and N. Zywucka  and V. Savchenko  and C. Ferrigno  and A. Bazzano  and R. Diehl  and E. Kuulkers  and P. Laurent  and S. Mereghetti  and L. Natalucci  and F. Panessa  and J. Rodi  and P. Ubertini  and Tomoki Morokuma  and Kouji Ohta  and Yasuyuki T. Tanaka  and Hiroki Mori  and Masayuki Yamanaka  and Koji S. Kawabata  and Yousuke Utsumi  and Tatsuya Nakaoka  and Miho Kawabata  and Hiroki Nagashima  and Michitoshi Yoshida  and Yoshiki Matsuoka  and Ryosuke Itoh  and William Keel  and Christopher Copperwheat  and Iain Steele  and S. Bradley Cenko  and Philip Evans  and Derek Fox  and Jamie Kennea  and Francis Marshall  and Julian Osborne  and Aaron Tohuvavohu  and Colin Turley  and Douglas Cowen  and James DeLaunay  and Azadeh Keivani  and Marcos Santander  and Anushka Abeysekara  and Avery Archer  and Wystan Benbow  and Ralph Bird  and Aryeh Brill  and Robert Brose  and Matthew Buchovecky  and James Buckley  and Viatcheslav Bugaev  and Jodi Christiansen  and Michael Connolly  and Wei Cui  and Michael Daniel  and Manel Errando  and Abraham Falcone  and Qi Feng  and John Finley  and Lucy Fortson  and Amy Furniss  and Orel Gueta  and Moritz Hütten  and Olivier Hervet  and Gareth Hughes  and Thomas Humensky  and Caitlin Johnson  and Philip Kaaret  and Payel Kar  and Nathan Kelley-Hoskins  and Mary Kertzman  and David Kieda  and Maria Krause  and Frank Krennrich  and Sajan Kumar  and Mark Lang  and Tony Lin  and Gernot Maier  and Steven McArthur  and Patrick Moriarty  and Reshmi Mukherjee  and Daniel Nieto  and Stephen O'Brien  and Rene Ong  and Adam Otte  and Nahee Park  and Andriy Petrashyk  and Martin Pohl  and Alexis Popkow  and Elisa Pueschel  and John Quinn  and Kenneth Ragan  and Paul Reynolds  and Gregory Richards  and Emmet Roache  and Cameron Rulten  and Iftach Sadeh  and Marcos Santander  and Skyler Scott  and Glenn Sembroski  and Karlen Shahinyan  and Iurii Sushch  and Samuel Trépanier  and Jonathan Tyler  and Vladimir Vassiliev  and Scott Wakely  and Amanda Weinstein  and Rita Wells  and Patrick Wilcox  and Alina Wilhelm  and David Williams  and Benjamin Zitzer  and Alexandra Tetarenko  and Amy Kimball  and James Miller-Jones  and Gregory Sivakoff},
title = {Multimessenger observations of a flaring blazar coincident with high-energy neutrino IceCube-170922A},
journal = {Science},
volume = {361},
number = {6398},
pages = {eaat1378},
year = {2018},
doi = {10.1126/science.aat1378},
URL = {https://www.science.org/doi/abs/10.1126/science.aat1378},
eprint = {https://www.science.org/doi/pdf/10.1126/science.aat1378},
}

@article{Bottcher:2019ole,
    author = {B\"ottcher, Markus and Baring, Matthew G.},
    title = "{Spectral Variability Signatures of Relativistic Shocks in Blazars}",
    eprint = "1903.12381",
    archivePrefix = "arXiv",
    primaryClass = "astro-ph.HE",
    journal = "PoS",
    volume = "HEASA2018",
    pages = "031",
    year = "2019"
}

@ARTICLE{Dermer1993,
       author = {{Dermer}, Charles D. and {Schlickeiser}, Reinhard},
        title = "{Model for the High-Energy Emission from Blazars}",
      journal = {\apj},
         year = 1993,
        month = oct,
       volume = {416},
        pages = {458},
          doi = {10.1086/173251},
       adsurl = {https://ui.adsabs.harvard.edu/abs/1993ApJ...416..458D}
}

@ARTICLE{Maraschi1992,
       author = {{Maraschi}, L. and {Ghisellini}, G. and {Celotti}, A.},
        title = "{A Jet Model for the Gamma-Ray--emitting Blazar 3C 279}",
      journal = {\apjl},
         year = 1992,
        month = sep,
       volume = {397},
        pages = {L5},
          doi = {10.1086/186531},
       adsurl = {https://ui.adsabs.harvard.edu/abs/1992ApJ...397L...5M}
}

@article{Blazejowski_2000,
doi = {10.1086/317791},
url = {https://dx.doi.org/10.1086/317791},
year = {2000},
month = {dec},
publisher = {},
volume = {545},
number = {1},
pages = {107},
author = {{B{\l}a{\.z}ejowski}, M. and Sikora, M. and Moderski, R. and Madejski, G. M.},
title = {Comptonization of Infrared Radiation from Hot Dust by Relativistic
Jets in
Quasars},
journal = {The Astrophysical Journal},
}

@ARTICLE{Bloom1996,
       author = {{Bloom}, Steven D. and {Marscher}, Alan P.},
        title = "{An Analysis of the Synchrotron Self-Compton Model for the Multi--Wave Band Spectra of Blazars}",
      journal = {\apj},
         year = 1996,
        month = apr,
       volume = {461},
        pages = {657},
          doi = {10.1086/177092},
       adsurl = {https://ui.adsabs.harvard.edu/abs/1996ApJ...461..657B}
}

@article{AHARONIAN2000,
title = {TeV gamma rays from BL Lac objects due to synchrotron radiation of extremely high energy protons},
journal = {New Astronomy},
volume = {5},
number = {7},
pages = {377-395},
year = {2000},
issn = {1384-1076},
doi = {https://doi.org/10.1016/S1384-1076(00)00039-7},
url = {https://www.sciencedirect.com/science/article/pii/S1384107600000397},
author = {F.A. Aharonian},
}

@ARTICLE{Mastichiadis_Kirk1995,
       author = {{Mastichiadis}, A. and {Kirk}, J.~G.},
        title = "{Self-consistent particle acceleration in active galactic nuclei.}",
      journal = {\aap},
         year = 1995,
        month = mar,
       volume = {295},
        pages = {613},
       adsurl = {https://ui.adsabs.harvard.edu/abs/1995A&A...295..613M}
}

@ARTICLE{Mucke2003,
       author = {{M{\"u}cke}, A. and {Protheroe}, R.~J. and {Engel}, R. and {Rachen}, J.~P. and {Stanev}, T.},
        title = "{BL Lac objects in the synchrotron proton blazar model}",
      journal = {Astroparticle Physics},
         year = 2003,
        month = mar,
       volume = {18},
       number = {6},
        pages = {593-613},
          doi = {10.1016/S0927-6505(02)00185-8},
archivePrefix = {arXiv},
       eprint = {astro-ph/0206164},
 primaryClass = {astro-ph},
       adsurl = {https://ui.adsabs.harvard.edu/abs/2003APh....18..593M}
}

@article{Boettcher1997,
       author = {{Boettcher}, M. and {Schlickeiser}, R.},
        title = "{The pair production spectrum from photon-photon annihilation.}",
      journal = {\aap},
         year = 1997,
        month = sep,
       volume = {325},
        pages = {866-870},
          doi = {10.48550/arXiv.astro-ph/9703069},
archivePrefix = {arXiv},
       eprint = {astro-ph/9703069},
 primaryClass = {astro-ph},
       adsurl = {https://ui.adsabs.harvard.edu/abs/1997A&A...325..866B}
}

@article{Kelner_Aha_2008,
  title = {Energy spectra of gamma rays, electrons, and neutrinos produced at interactions of relativistic protons with low energy radiation},
  author = {Kelner, S. R. and Aharonian, F. A.},
  journal = {Phys. Rev. D},
  volume = {78},
  issue = {3},
  pages = {034013},
  numpages = {16},
  year = {2008},
  month = {Aug},
  publisher = {American Physical Society},
  doi = {10.1103/PhysRevD.78.034013},
  url = {https://link.aps.org/doi/10.1103/PhysRevD.78.034013}
}

@article{Hummer_2010,
doi = {10.1088/0004-637X/721/1/630},
url = {https://dx.doi.org/10.1088/0004-637X/721/1/630},
year = {2010},
month = {aug},
publisher = {The American Astronomical Society},
volume = {721},
number = {1},
pages = {630},
author = {H\"ummer, S. and R\"uger, M. and Spanier, F. and Winter, W.},
title = {SIMPLIFIED MODELS FOR PHOTOHADRONIC INTERACTIONS IN COSMIC ACCELERATORS},
journal = {The Astrophysical Journal},
}

@article{UPRETI2024146,
title = {Bridging simulations of kink instability in relativistic magnetized jets with radio emission and polarisation},
journal = {Journal of High Energy Astrophysics},
volume = {44},
pages = {146-163},
year = {2024},
issn = {2214-4048},
doi = {https://doi.org/10.1016/j.jheap.2024.09.007},
url = {https://www.sciencedirect.com/science/article/pii/S2214404824000892},
author = {Nikita Upreti and Bhargav Vaidya and Amit Shukla},
}

@ARTICLE{English2016,
       author = {English, W. and Hardcastle, M.~J. and Krause, M.~G.~H.},
        title = "{Numerical modelling of the lobes of radio galaxies in cluster environments - III. Powerful relativistic and non-relativistic jets}",
      journal = {\mnras},
         year = 2016,
        month = sep,
       volume = {461},
       number = {2},
        pages = {2025-2043},
          doi = {10.1093/mnras/stw1407},
archivePrefix = {arXiv},
       eprint = {1606.03374},
 primaryClass = {astro-ph.HE},
       adsurl = {https://ui.adsabs.harvard.edu/abs/2016MNRAS.461.2025E}
}

@article{Mukherjee2020,
    author = {Mukherjee, Dipanjan and Bodo, Gianluigi and Mignone, Andrea and Rossi, Paola and Vaidya, Bhargav},
    title = {Simulating the dynamics and non-thermal emission of relativistic magnetized jets I. Dynamics},
    journal = {Monthly Notices of the Royal Astronomical Society},
    volume = {499},
    number = {1},
    pages = {681-701},
    year = {2020},
    month = {09},
    issn = {0035-8711},
    doi = {10.1093/mnras/staa2934},
    url = {https://doi.org/10.1093/mnras/staa2934},
    eprint = {https://academic.oup.com/mnras/article-pdf/499/1/681/33857350/staa2934.pdf},
}

@article{Reimer_2019,
doi = {10.3847/1538-4357/ab2bff},
url = {https://dx.doi.org/10.3847/1538-4357/ab2bff},
year = {2019},
month = {aug},
publisher = {The American Astronomical Society},
volume = {881},
number = {1},
pages = {46},
author = {Reimer, Anita and Böttcher, Markus and Buson, Sara},
title = {Cascading Constraints from Neutrino-emitting Blazars: The Case of TXS 0506+056},
journal = {The Astrophysical Journal},
}

@article{Schlickeiser_2010,
doi = {10.1088/1367-2630/12/3/033044},
url = {https://dx.doi.org/10.1088/1367-2630/12/3/033044},
year = {2010},
month = {mar},
publisher = {},
volume = {12},
number = {3},
pages = {033044},
author = {Schlickeiser, R and Ruppel, J},
title = {Klein–Nishina steps in the energy spectrum of galactic cosmic-ray electrons},
journal = {New Journal of Physics},
}

@article{Khangulyan_2014,
doi = {10.1088/0004-637X/783/2/100},
url = {https://dx.doi.org/10.1088/0004-637X/783/2/100},
year = {2014},
month = {feb},
publisher = {The American Astronomical Society},
volume = {783},
number = {2},
pages = {100},
author = {Khangulyan, D. and Aharonian, F. A. and Kelner, S. R.},
title = {SIMPLE ANALYTICAL APPROXIMATIONS FOR TREATMENT OF INVERSE COMPTON SCATTERING OF RELATIVISTIC ELECTRONS IN THE BLACKBODY RADIATION FIELD},
journal = {The Astrophysical Journal},
}

@article{Jones1968,
  title = {Calculated Spectrum of Inverse-Compton-Scattered Photons},
  author = {Jones, Frank C.},
  journal = {Phys. Rev.},
  volume = {167},
  issue = {5},
  pages = {1159--1169},
  numpages = {0},
  year = {1968},
  month = {Mar},
  publisher = {American Physical Society},
  doi = {10.1103/PhysRev.167.1159},
  url = {https://link.aps.org/doi/10.1103/PhysRev.167.1159}
}

@article{Xue_2019,
doi = {10.3847/1538-4357/ab4b44},
url = {https://dx.doi.org/10.3847/1538-4357/ab4b44},
year = {2019},
month = {nov},
publisher = {The American Astronomical Society},
volume = {886},
number = {1},
pages = {23},
author = {Xue, Rui and Liu, Ruo-Yu and Petropoulou, Maria and Oikonomou, Foteini and Wang, Ze-Rui and Wang, Kai and Wang, Xiang-Yu},
title = {A Two-zone Model for Blazar Emission: Implications for TXS 0506+056 and the Neutrino Event IceCube-170922A},
journal = {The Astrophysical Journal},

}

@article{Zacharias_2022,
    author = {Zacharias, M and Reimer, A and Boisson, C and Zech, A},
    title = {ExHaLe-jet: an extended hadro-leptonic jet model for blazars – I. Code description and initial results},
    journal = {Monthly Notices of the Royal Astronomical Society},
    volume = {512},
    number = {3},
    pages = {3948-3971},
    year = {2022},
    month = {03},
    issn = {0035-8711},
    doi = {10.1093/mnras/stac754},
    url = {https://doi.org/10.1093/mnras/stac754},
    eprint = {https://academic.oup.com/mnras/article-pdf/512/3/3948/43290956/stac754.pdf},
}

@book{Kai_Zuber,
author = {Zuber, Kai},
year = {2020},
month = {05},
pages = {},
title = {Neutrino Physics},
isbn = {9781315195612},
doi = {10.1201/9781315195612}
}

@ARTICLE{Gao_2019,
       author = {Gao, Shan and Fedynitch, Anatoli and Winter, Walter and Pohl, Martin},
        title = "{Modelling the coincident observation of a high-energy neutrino and a bright blazar flare}",
      journal = {Nature Astronomy},
         year = 2019,
        month = jan,
       volume = {3},
        pages = {88-92},
          doi = {10.1038/s41550-018-0610-1},
archivePrefix = {arXiv},
       eprint = {1807.04275},
 primaryClass = {astro-ph.HE},
       adsurl = {https://ui.adsabs.harvard.edu/abs/2019NatAs...3...88G}
}

@article{Marcowith:2020,
    author = "Marcowith, A. and Ferrand, G. and Grech, M. and Meliani, Z. and Plotnikov, I. and Walder, R.",
    title = "{Multi-scale simulations of particle acceleration in astrophysical systems}",
    eprint = "2002.09411",
    archivePrefix = "arXiv",
    primaryClass = "astro-ph.HE",
    reportNumber = "RIKEN-iTHEMS-Report-20",
    doi = "10.1007/s41115-020-0007-6",
    journal = "Liv. Rev. Comput. Astrophys.",
    volume = "6",
    pages = "1",
    year = "2020"
}

@article{MATTHEWS2020,
title = {Particle acceleration in astrophysical jets},
journal = {New Astronomy Reviews},
volume = {89},
pages = {101543},
year = {2020},
issn = {1387-6473},
doi = {https://doi.org/10.1016/j.newar.2020.101543},
url = {https://www.sciencedirect.com/science/article/pii/S1387647320300208},
author = {James H. Matthews and Anthony R. Bell and Katherine M. Blundell},
}

@article{Liu_2017,
doi = {10.3847/1538-4357/aa7410},
url = {https://dx.doi.org/10.3847/1538-4357/aa7410},
year = {2017},
month = {jun},
publisher = {The American Astronomical Society},
volume = {842},
number = {1},
pages = {39},
author = {Liu, Ruo-Yu and Rieger, F. M. and Aharonian, F. A.},
title = {Particle Acceleration in Mildly Relativistic Shearing Flows: The Interplay of Systematic and Stochastic Effects, and the Origin of the Extended High-energy Emission in AGN Jets},
journal = {The Astrophysical Journal},
}

@article{Kundu_2021,
doi = {10.3847/1538-4357/ac1ba5},
url = {https://dx.doi.org/10.3847/1538-4357/ac1ba5},
year = {2021},
month = {nov},
publisher = {The American Astronomical Society},
volume = {921},
number = {1},
pages = {74},
author = {Kundu, Sayan and Vaidya, Bhargav and Mignone, Andrea},
title = {Numerical Modeling and Physical Interplay of Stochastic Turbulent Acceleration for Nonthermal Emission Processes},
journal = {The Astrophysical Journal},
}

@unknown{Jerrim2025,
author = {Jerrim, Larissa and Shabala, Stas and Yates-Jones, Patrick and Krause, Martin and Turner, Ross and Stewart, Georgia and Power, Chris},
year = {2025},
month = {06},
pages = {},
title = {BRAiSE: synthetic polarisation in RMHD AGN jet simulations},
doi = {10.48550/arXiv.2506.19541}
}

@INPROCEEDINGS{Winter_2019,
       author = {{Winter}, W. and {Gao}, S.},
        title = "{Multi-messenger interpretation of neutrinos from TXS 0506+056}",
    booktitle = {36th International Cosmic Ray Conference (ICRC2019)},
         year = 2019,
       series = {International Cosmic Ray Conference},
       volume = {36},
        month = jul,
          eid = {1032},
        pages = {1032},
          doi = {10.22323/1.358.01032},
archivePrefix = {arXiv},
       eprint = {1909.06289},
 primaryClass = {astro-ph.HE},
       adsurl = {https://ui.adsabs.harvard.edu/abs/2019ICRC...36.1032W}
}

@article{Lipari2007,
  title = {Flavor composition and energy spectrum of astrophysical neutrinos},
  author = {Lipari, Paolo and Lusignoli, Maurizio and Meloni, Davide},
  journal = {Phys. Rev. D},
  volume = {75},
  issue = {12},
  pages = {123005},
  numpages = {24},
  year = {2007},
  month = {Jun},
  publisher = {American Physical Society},
  doi = {10.1103/PhysRevD.75.123005},
  url = {https://link.aps.org/doi/10.1103/PhysRevD.75.123005}
}

@article{Mattia2025, 
doi = {10.21105/joss.08448}, 
url = {https://doi.org/10.21105/joss.08448}, 
year = {2025}, 
publisher = {The Open Journal}, 
volume = {10}, 
number = {113}, 
pages = {8448}, 
author = {Mattia, Giancarlo and Crocco, Daniele and Melon Fuksman, David and Bugli, Matteo and Berta, Vittoria and Puzzoni, Eleonora and Mignone, Andrea and Vaidya, Bhargav}, 
title = {PyPLUTO: a data analysis Python package for the PLUTO code}, 
journal = {Journal of Open Source Software} 
}

@article{Cleary_2007,
doi = {10.1086/511969},
url = {https://doi.org/10.1086/511969},
year = {2007},
month = {may},
publisher = {},
volume = {660},
number = {1},
pages = {117},
author = {Cleary, K. and Lawrence, C. R. and Marshall, J. A. and Hao, L. and Meier, D.},
title = {Spitzer Observations of 3C Quasars and Radio Galaxies: Mid-Infrared Properties of Powerful Radio Sources},
journal = {The Astrophysical Journal},
}

@article{Tavecchio2013,
    author = {Tavecchio, F. and Pacciani, L. and Donnarumma, I. and Stamerra, A. and Isler, J. and MacPherson, E. and Urry, C. M.},
    title = {The far emission region of the γ-ray blazar PKS B1424–418},
    journal = {Monthly Notices of the Royal Astronomical Society: Letters},
    volume = {435},
    number = {1},
    pages = {L24-L28},
    year = {2013},
    month = {07},
    issn = {1745-3925},
    doi = {10.1093/mnrasl/slt087},
    url = {https://doi.org/10.1093/mnrasl/slt087},
    eprint = {https://academic.oup.com/mnrasl/article-pdf/435/1/L24/54658031/mnrasl_435_1_l24.pdf},
}

@article{Ghisellini_Tavecchio_2009,
    author = {Ghisellini, G. and Tavecchio, F.},
    title = {Canonical high-power blazars},
    journal = {Monthly Notices of the Royal Astronomical Society},
    volume = {397},
    number = {2},
    pages = {985-1002},
    year = {2009},
    month = {07},
    issn = {0035-8711},
    doi = {10.1111/j.1365-2966.2009.15007.x},
    url = {https://doi.org/10.1111/j.1365-2966.2009.15007.x},
    eprint = {https://academic.oup.com/mnras/article-pdf/397/2/985/2943748/mnras0397-0985.pdf},
}

@article{Ghisellini2008,
    author = {Ghisellini, G. and Tavecchio, F.},
    title = {The blazar sequence: a new perspective},
    journal = {Monthly Notices of the Royal Astronomical Society},
    volume = {387},
    number = {4},
    pages = {1669-1680},
    year = {2008},
    month = {07},
    issn = {0035-8711},
    doi = {10.1111/j.1365-2966.2008.13360.x},
    url = {https://doi.org/10.1111/j.1365-2966.2008.13360.x},
    eprint = {https://academic.oup.com/mnras/article-pdf/387/4/1669/3815047/mnras0387-1669.pdf},
}

\appendix
\twocolumngrid
%--------------------------------------------------------------------------------------------------------------------
\section{Initial profiles for jet simulations and multi-zone parameters}
\label{A_Profiles}
For the initial profiles of the different MHD variables in our jet simulations, we adopt the steady-state axisymmetric solutions in cylindrical coordinates for the
ideal RMHD equations from \citet{Bodo_2019}. 
Here, for convenience, we provide the profiles of the important dynamical quantities. 
These profiles have been obtained by ensuring that the initial plasma column is in radial equilibrium.

The radial profile for the azimuthal velocity is expressed as follows,
\begin{equation}
    v_\phi^2 = \frac{r^2 \Gamma_{\rm c}^2 \Omega_{\rm c}^2}{\Gamma_z^2 \left[1 + r^2 \Gamma_{\rm c}^2 \Omega_{\rm c}^2 \exp\left(-r^4/a^4\right)\right]} \exp\left(-\frac{r^4}{a^4}\right),
\end{equation}
where $r$ is the cylindrical radius and the characteristic concentration radius for the electric current in the jet is taken as $a = 0.6$. 
The initial magnetic field configuration with a pitch parameter $\mathcal{P} = rB_z/B_{\phi}$ is defined as,
\begin{equation}
   B_z^2 = B_{\rm zc}^2 - (1 - \alpha) H_{\rm c}^2 \sqrt{\pi} a^{-2} \, \text{erf}\left(\frac{r^2}{a^2}\right),
\end{equation}
\begin{equation}
    B_\phi = \frac{-v_\phi v_z B_z \mp \sqrt{v_\phi^2 B_z^2 + H^2 (1 - v_z^2)}}{1 - v_z^2},
\end{equation}
where $B_z$ and $B_{\phi}$ are the poloidal and the toroidal magnetic fields, respectively, while the radial field component is $B_r = 0$.
With $B_{\rm zc}$ we denote the poloidal magnetic field along the central axis, \text{erf} is the error function, and $H$ is defined as $H^2 = B_\phi^2 - E_r^2 = [1 - \exp(- r^4/a^4)]H_{\rm c}^2/r^2$ where $H_{\rm c}$ is the value of $H$ along the central axis and $E_r = v_z B_\phi - v_\phi B_z$. We take $\alpha = 1$ where the parameter $\alpha$ represents the strength of rotation for the flow.

Using the aforementioned variable profiles, as well as the setup and initial conditions specified in Section~\ref{Numerical_Approach}, the RMHD jet simulations were carried out.  
%--------------------------------------------------------------------------
\section{Details of the EC simulations}
\label{Appendix_EC}
For the EC simulations considering BLR (\textit{Ref\_g5\_BLR}) and DT (\textit{Ref\_g5\_DT}) as the respective external photon fields, the adopted dilution factor is motivated by \cite{Khangulyan_2014}. We approximate the BLR to be a spherical photon field with a characteristic radius of $R^{\rm AGN}_{\rm BLR}$ from the SMBH, while we consider the DT to be an annular ring at a characteristic radius $R^{\rm AGN}_{\rm DT}$ (at the midpoint of the annulus). $R^{\rm AGN}_{\rm BLR}$ and $R^{\rm AGN}_{\rm DT}$ scale proportional to $(L^{\rm AGN}_{\rm \, disc})^{1/2}$ as defined in \citet{Ghisellini2008}. Considering an emission zone at a distance $D^{\rm AGN}_z$, the dilution factor multiplied to the AGN frame energy density for BLR photon field can be expressed as,
\begin{equation}
    \kappa^{\rm AGN}_{\rm BLR} =  \frac{\Delta \Omega}{4 \pi} = \frac{(R
    ^{\rm AGN}_{\rm BLR})^2}{4 (D_z^{\rm AGN})^2},
\end{equation}
considering $D_z^{\rm AGN} >> R^{\rm AGN}_{\rm BLR}$.
\begin{figure}
    \centering
    \includegraphics[width=1\linewidth]{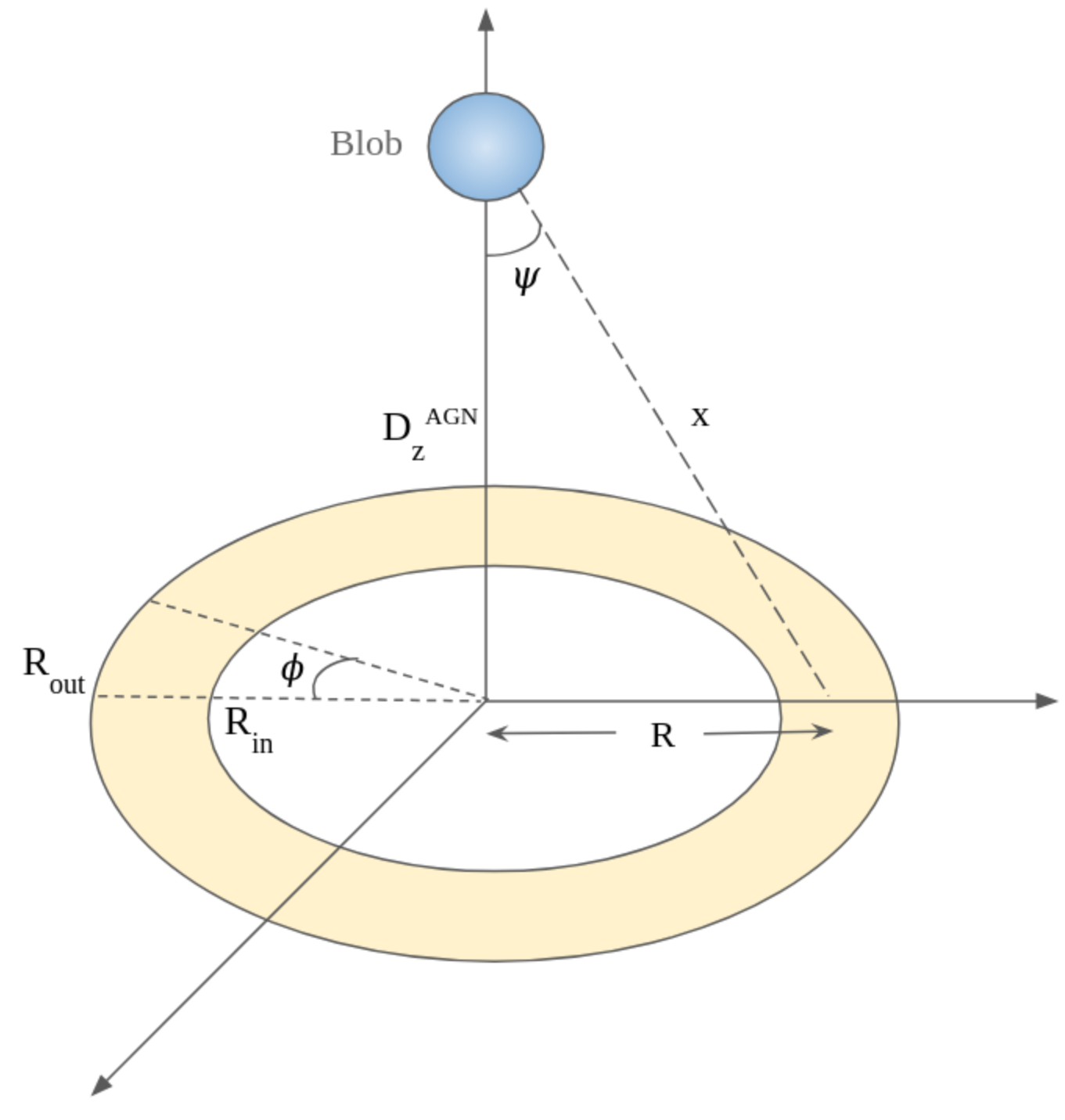}
    \caption{Illustration of the geometry used for dusty Torus.}
    \label{fig:placeholder}
\end{figure}

 For the DT photon field, distance of the zone from any point on the Dusty torus is $x \approx [ \, (R)^2 + (D_z^{\rm AGN})^2 \, ]^{1/2}$ (R is the radial distance to the DT element), and

\begin{equation}
    d\Omega = \frac{\cos\psi}{x^2}\,dA
= \frac{(D_z^{\rm AGN}/x)}{x^2}\,\big(R\,d\phi\,dR\big),
\end{equation} 
where $\psi$ is the angle between $D_z^{\rm AGN}$ and $x$. Doing the necessary integrations and approximating for $D_z^{\rm AGN} >> R_{\rm DT}$, the corresponding dilution factor can be expressed as \footnote{We take inner and outer radius of DT to be $R_{in}\approx R^{\rm AGN}_{\rm DT} - (R^{\rm AGN}_{\rm DT}/2)$ and $R_{out}\approx R^{\rm AGN}_{\rm DT} + (R^{\rm AGN}_{\rm DT}/2)$} 
\begin{equation}
    \kappa^{\rm AGN}_{\rm DT} =  \frac{\Delta \Omega}{4 \pi} \, \approx \, \frac{R_{out}^2 - R_{in}^2}{4\, (D_z^{\rm AGN})^2} \, \approx \, \frac{(R^{\rm AGN}_{\rm DT})^2}{2\, (D_z^{\rm AGN})^2}
\end{equation}
The dilution factor provides an effective estimate of the target photon field available for inverse-Compton scattering at the emission site, by rescaling the photon field energy density, accounting for the external photon field geometry and its separation from the dissipation region. For computational simplicity, we treat the resulting photon field as a geometrically diluted but isotropic radiation field and use the corresponding Lorentz transformation (as described in Equation (A.9) in \cite{Nigro2022}). A full anisotropic treatment of EC scattering however, is currently out of scope for this work. 

%---------------------------------------------------------------------------
\section{SED Tests} 
\label{Appendix}
To validate the multi-zone framework algorithm (Section~\ref{MZ_sec}), it is necessary to first confirm that the SEDs generated for the individual zones following this algorithm are accurate. In particular, we want to corroborate that extracting average fluid parameters and the net electron distribution from the RMHD simulations for a given zone and using them as inputs to a one zone lepto-hadronic (OZLH) code, can produce viable SEDs for individual emission zones. A detailed consistency test of the inverse Compton component for a spherical blob will be presented in a subsequent work (Sharma et. al. - \textit{In prep}).\\

\noindent We perform two test simulations with different initial conditions:
\begin{enumerate}
    \item \textbf{\textit{blob} simulation} - In this test, the blob is initialized with an initial uniform bulk Lorentz factor of $\Gamma_{\rm blob} = 10$ inside the sphere with velocity of the form $v=(0,0,v_z)$. The blob has a homogeneous poloidal magnetic field along z-axis with a magnitude of $B_{\rm z,blob} = 2.92 \times 10^{-2}$ G and has a density contrast of $\rho_{\rm blob}/\rho_{\rm amb} = 10^{-3}$ where $\rho_{\rm blob} = \rho_{0} = 1.661 \times 10^{-26}$\ g cm$^{-2}$. Its Doppler factor is $\delta_{\rm blob} \approx 19.36$ with $\theta_{\rm los} = 1^\circ$. The blob has a radius $r_{\rm blob} = 1$\ pc in a simulation domain with spatial extents of $x \in [-L_x,L_x]$, $y \in [-L_y,L_y] $, $z \in [0,L_z]$ where $L_x = L_y = L_z = 10\ l_{\rm b} = 10$\ pc with a domain spatial grid resolution of $(400 \times 400 \times 200)$. The net electron distribution inside the blob is a perfect power-law with an index of $\alpha^{\rm e} = 2.4$, $\gamma_{\rm min}^{\rm e} = 10^2$, and $\gamma_{\rm max}^{\rm e} = 10^8$. 
    
    \item \textbf{\textit{zone} simulation} - This test is performed for a spherical zone taken at the base of the $\Gamma_{\rm c,ini} = 10$ jet simulation at time $t=743$ years. This zone has non uniform density, an an average zonal magnetic field of $B_{\rm zone} \approx 2.46 \times 10^{-2}$\, G, and average Doppler factor $\delta_{\rm zone} \approx 2.43$ with $\theta_{\rm los} = 1^\circ$. The net electron distribution for the zone, obtained from Equation \ref{Eq_net_dist} is fitted with a PLEC distribution  having $\alpha^{\rm e}_{\rm fit} = 2.62 $, $\gamma_{\rm min}^{\rm e} = 50.5$, $\gamma_{\rm max}^{\rm e} = 3.09 \times 10^8$ and $\gamma^{\rm e}_{\rm break} = 8.19 \times 10^{7}$. 
\end{enumerate}

For the \textit{blob} and the \textit{zone} simulations, we assume the proton-to-electron number density ratio to be very small, 
$\eta_{\rm zone} = 10^{-33}$, since we are only interested in generating the electron synchrotron component. 

In Figure~\ref{Appendix_blob_zone} (\textit{top panel}), we compute the synthetic synchrotron SEDs for the \textit{blob} simulations for different codes typically used for OZLH modeling for AGN jets 
like \texttt{Katu} \citep{Katu2020}, \texttt{agnpy} \citep{Nigro2022} and compare with the synthetic synchrotron SED obtained from hybrid PLUTO simulations \citep{Vaidya_2018}. 
The fluxes in these simulations are obtained using the formulations provided in \citep{DermerMenon, Aharonian2010}.
We see that in the \textit{blob} simulation, the electron distribution has an almost perfect power-law fit and uniform Doppler factors across all cells, 
the resultant synchrotron SED from \texttt{Katu} and \texttt{agnpy} match very well with the synthetic PLUTO SED.

\begin{figure}
    \centering
    \begin{minipage}{0.49\textwidth} 
        \centering
        \includegraphics[width=0.95\linewidth]{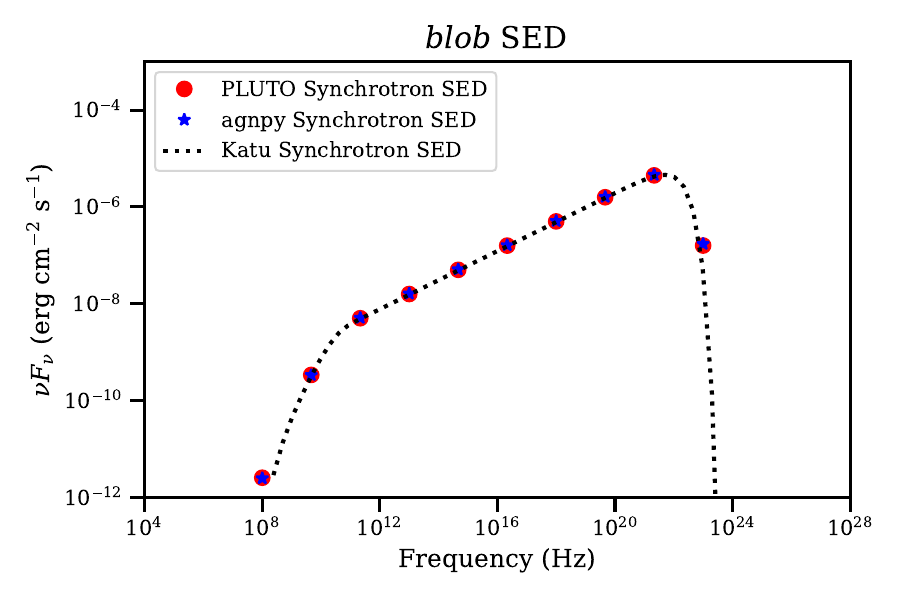}
    \end{minipage}
    \begin{minipage}{0.49\textwidth} 
        \centering
        \includegraphics[width=0.95\linewidth]{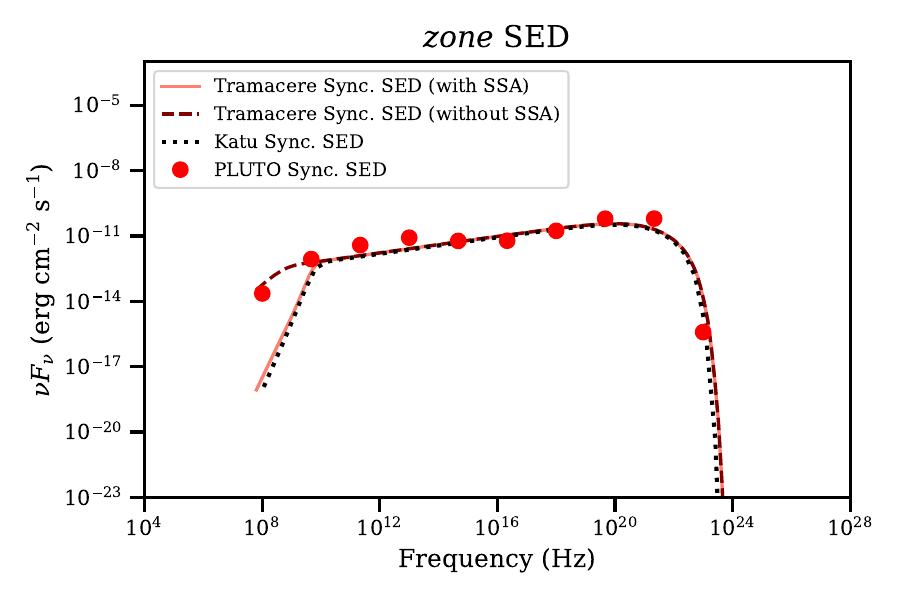}
    \end{minipage}
   \caption{\small \textit{Top panel:} Comparison of synchrotron SEDs for the \textit{blob} simulations generated by PLUTO's particle module (red dots), \texttt{Katu} (black dashed) and the \texttt{agnpy} (blue start). \textit{Bottom panel:} Synchrotron SED comparison for the \textit{zone} simulations generated by PLUTO's particle module (red dots), \texttt{Katu} ( black dashed), and the Tramacere tool (orange solid for with SSA, and brown dashed for without SSA case). Note, the OLZH code \texttt{Katu} also accounts for SSA.
   }
   \label{Appendix_blob_zone}
\end{figure}

The \textit{bottom panel} of Figure~\ref{Appendix_blob_zone}, we compare the synthetic synchrotron SEDs generated from \textit{zone} simulations. Here, synthetic SEDs obtained from OZLH codes like \texttt{Katu} and Tramacere \citep{Trama3, Trama2, Trama1} are compared with the SED obtained from hybrid PLUTO simulation. 
We observe that the SED from PLUTO (red dots) deviates for certain points from the SEDs made with the OZLH codes. 
At the low frequency $\nu \lesssim 10^{11}\,\rm{Hz}$, the deviation is primarily due to absence of synchrotron self absorption (SSA) in the hybrid PLUTO framework. 
Further, as the hybrid PLUTO framework, uses the exact spectra to compute the SED instead of a fitted one, we see slight deviations from those obtained from OLZH codes which uses a well-defined functional form for particle distribution. 

Additionally, there are turbulent signatures within the zones which leads to variations in the Doppler factor, causing the emissivity from different parts of the zone being boosted somewhat differently.
Wn contrast, the SEDs produced by OZLH codes assume a uniform Doppler factor, which we determine by computing a weighted average within the zone. 
Such an approximation results in some mismatch, between the flux generated using PLUTO which can resolve the variations within the zone, and the OZLH codes which assumes a uniform zone.

\end{document}